\documentclass{lmcs}
\pdfoutput=1

\usepackage{lastpage}
\lmcsdoi{20}{4}{22}
\lmcsheading{}{\pageref{LastPage}}{}{}%
{Mar.~30,~2023}{Dec.~10,~2024}{}

\usepackage[utf8]{inputenc}

\usepackage{amsmath}
\usepackage[sort, noadjust]{cite}
\usepackage{pb-diagram}
\usepackage{stmaryrd}
\usepackage{tikz}
\usetikzlibrary{matrix,arrows,decorations.pathmorphing,cd,automata,fadings}
\tikzset{->, auto, >=stealth', font=\small}
\tikzset{state/.style={shape=circle, draw, fill=black, initial text=,
    initial distance=2.3ex, accepting distance=2.3ex,
    inner sep=.4mm, minimum size=1.3mm}}
\tikzset{accepting/.style=accepting by arrow}
\usepackage{tikz-cd}

\DeclareMathOperator{\colim}{colim}
\DeclareMathOperator{\id}{id}
\DeclareMathOperator{\im}{im}

\newcommand*\cat[1]{\text{\textup{\textsf{#1}}}}
\newcommand*\Set{\cat{Set}} 
\newcommand*\HDA{\cat{HDA}}
\newcommand*\IHDA{\cat{iHDA}}

\newcommand*\evord{\dashrightarrow}
\newcommand*\ibullet{\vcenter{\hbox{\tiny $\bullet$}}}

\DeclareMathOperator\iPoms{\mathsf{iPoms}}
\DeclareMathOperator\iiPoms{\mathsf{iiPoms}}
\DeclareMathOperator{\Path}{\mathsf{P}}
\newcommand*\Qath[1]{\Path_{#1}} 
\newcommand*\Qathft[3]{\Qath{#1}(#2,#3)} 

\DeclareMathOperator{\Weld}{\mathsf{W}}
\DeclareMathOperator{\Spider}{\mathsf{Sp}}
\DeclareMathOperator{\CSpider}{\mathsf{CSp}}

\DeclareMathOperator{\Cell}{\mathsf{Cell}}
\DeclareMathOperator{\Lang}{\mathsf{Lang}}
\DeclareMathOperator{\Id}{\mathsf{Id}}
\DeclareMathOperator{\Cl}{\mathsf{Cl}}
\DeclareMathOperator{\Res}{\mathsf{Res}}
\DeclareMathOperator{\Cyl}{C}

\DeclareMathOperator{\size}{\#}
\DeclareMathOperator{\wid}{\mathsf{wid}}
\DeclareMathOperator{\ev}{\mathsf{ev}}
\DeclareMathOperator{\Iev}{\mathsf{iev}}

\newcommand*\conceq{\cong}

\DeclareMathOperator{\ZZ}{\mathsf{Z}}
\DeclareMathOperator{\VV}{\mathsf{V}}

\def\Snake{\mathfrak{S}}

\newcommand*\isquare{{\textup{I}}\mkern-1mu \square}
\newcommand*\sq{\square}
\newcommand*\isq{\isquare}
\newsavebox{\lmcsQEDSymbolBold}
\savebox{\lmcsQEDSymbolBold}{\begin{tikzpicture}[baseline=-0.5pt]
  \draw [line width=1pt] (0,0) rectangle (6.5pt,6.5pt);
\end{tikzpicture}}
\newcommand*\fullsq{\usebox{\lmcsQEDSymbolBold}}
\newcommand*\fullisq{{\textup{I}}\fullsq}
\newcommand*\arrO[1]{\mathrel{\nearrow^{#1}}}
\newcommand*\arrI[1]{\mathrel{\searrow_{#1}}}
\newcommand*\intord{\dashrightarrow}
\newcommand*\subsu{\sqsubseteq}

\newcommand*\ilo[3]{{}_{#1}{#2}_{#3}}

\newcommand*\ie{\textit{i.e.}}

\newcommand*\beInt{b}
\newcommand*\enInt{e}

\newcommand*\down{\mathord{\downarrow}}

\newcommand*\op{\textup{op}}
\newcommand*\fullsqz{\fullsq_\textup{Z}}
\newcommand*\exec{%
  \raisebox{1pt}{%
    \begin{tikzpicture}[x=.8ex,y=1ex,-]
      \draw (0,0) -- (1,0) -- (1,1) -- (2,1);
    \end{tikzpicture}}}

\newcommand*\sqz{\sq_\textup{Z}}
\newcommand*\bang{\mathord{!}}

\newcommand*\longrightrightarrows{\mathrel{\substack{\textstyle\longrightarrow\\[-0.6ex]
      \textstyle\longrightarrow}}}

\newcommand*\src{\mathsf{src}}
\newcommand*\tgt{\mathsf{tgt}}

\let\phi\varphi
\let\epsilon\varepsilon

\newcommand*\sqgra{\square_\textup{G}}

\def\sfF{\mathsf{F}}

\newcommand*\spO[2]{[{#1}{\backslash}{#2}]}
\newcommand*\spI[2]{[{#1}{/}{#2}]}

\newcommand{\sbt}{\,\begin{picture}(-1,1)(-1,-3)\circle*{3}\end{picture}\ }

\title{Kleene Theorem for Higher-Dimensional Automata}

\author[U.~Fahrenberg]{Uli Fahrenberg}[a]
\author[C.~Johansen]{Christian Johansen}[b]
\author[G.~Struth]{Georg Struth}[c,d]
\author[K.~Ziemia{\'n}ski]{Krzysztof Ziemia{\'n}ski}[e]

\address{EPITA Research Laboratory (LRE), France}
\email{uli@lrde.epita.fr}

\address{NTNU, Norway}
\email{christian.johansen@ntnu.no}

\address{University of Sheffield, UK}
\email{g.struth@sheffield.ac.uk}
\address{Collegium de Lyon, France}

\address{University of Warsaw, Poland}
\email{ziemians@mimuw.edu.pl}

\begin{document}
\begin{abstract}
  We prove a Kleene theorem for higher-dimensional automata. It states
  that the languages they recognise are precisely the rational
  subsumption-closed sets of finite interval pomsets. The rational
  operations on these languages include a gluing composition, for
  which we equip pomsets with interfaces. For our proof, we introduce
  higher-dimensional automata with interfaces, which are modelled as
  presheaves over labelled precube categories, and develop tools and
  techniques inspired by algebraic topology, such as cylinders and
  (co)fibrations.  Higher-dimensional automata form a general model of
  non-interleaving concurrency, which subsumes many other approaches.
  Interval orders are used as models for concurrent and distributed
  systems where events extend in time. Our tools and techniques may
  therefore yield templates for Kleene theorems in various models and
  applications.
\end{abstract}

\maketitle


\section{Introduction}
\label{s:intro}

Higher-dimensional automata (HDAs) were introduced by Pratt and van
Glabbeek as a general geometric model for non-interleaving concurrency
\cite{Pratt91-geometry, Glabbeek91-hda}.  They support autoconcurrency
and events with duration or structure, whereas events in interleaving
models must be instantaneous. They subsume, for example, event
structures and safe Petri nets \cite{DBLP:journals/tcs/Glabbeek06},
while asynchronous transition systems and standard automata correspond
to two-dimensional and one-dimensional HDAs,
respectively~\cite{Goubault02-cmcim}.  We have recently used van
Glabbeek's (execution) paths~\cite{DBLP:journals/tcs/Glabbeek06} to
relate HDAs with certain languages of interval pomsets~\cite{Hdalang}.
Yet a precise description of the relationship between HDAs and these
languages in terms of a Kleene theorem -- a key theorem for any type of
automaton -- has so far been missing. Our main contribution lies in the
formalisation and proof of such a theorem.

HDAs consist of cells and lists of concurrent events that are active
in them. Zero-dimensional cells represent states in which no event is
active, while $1$-dimensional cells represent transitions in which
exactly one event is active -- as for standard automata. Higher
$n$-dimensional cells model higher transitions in which $n$ concurrent
events are active.  Figure~\ref{fi:hda-cylinder} shows an example of
an HDA with cells of dimension $\le 2$.  In its $2$-dimensional cells
$x$ and $y$, the concurrent events
$\left[ \begin{smallmatrix} a\\b \end{smallmatrix}\right]$ and
$\left[ \begin{smallmatrix} a\\\vphantom{b}c \end{smallmatrix}\right]$
are active, respectively. Cells at any dimension may serve as start
and accept cells. In Figure \ref{fi:hda-cylinder}, these are labelled
with $\bot$ and $\top$.

Lower-dimensional cells or faces are attached to higher-dimensional
ones by face maps. These maps also indicate when particular events
start or end their activity. In Figure \ref{fi:hda-cylinder}, the
lower face $\delta^0_a(x)$ of $x$ forms the lower $b$-transition in
which $a$ is not yet active; its upper face $\delta^1_a(x)$ forms the
upper $b$-transition in which $a$ is no longer active.  Intuitively,
events can thus be terminated in upper faces and unstarted in lower
faces, where ``unstarted'' refers to the dual of ``terminated''. The
cubical structure of cells is determined by relations between faces.

\begin{figure}
  \centering
  \begin{tikzpicture}[x=1.8cm, y=1.8cm]
    \begin{scope}
      \fill[-, fill=black!12] (-0.7,0)--(-0.7,1)--(0.7,1)--(0.7,0)--(-0.7,0);
      \draw[fill=black!18] (0,0) ellipse (0.7 and 0.2);
      \draw[fill=black!6] (0,1) ellipse (0.7 and 0.2);
      \draw[-] (0.7,0) -- (0.7,1);
      \draw[-] (-0.7,0) -- (-0.7,1);
      \node[state] (00) at (-0.45,-0.14) {};
      \node [below] at (00)  {$\bot$}; 
      \node[state] (10) at (0.45,0.14) {};
      \node[state] (01) at (-0.45,1-0.14) {};
      \node[above] at (01) {$\top$}; 
      \node[state] (11) at (0.45,1+0.14) {};
      \path (00) edge node[left] {$a$} (01);
      \path (10) edge node[left] {$a$} (11);
      \node at (0.01,-0.34) {$b$};
      \node at (-0.01,0.34) {$c$};
      \node at (0.01,1-0.07) {$b$};
      \node at (-0.01,1+0.34) {$c$};
      \draw (-0.49,-0.142)--(-0.48,-0.147);
      \draw (0.49,0.142)--(0.48,0.147);
      \draw (-0.49,1-0.142)--(-0.48,1-0.147);
      \draw (0.49,1.142)--(0.48,1.147);
    \end{scope}
  
    \begin{scope}[shift={(1.8,0)}]
      \path[fill=black!10] (0,0) to (2,0) to (2,1) to (0,1);
      \fill[-, fill=black!10, path fading=west] (0,0) -- (-.3,0) -- (-.3,1) --
      (0,1);
      \draw[-, path fading=west] (0,0) -- (-.3,0);
      \draw[-, path fading=west] (0,1) -- (-.3,1);
      \fill[-, fill=black!10, path fading=east] (2,0) -- (2.3,0) -- (2.3,1) --
      (2,1);
      \draw[-, path fading=east] (2,0) -- (2.3,0);
      \draw[-, path fading=east] (2,1) -- (2.3,1);
      \node[state] (00) at (0,0) {};
        \node[below] at (00) {$\bot$};
      \node[state] (10) at (1,0) {};
      \node[state] (20) at (2,0) {};
      \node[below] at (20) {$\bot$};
      \node[state] (01) at (0,1) {};
      \node[above] at (01) {$\top$};
      \node[state] (11) at (1,1) {};
      \node[state] (21) at (2,1) {};
      \node[above] at (21) {$\top$}; 
      \node[align=center] at (.5,.45) {$\left[ \begin{smallmatrix}
            a\\b \end{smallmatrix}\right]$ \\[1ex] $\vphantom{y}x$};
      \node[align=center] at (1.5,.45) {$\left[ \begin{smallmatrix}
            a\\\vphantom{b}c \end{smallmatrix}\right]$ \\[1ex] $y$};
      \path (10) edge node[below] {$\vphantom{b}c$} (20);
      \path (00) edge node[below] {$b$} (10);
      \path (11) edge node[above] {$\vphantom{b}c$} (21);
      \path (01) edge node[above] {$b$} (11);
      \path (00) edge node[left] {$a$} (01);
      \path (10) edge (11);
      \path (20) edge node[right] {$a$} (21);
    \end{scope}
    
    \begin{scope}[shift={(5.1,0)}]
      \path[fill=black!10] (0,0) to (2,0) to (2,1) to (0,1);
      \fill[-, fill=black!10, path fading=west] (0,0) -- (-.3,0) -- (-.3,1) --
      (0,1);
      \draw[-, path fading=west] (0,0) -- (-.3,0);
      \draw[-, path fading=west] (0,1) -- (-.3,1);
      \fill[-, fill=black!10, path fading=east] (2,0) -- (2.3,0) -- (2.3,1) --
      (2,1);
      \draw[-, path fading=east] (2,0) -- (2.3,0);
      \draw[-, path fading=east] (2,1) -- (2.3,1);
      \node[state,] (00) at (0,0) {};
         \node[below] at (00) {$\bot$};
      \node[state] (10) at (1,0) {};
      \node[state] (20) at (2,0) {};
        \node[below] at (20) {$\bot$};
        \node[state] (01) at (0,1) {};
        \node[above] at (01) {$\top$};
      \node[state] (11) at (1,1) {};
      \node[state] (21) at (2,1) {};
      \node[above] at (21) {$\top$};
      \path (10) edge node[below] {$\vphantom{b}c$} (20);
      \path (00) edge node[below] {$b$} (10);
      \path (11) edge node[above] {$\vphantom{b}c$} (21);
      \path (01) edge node[above] {$b$} (11);
      \path (00) edge node[left] {$a$} (01);
      \path (10) edge (11);
      \path (20) edge node[right] {$a$} (21);
      \path[-, very thick, orange] (00) edge node[right, pos=.6] {$\alpha_1$} (01);
      \path[-, very thick, cyan] (00) edge node[below, pos=.35] {$\alpha_2$} (21);
      \path[-, very thick, teal] (00) edge (10) (10) edge
      node[right, pos=.4] {$\alpha_3$} (21);
    \end{scope}
  \end{tikzpicture}
  \caption{HDA with two 2-dimensional cells $x$ and $y$ modelling the
    parallel execution of $a$ and $(b c)^*$ on the left, an unfolded view in the
    middle and three accepting paths of this automaton on the right.}
  \label{fi:hda-cylinder}
\end{figure}

Executions of HDAs are (higher-dimensional) paths
\cite{DBLP:journals/tcs/Glabbeek06}: sequences of cells, which
indicate where events start or terminate.  Every path $\alpha$ is
characterised by the temporal precedences between the intervals of
activity of the concurrent events $\ev(\alpha)$ that occur in it. This
naturally induces interval orders, as is further explained in
Section~\ref{s:overview}.  In addition, $\ev(\alpha)$ is equipped with
source and target interfaces, which model events that are already
active in the initial cell of $\alpha$ or still active in its final
cell, respectively, and a secondary event order, which captures the
list order of events in cells and is useful for coordinating the
composition of paths along the interfaces at their ends.

The isomorphism classes of such labelled posets with interfaces and an
event order form \emph{ipomsets}.  The language of an HDA is then
related to the set of (interval) ipomsets associated with all its
accepting paths -- from start cells to accept cells~\cite{Hdalang}.
Languages of HDAs must in particular be down-closed with respect to
less concurrent executions, modelled by a subsumption preorder and
restricted to interval ipomsets. This motivates the definition of
languages as subsumption-closed sets of interval ipomsets.

Kleene theorems usually require a notion of \emph{rational}
language. Ours is based on the union $\cup$, gluing (serial)
composition~$*$, parallel composition $\|$, and (serial) Kleene plus
$^+$ of languages. These definitions are not entirely straightforward,
as down-closure and the interval property must be preserved. In
particular, in the presence of interfaces, gluing composition is more
complicated than, for instance, the standard series composition of
pomsets.  We consider finite HDAs only and thus can neither include
the parallel Kleene star nor the full serial Kleene star as a rational
operation: the latter contains the identity language, which would
require an HDA of infinite dimension.

Our Kleene theorem shows that the rational languages are precisely the
regular ones (recognised by finite HDAs).  To show that regular
languages are rational, we translate the cells of an HDA into a
standard automaton and reuse one direction of the classical Kleene
theorem. Proving that rational languages are regular is harder.
Regularity of $\cup$ is straightforward, and for $\|$, the
corresponding operation on HDAs is a tensor product. Yet $*$ and $^+$
require an intricate gluing operation on HDAs along higher-dimensional
cells and ultimately a new variant of HDAs.

Beyond the Kleene theorem for HDAs, three contributions seem of
independent interest.  We model HDAs as presheaves on novel precube
categories, where events and labels feature in the base category.
These are equivalent to standard
HDAs~\cite{DBLP:journals/tcs/Glabbeek06}, but constructions related to
the Kleene theorem become simpler, the precedence ordering of events
with respect to the beginning and end of their activity becomes more
transparent, and the relationship between iposets and precubical sets
becomes clearer.

We also introduce HDAs with interfaces (iHDAs), which may assign
events to source or target interfaces.
This allows us to indicate
events that cannot terminate in a given iHDA by
assigning them to a target interface, or those that
cannot be unstarted by
assigning them to a source interface, and to
keep track of them across the iHDA.

Using operations of resolution and closure, we show
that any HDA can be converted into an equivalent iHDA with respect to
language recognition and vice versa.  Both variants play a role in our
proofs, and we frequently switch between them.

Another tool in our proof of the Kleene theorem is motivated by
algebraic topology.  We introduce cylinder objects and show that each
map between (i)HDAs can be decomposed into an (initial or final)
inclusion followed by a (future or past) path-lifting map.  This
allows us to pull apart start and accept cells of iHDAs when dealing
with serial compositions and loops -- we refer to the resulting iHDAs
as ``proper''.

The remainder of this article has three main parts. In its first part,
Sections~\ref{s:intro} to~\ref{s:languages}, we introduce HDAs and
their languages to the point where we can state and discuss the Kleene
theorem for HDAs in its second part, Section~\ref{s:KleeneTheorem}. In
the third part, Sections~\ref{s:reg2rat} to~\ref{s:Spider}, we
introduce more advanced concepts and develop our main proofs.  More
specifically, Section \ref{s:overview} contains a detailed semi-formal
overview of the relationship between HDAs and their languages. In
Section \ref{s:hdas} we introduce precube categories and formalise
HDAs as presheaves on them. In Section \ref{s:ipomsets} we define
ipomsets and their languages, while in Section \ref{s:languages} we
define executions of HDAs and languages recognised by them.

Section
\ref{s:KleeneTheorem} constitutes the central part of this paper.
Here, we formulate the Kleene theorem for HDAs and provide a roadmap
towards its proof.  We then show in Section \ref{s:reg2rat} how HDAs
can be converted into classical finite state automata over an extended
alphabet and use this construction together with the standard Kleene
theorem to prove that regular languages are rational.  In Section
\ref{s:Tracks} we introduce track objects, which provide an
alternative description of executions of HDAs.  Tensor products of
HDAs are defined in Section \ref{s:tensor}, and they are used to show
that parallel compositions of regular languages are regular.

Higher-dimensional automata with interfaces are introduced in Section
\ref{s:ihda}, and translations between HDAs and iHDAs are discussed in
Section \ref{s:hda-ihda}.  In Section \ref{s:cofib} we introduce
cylinders for iHDAs. This construction allows us to replace iHDAs by
proper ones without changing their languages.  Finally, in Sections
\ref{s:Toolbox} and \ref{s:SeqComp}, we use proper iHDAs to prove that
gluings of regular languages yields regular languages. These arguments
are further refined in Section \ref{s:Spider} to show an analogous
result for the Kleene plus.

This article is a complete revision of a previous conference
paper~\cite{DBLP:conf/concur/FahrenbergJSZ22}, published at CONCUR,
with concepts and notation reconsidered, an overview section
(Section~\ref{s:overview}), pictures and examples added, and in
particular the complete technical development leading to the Kleene
theorem and its proof, which could only be sketched in the CONCUR
paper.


\section{Overview}
\label{s:overview}

Higher-dimensional automata generalise standard finite state automata.
Let $\Sigma$ stand for a fixed alphabet of actions (of a
concurrent system).

A \emph{higher-dimensional automaton} (HDA) $X$ is defined by the following data:
\begin{itemize}
\item a set $\Cell(X)$ of \emph{cells};
\item for each cell $x\in \Cell(X)$ a totally ordered set of
  $\Sigma$-labelled events $\ev(x)$;
\item for each cell $x\in \Cell(X)$ and disjoint subsets
  $A,B\subseteq \ev(x)$ a cell $\delta_{A,B}(x)$, called a \emph{face}
  of $x$, with set of events $\ev(x)\setminus(A\cup B)$;
\item the identity $\delta_{\emptyset,\emptyset}(x)=x$ for each $x\in\Cell(X)$;
\item the equality
  $\delta_{A,B}(\delta_{C,D}(x))=\delta_{A\cup C, B\cup D}(x)$
  whenever $A,B,C,D\subseteq \ev(x)$ are disjoint;
  \item sets $X_\bot,X_\top\subseteq \Cell(X)$ of \emph{start cells}
    and \emph{accept cells}, respectively.
  \end{itemize}
  HDAs without start and accept cells are known as \emph{precubical
    sets}.

  Cells correspond to transitions of concurrent events in a concurrent
  system. These include degenerate transitions where no event is
  active, corresponding to states of a classical automaton,
  transitions where one single event is active, as in a classical
  automaton, but also higher transitions in which more than one event
  is active.  The name ``cell'' also emphasises the roots of HDAs in
 topology and geometry.

  The set $\ev(x)$ of events records the concurrent events that are
  active in the cell $x$. The total \emph{event order} $\intord$ on
  $\ev(x)$ can be seen as an order on indices of concurrent events. We
  also use it to identify events across cells and their faces and to
  relate HDAs with the ipomsets that model their behaviour. The
  labelling function $\lambda:\ev(x)\to \Sigma$ associates events with
  the actions they perform. We call $(\ev(x),\intord,\lambda)$ the
  \emph{concurrency list} of $x$.

  The faces $\delta_{A,B}(x)$ of the cell $x$ keep track of the
  intervals of activity of concurrent events in an HDA. Each cell
  $\delta_{A,\emptyset}(x)$, also written $\delta^0_A(x)$, forms a
  \emph{lower face} of $x$; each cell $\delta_{\emptyset,B}(x)$, also
  written $\delta^1_B(x)$, forms an \emph{upper face} of $x$.  In
  $\delta^0_A(x)$, the events in $A$, which are active in $x$, have
  not yet started; in $\delta^1_B(x)$, the events in $B$, which are
  active in $x$, have terminated.  All events in
  $\ev(x)\setminus A$ remain active in $\delta^0_A(x)$ and those in $\ev(x)\setminus B$ remain active in
  $\delta^1_B(x)$. The functional relationship between each cell $x$
  and its faces, for each $A,B\subseteq \ev(x)$, allows us to view the
  $\delta_{A,B}$ as \emph{face maps} for $x$, which attach faces to
  cells.  The identity in the penultimate bullet point above states
  that the result of removing disjoint sets of events (terminating or
  does not depend on the order in which these removals occur; it implies $\delta_{A, B}=\delta_A^0 \delta_B^1=\delta_B^1 \delta_A^0$.

  \begin{figure}[tbp]
    \centering
    \begin{tikzpicture}[x=.72cm, y=.72cm]
      \begin{scope}
    \node[circle,draw=black,fill=black!10,inner sep=0pt,minimum size=15pt]
    (aa) at (0,0) {$x_1$};			
    \node[circle,draw=black,fill=black!10,inner sep=0pt,minimum size=15pt]
    (ac) at (0,4) {$x_6$};			
    \node[circle,draw=black,fill=black!10,inner sep=0pt,minimum size=15pt]
    (ca) at (4,0) {$x_3$};			
    \node[circle,draw=black,fill=black!10,inner sep=0pt,minimum size=15pt]
    (cc) at (4,4) {$x_8$};			
    \node[circle,draw=black,fill=red!30,inner sep=0pt,minimum size=15pt]
    (ba) at (2,0) {$x_2$};			
    \node[circle,draw=black,fill=red!30,inner sep=0pt,minimum size=15pt]
    (bc) at (2,4) {$x_7$};			
    \node[circle,draw=black,fill=green!30,inner sep=0pt,minimum size=15pt]
    (ab) at (0,2) {$x_4$};			
    \node[circle,draw=black,fill=green!30,inner sep=0pt,minimum size=15pt]
    (cb) at (4,2) {$x_5$};			
    \node[circle,draw=black,fill=yellow!60,inner sep=0pt,minimum size=15pt]
    (bb) at (2,2) {$x$};
    \path (ba) edge node[below=-1mm] {\scriptsize{$\delta^0_a$}} (aa);
    \path (ba) edge node[below=-1mm] {\scriptsize{$\delta^1_a$}} (ca);
    \path (bb) edge node[below=-1mm] {\scriptsize{$\delta^0_a$}} (ab);
    \path (bb) edge node[above=-1mm] {\scriptsize{$\delta^1_a$}} (cb);
    \path (bc) edge node[above=-1mm] {\scriptsize{$\delta^0_a$}} (ac);
    \path (bc) edge node[above=-1mm] {\scriptsize{$\delta^1_a$}} (cc);
    \path (ab) edge node[left=-1mm] {\scriptsize{$\delta^0_b$}} (aa);
    \path (ab) edge node[left=-1mm] {\scriptsize{$\delta^1_b$}} (ac);
    \path (bb) edge node[right=-1mm] {\scriptsize{$\delta^0_b$}} (ba);
    \path (bb) edge node[left=-1mm] {\scriptsize{$\delta^1_b$}} (bc);
    \path (cb) edge node[right=-1mm] {\scriptsize{$\delta^0_b$}} (ca);
    \path (cb) edge node[right=-1mm] {\scriptsize{$\delta^1_b$}} (cc);
    \path (bb) edge node[above left=-1mm] {\scriptsize{$\delta^1_{ab}\!\!$}} (cc);
    \path (bb) edge node[below right=-1mm]  {\!\!\scriptsize{$\delta^0_{ab}$}} (aa);
     \path (bb) edge node[above right=-1mm]
     {\scriptsize{$\!\!\!\delta_{b,a}$}} (ca);
       \path (bb) edge node[below left=-1mm]
       {\scriptsize{$\delta_{a,b}\!\!\!\!$}} (ac);
    \node[below left] at (aa) {$\bot\;$};
    \node[right] at (cb) {$\,\,\, \top$};
    \node[above right] at (cc) {$\;\top$};
  \end{scope}
  
  \begin{scope}[shift={(6.5,0)}]
      \path[fill=black!10] (0,0) to (4,0) to (4,4) to (0,4);
    \node[state] (a) at (0,0) {};			
    \node[state] (b) at (0,4) {};			
    \node[state] (c) at (4,0) {};			
    \node[state] (d) at (4,4) {};
    \path (a) edge  node [left] {$b$} (b);
    \path (a) edge  node [below] {$a$} (c);
    \path (b) edge  node [above] {$a$} (d);
    \path (c) edge  node [above right] {$b$} node [below right]
    {$\top$} (d);
    \node (e) at (2,2) {$ab$};
    \node[below left] at (a) {$\bot$};
    \node[above right] at (d) {$\top$};
  \end{scope}
  \begin{scope}[shift={(14,0)}]
      \path[fill=black!10] (0,0) to (4,0) to (4,4) to (0,4);
    \node[state] (a) at (0,0) {};			
    \node[state] (b) at (0,4) {};			
    \node[state] (c) at (4,0) {};			
    \node[state]
    (d) at (4,4) {};
    \path (a) edge  node [left] {$[a | ab | \emptyset]$} (b);
    \path (a) edge  node [below] {$[b | ab | \emptyset]$} (c);
    \path (b) edge  node [above] {$[\emptyset | ab | b]$} (d);
    \path (c) edge  node [above right] {$[\emptyset | ab | a]$} node
    [below right] {$\top$} (d);
    \node (e) at (2,2) {$[\emptyset | ab | \emptyset]$};
    \node[below left] at (0,0) {$[ab | ab | \emptyset]\,\,$};
    \node[below] at (0,0) {$\bot$};
    \node[below right] at (4,0) {$\,\, [b | ab | a]$};
    \node[above left] at (0,4) {$[a | ab | b]\,\,$};
    \node[above right] at (4,4) {$\,\, [\emptyset | ab | ab]$};
    \node[right] at (4,3.9) {$\top$};
    \end{scope}
  \end{tikzpicture}
  \caption{Three representations of a two-dimensional HDA.}
  \label{fig:abcube}
\end{figure}

\begin{exa}
  The diagram on the left of Figure~\ref{fig:abcube} shows an HDA with
  cells $x,x_1,\dots, x_8$, where $x_1,\dots, x_8$ are faces of
  $x$. The concurrent events active in $x$ are $\ev(x)=\{a,b\}$, where
  we assume that $a\intord b$. Here, and henceforth in this article,
  we identify events with their actions, whenever suitable. We call
  the cells $x_1$, $x_3$, $x_6$ and $x_8$ $0$-dimensional and $x_2$,
  $x_4$, $x_5$ and $x_7$ $1$-dimensional, while $x$ is a
  $2$-dimensional cell.  The arrows in this diagram indicate the face
  maps of $x$. For simplicity, we also write $\delta^0_{ab}$ instead
  of $\delta^0_{\{a,b\}}$ and likewise. The faces of $x$ are given by
\begin{align*}
  x_1 & = \delta_{ab}^0(x)=\delta_a^0(x_2)=
        \delta_b^0(x_4)=\delta_a^0(\delta_b^0(x))=\delta_b^0(\delta_a^0(x))
        =\delta_{ab,\emptyset}(x),\\
  x_2 &=\delta_b^0(x),\\
  x_3 &=
        \delta_a^1(x_2)=\delta_b^0(x_5)=\delta_a^1(\delta_b^0(x))=\delta_b^0(\delta_a^1(x)) =\delta_{b,a}(x),\\
  x_4 &= \delta_a^0(x),\\
  x_5&= \delta_a^1(x),\\
  x_6&=\delta_b^1(x_4)=\delta_a^0(x_7)=\delta_b^1(\delta_a^0(x))=\delta_a^0(\delta_b^1(x))=\delta_{a,b}(x),\\
  x_7&= \delta_b^1(x),\\
  x_8 &= \delta_{ab}^1(x)=\delta_b^1(x_5)=\delta_a^1(x_7)=\delta_b^1(\delta_a^1(x))=\delta_a^1(\delta_b^1(x))=\delta_{\emptyset,ab}(x).
\end{align*}
The face $x_1$, for instance, is a lower face of $x$, $x_2$ and
$x_4$. Neither $a$ nor $b$ is active in $x_1$, whereas $a$, but not
$b$, is active in $x_2$ and $b$, but not $a$, is active in
$x_4$. Indeed, the result of removing first $a$ and then $b$, or first
$b$ and then $a$, lead to the same face of $x$, namely $x_1$. The cell
$x_3$ is neither a lower nor an upper face of $x$, though it is the upper
face of $x_2$ and the lower face of $x_5$. Yet it is a face of $x$ as
$x_3=\delta_{b,a}(x)$, which indicates that $b$ has not yet started,
but $a$ has terminated in $x_3$. The remaining faces satisfy similar
relationships.

The cell $x_1$ is the start cell of the HDA, while $x_5$ and $x_8$ are
accept cells. As in the introduction, we label such cells $\bot$ and
$\top$, respectively.

The $0$-cells of the HDA, where no event is active, are grey, and its
$1$-cells, where precisely one event is active, $a$ in the pink cells
and $b$ in the green ones, correspond to the states and transitions of
a classical automaton.  The $2$-cell $x$, in yellow, where $a$ and $b$
are concurrently active, models a higher transition. It has no
classical analogue. Similarly, the HDA has the $1$-cell $x_5$ as an
accept cell, while classical automata can only accept in $0$-cells.

The diagram in the middle represents the HDA geometrically in terms of
the actions that are concurrently active in each cell. The $0$- and
$1$-cells are represented as states and arrows. The $2$-cell is
represented as a filled-in square.  Cells of dimension strictly greater than $0$ are
labelled with their active events or actions. This geometric
view allows depicting events as paths or trajectories that
traverse the HDA along the directions of the arrows.

Relative to the cell $x$ and its faces, the diagram on the right uses
triples $[-|-|-]$ to list the events or actions that are not yet
active in a face in the first component, those that are active in the
``top cell'' $x$ in the second, and those that have already terminated
in the face in the third. This notation is local to the faces of
particular top cells.
\end{exa}

\begin{rem}
  Precubical sets and HDAs can be represented by geometric objects,
  which are higher-dimensional cubes in topological spaces. Details of
  these geometric realisations of precubical sets and HDAs can be
  found in the
  literature~\cite{book/Grandis09,DBLP:books/sp/FajstrupGHMR16,
    DBLP:journals/tcs/FajstrupRG06}.  Geometric realisations provide
  intuitions for concurrent systems evolving continuously across
  higher-dimensional cells, while their events start, are active and
  terminate. Yet our results do not require their formalisation.
\end{rem}

\begin{exa}
  \label{ex:fsa}
  The classical finite state automata are one-dimensional HDAs in
  which all start and accept cells have dimension $0$ and hence no
  active events. The $0$-cells without active events form the states of
  such automata, $1$-cells with precisely one active, $a$ say,
  correspond to $a$-labelled transitions.  The face maps
  $\delta_{a}^0$ and $\delta_{a}^1$ attach source and target states to
  transitions.
\end{exa}

A \emph{path} on an HDA $X$ is a sequence of \emph{steps}, formed by
triples of two cells and a step between them, so that each
subsequent step must start in the cell in which the previous one has
terminated.  We distinguish two kinds of steps:
\begin{itemize}
\item
  \emph{up-steps} ($\delta^0_A(x)\arrO{A}x$) from a lower face of
  the cell $x$ to $x$;
\item \emph{down-steps} ($x\arrI{B}\delta^1_B(x)$) from $x$ to
  one of its upper faces.
\end{itemize}
Up-steps start events while down-steps terminate them.  Steps
thus keep track of the events that start or terminate in them. If $y=\delta^0_A(x)=\delta^0_C(x)$ for $A,C\subseteq \ev(x)$
and $A\neq C$, then $(y\arrO{A}x)$ and $(y\arrO{C}x)$ are distinct.
While the termination of a specific set of events that are active in a
given cell is a deterministic operation, starting new events in a
given cell can be nondeterministic, as any cell may be a lower face of
several cells.

In finite state automata as in Example \ref{ex:fsa}, every transition consists of an up-step followed by a down-step.

\begin{figure}[tbp]
  \centering
  \begin{tikzpicture}[x=.6cm, y=.6cm]
    \begin{scope}
    \node[circle,draw=black,fill=black!10,inner sep=0pt,minimum size=10pt]
    (00) at (0,0) {};			
    \node[circle,draw=black,fill=black!10,inner sep=0pt,minimum size=10pt]
    (04) at (0,4) {};			
    \node[circle,draw=black,fill=black!10,inner sep=0pt,minimum size=10pt]
    (40) at (4,0) {};			
    \node[circle,draw=black,fill=black!10,inner sep=0pt,minimum size=10pt]
    (44) at (4,4) {};			
    \node[circle,draw=black,fill=black!10,inner sep=0pt,minimum size=10pt]
    (20) at (2,0) {};			
    \node[circle,draw=black,fill=black!10,inner sep=0pt,minimum size=10pt]
    (24) at (2,4) {};			
    \node[circle,draw=black,fill=black!10,inner sep=0pt,minimum size=10pt]
    (02) at (0,2) {};			
    \node[circle,draw=black,fill=black!10,inner sep=0pt,minimum size=10pt]
    (42) at (4,2) {};			
    \node[circle,draw=black,fill=black!10,inner sep=0pt,minimum size=10pt]
    (22) at (2,2) {};
		
     \node[circle,draw=black,fill=black!10,inner sep=0pt,minimum size=10pt]
     (46) at (4,6) {};	
      \node[circle,draw=black,fill=black!10,inner sep=0pt,minimum size=10pt]
    (48) at (4,8) {};			
    \node[circle,draw=black,fill=black!10,inner sep=0pt,minimum size=10pt]
    (64) at (6,4) {};			
    \node[circle,draw=black,fill=black!10,inner sep=0pt,minimum size=10pt]
    (66) at (6,6) {};			
    \node[circle,draw=black,fill=black!10,inner sep=0pt,minimum size=10pt]
    (68) at (6,8) {};			
    \node[circle,draw=black,fill=black!10,inner sep=0pt,minimum size=10pt]
    (84) at (8,4) {};			
    \node[circle,draw=black,fill=black!10,inner sep=0pt,minimum size=10pt]
    (86) at (8,6) {};			
    \node[circle,draw=black,fill=black!10,inner sep=0pt,minimum size=10pt]
    (88) at (8,8) {};

        \node[circle,draw=black,fill=black!10,inner sep=0pt,minimum size=10pt]
     (60) at (6,0) {};	
      \node[circle,draw=black,fill=black!10,inner sep=0pt,minimum size=10pt]
    (62) at (6,2) {};			
    \node[circle,draw=black,fill=black!10,inner sep=0pt,minimum size=10pt]
    (80) at (8,0) {};			
    \node[circle,draw=black,fill=black!10,inner sep=0pt,minimum size=10pt]
    (82) at (8,2) {};			
  		  
    \path (20) edge node[below] {$\delta^0_a$} (00);
    \path (20) edge[cyan, line width=1pt] node[below, black] {$\delta^1_a$} (40);
    \path (22) edge node[below] {} (02);
    \path (22) edge[orange, line width=1pt] node[above] {} (42);
    \path (24) edge node[above] {$\delta^0_a$} (04);
    \path (24) edge node[above] {$\delta^1_a$} (44);
    \path (02) edge node[left] {$\delta^0_b$} (00);
    \path (02) edge node[left] {$\delta^1_b$} (04);
    \path (22) edge[orange, line width = 1pt] node[right] {} (20);
    \path (22) edge node[left] {} (24);
    \path (42) edge node[right] {$\delta^0_b$} (40);
    \path (42) edge node[right] {$\delta^1_b$} (44);
    \path (22) edge[teal, line width = 1pt] node[above left] {} (44);
    \path (22) edge[teal, line width = 1pt] node[below right] {} (00);
  \path (60) edge[cyan, line width = 1pt] node[below, black] {$\delta^0_c$} (40);
    \path (62) edge[orange, line width = 1pt] (42);
    \path (64) edge[teal, line width = 1pt] node[above, black] {$\delta^0_c$} (44);
    \path (66) edge (46);
    \path (68) edge node[above] {$\delta^0_c$} (48);
    \path (60) edge node[below] {$\delta^1_c$} (80);
    \path (62) edge[orange, line width=1pt] (82);
    \path (64) edge node[above] {$\delta^1_c$} (84);
       \path (66) edge (86);
       \path (68) edge node[above] {$\delta^1_c$}  (88);
       \path (46) edge node[left] {$\delta^0_d$} (44);
       \path (46) edge node[left] {$\delta^1_d$}  (48);
             \path (66) edge[teal, line width = 1pt] (64);
             \path (66) edge[teal, line width =1pt] (68);
                         \path (86) edge node[right] {$\delta^0_d$}  (84);
                         \path (86) edge node[right] {$\delta^1_d$}  (88);
                                      \path (62) edge (60);
             \path (62) edge (64);
                         \path (82) edge node[right] {$\delta^0_b$}  (80);
        \path (82) edge[orange, line width=1pt] node[right, black] {$\delta^1_b$}  (84); 
        \path (62) edge (40);
        \path (62) edge (84);
        \path (66) edge (44);
        \path (66) edge (88);
        \node[below left] at (00) {$\bot$};
        \node[right] at (8,6) {$\,\,\top$};
  \end{scope}

  \begin{scope}[shift={(12,0)}]
    \path[fill=black!10] (0,0) to (4,0) to (4,4) to (0,4);
    \path[fill=black!10] (4,0) to (8,0) to (8,4) to (4,4);
    \path[fill=black!10] (4,4) to (8,4) to (8,8) to (4,8);
    \node[state] (00) at (0,0) {};
    \node[state] (04) at (0,4) {};			
    \node[state] (40) at (4,0) {};			
    \node[state] (44) at (4,4) {};
    \node[state] (80) at (8,0) {};
    \node[state] (84) at (8,4) {};
    \node[state] (88) at (8,8) {};
    \node[state] (48) at (4,8) {};
    \path (00) edge node [left] {$b$} (04);
    \path (00) edge[-]  node [below] {$a$} (40);
    \path (04) edge  node [above] {$a$} (44);
    \path (40) edge node [right] {$b$} (44);
    \path (40) edge node [below] {$c$} (80);
    \path (44) edge node [above] {$c$} (84);
    \path (48) edge node [above] {$c$} (88);
    \path (80) edge[-] node [right] {$b$} (84);
    \path (44) edge node [left] {$d$} (48);
      \path (84) edge node [above right] {$d$} node [below right] {$\top$} (88);
    \node (22) at (2,2) {$ab$};
    \node (62) at (6,2) {$bc$};
    \node (66) at (6,6) {$cd$};
    \node[below left] at (00) {$\bot$};

    \draw [-,cyan, line width = 1pt] plot [smooth, tension=0] coordinates { (1,0) (5,0)};
    \draw [-,orange, line width = 1pt] plot [ smooth, tension=.5] coordinates { (2,0)
      (2.2,0.5) (2.6,0.7) (3,1.2) (3.6,1.5) (4,1.7) (5,1.8) (6,2.6)
      (6.8,2.7) (8,3)};
    \draw [-,orange, line width = 1pt] plot [smooth, tension=0] coordinates {(8,3)
      (8,4)};
     \draw [-,teal, line width = 1pt] plot [ smooth, tension=.4] coordinates { (0,0)
       (.5,.2) (.7,.6) (1.5,2) (1.8,2.3) (2.5,2.5) (3,2.9) (3.8,3.9) (4,4)};
      \draw [-,teal, line width = 1pt] plot [ smooth, tension=0] coordinates { (4,4)
        (5,4)};
         \draw [-,teal, line width = 1pt] plot [ smooth, tension=.4]
         coordinates { (5,4) (5.2,4.2) (5.5,5) (6.3,5.6)  (6.7,7) (7.2,8)};
  \end{scope}
   \end{tikzpicture}
  \caption{Paths in an HDA}
  \label{fig:abpath}
\end{figure}

Each path $\alpha$ is associated with its \emph{set of events}
$\ev(\alpha)$.  It contains the local events of all cells appearing in
$\alpha$, but certain events of consecutive cells are identified.  In a step
$(\delta^0_A(x)\arrO{A}x)$, for instance, the events of
$\delta^0_A(x)$ are identified with events in $x$ via the equivalence
$\cong$ induced by the unique event-order preserving bijection
between sets in
$\ev(\delta^0_A(x))\cong \ev(x)\setminus A\subseteq \ev(x)$.

\begin{exa}
  Figure~\ref{fig:abpath} shows three paths through an HDA with three
  2-cells that are glued along two 1-dimensional faces. The diagram on
  the left shows paths consisting of up-steps and down-steps. Note
  that the direction of up-steps goes against the direction of lower
  face maps, while the direction of down-steps and upper face maps
  coincides. The concurrency lists of the missing faces and the
  missing face maps can be reconstructed from the data shown. The
  diagram on the right shows a geometric realisation with piecewise
  smooth paths or trajectories crossing the HDA from bottom left to
  top right.
\end{exa}

The set $\ev(\alpha)$ carries additional structure:
\begin{itemize}
\item the \emph{precedence} order $<$, where $p<q$ holds if event $p$
  is active in some cell before event $q$ is active in a different
  cell, and if there is no cell in which they are both active;
\item the \emph{event order} $\intord$, which is constructed from the
  local event orders in the concurrency lists of individual cells,
\item the \emph{source interface} of $\alpha$, which contains all
  events of the source cell, the first cell of $\alpha$, and
  the \emph{target interface} of $\alpha$, which contains all events of
  the target cell, the last cell of $\alpha$.
\end{itemize}

The resulting structure, formed by events labelled with actions and
equipped with a precedence, an event order and source and target
interfaces, is called the (labelled) \emph{iposet} of the path
$\alpha$ (it satisfies some extra conditions omitted here: in
particular $<$ is an interval order, see Figure \ref{fig:abiposet}.) As
usual in concurrency theory, we define an \emph{ipomset} as an
isomorphism class of iposets, where isomorphisms are order preserving
and reflecting and label and interface preserving bijections between
iposets.

A path of an HDA is \emph{accepting} if its source is a
start cell and its target an accept cell of the HDA.  The set of all
ipomsets associated with accepting paths in $X$ is the \emph{language}
$\Lang(X)$ of $X$.

\begin{exa}
 \begin{figure}[tbp]
    \centering
    \begin{tikzpicture}[x=.6cm, y=.6cm]
  
    \node[circle,draw=black,fill=black!10,inner sep=0pt,minimum size=10pt]
    (00) at (0,0) {};			
    \node[circle,draw=black!10,fill=black!10,inner sep=0pt,minimum size=10pt]
    (04) at (0,4) {};			
    \node[circle,draw=black!10,fill=black!10,inner sep=0pt,minimum size=10pt]
    (40) at (4,0) {};			
    \node[circle,draw=black!10,fill=black!10,inner sep=0pt,minimum size=10pt]
    (44) at (4,4) {};			
    \node[circle,draw=black!10,fill=black!10,inner sep=0pt,minimum size=10pt]
    (20) at (2,0) {};			
    \node[circle,draw=black!10,fill=black!10,inner sep=0pt,minimum size=10pt]
    (24) at (2,4) {};			
    \node[circle,draw=black!10,fill=black!10,inner sep=0pt,minimum size=10pt]
    (02) at (0,2) {};			
    \node[circle,draw=black,fill=black!10,inner sep=0pt,minimum size=10pt]
    (42) at (4,2) {};			
    \node[circle,draw=black,fill=black!10,inner sep=0pt,minimum size=10pt]
    (22) at (2,2) {};
		
     \node[circle,draw=black!10,fill=black!10,inner sep=0pt,minimum size=10pt]
     (46) at (4,6) {};	
      \node[circle,draw=black!10,fill=black!10,inner sep=0pt,minimum size=10pt]
    (48) at (4,8) {};			
    \node[circle,draw=black,fill=black!10,inner sep=0pt,minimum size=10pt]
    (64) at (6,4) {};			
    \node[circle,draw=black,fill=black!10,inner sep=0pt,minimum size=10pt]
    (66) at (6,6) {};			
    \node[circle,draw=black!10,fill=black!10,inner sep=0pt,minimum size=10pt]
    (68) at (6,8) {};			
    \node[circle,draw=black!10,fill=black!10,inner sep=0pt,minimum size=10pt]
    (84) at (8,4) {};			
    \node[circle,draw=black!10,fill=black!10,inner sep=0pt,minimum size=10pt]
    (86) at (8,6) {};			
    \node[circle,draw=black!10,fill=black!10,inner sep=0pt,minimum size=10pt]
    (88) at (8,8) {};

        \node[circle,draw=black!10,fill=black!10,inner sep=0pt,minimum size=10pt]
     (60) at (6,0) {};	
      \node[circle,draw=black,fill=black!10,inner sep=0pt,minimum size=10pt]
    (62) at (6,2) {};			
    \node[circle,draw=black!10,fill=black!10,inner sep=0pt,minimum size=10pt]
    (80) at (8,0) {};			
    \node[circle,draw=black!10,fill=black!10,inner sep=0pt,minimum size=10pt]
    (82) at (8,2) {};			
  		  
    \path (20) edge[black!10] (00);
    \path (20) edge[black!10]  (40);
    \path (22) edge[black!10]  (02);
    \path (22) edge[orange, line width=1pt] node[above, black] {$\delta_a^1$} (42);
    \path (24) edge[black!10]  (04);
    \path (24) edge[black!10] (44);
    \path (02) edge[black!10] (00);
    \path (02) edge[black!10] (04);
    \path (22) edge[black!10] (20);
    \path (22) edge[black!10](24);
    \path (42) edge[black!10] (40);
    \path (42) edge[black!10] (44);
    \path (22) edge[black!10] (44);
    \path (22) edge[orange, line width = 1pt]  node[above left, black] {$\delta_{ab}^0\!\!\!$} (00);
    \path (60) edge[black!10] (40);
    \path (62) edge[orange, line width = 1pt] node[above, black] {$\delta_c^0$} (42);
    \path (64) edge[black!10] (44);
    \path (66) edge[black!10] (46);
    \path (68) edge[black!10] (48);
    \path (60) edge[black!10] (80);
    \path (62) edge[black!10] (82);
    \path (64) edge[black!10] (84);
    \path (66) edge[orange, line width = 1pt] node[above, black] {$\delta_c^1$} (86);
    \path (68) edge[black!10] (88);
    \path (46) edge[black!10] (44);
    \path (46) edge[black!10] (48);
    \path (66)  edge[orange, line width = 1pt] node[right, black] {$\delta_d^0$} (64);
    \path (66) edge[black!10] (68);
    \path (86) edge[black!10] (84);
    \path (86) edge[black!10] (88);
    \path (62) edge[black!10] (60);
    \path (62) edge[orange, line width = 1pt] node[right, black] {$\delta_b^1$} (64);
    \path (82) edge[black!10] (80);
    \path (82) edge[black!10] (84);
    \path (62) edge[black!10] (40);
    \path (62) edge[black!10] (84);
    \path (66) edge[black!10] (44);
    \path (66) edge[black!10] (88);
      \node[below left] at (00) {$\bot$};
        \node[right] at (86) {$\,\,\top$};

    \path (10,0) edge[-, line width = 1pt]  (10,2);
    \path (10.5,0) edge[-, line width = 1pt] (10.5,4);
    \path (11,2) edge[-, line width = 1pt] (11,6);
    \path (11.5,4) edge[-, line width = 1pt] (11.5,6);
    \path (11.5,6) edge[-,dashed , line width = 1pt] (11.5,8);

    \node at (10,-.5) {$a$};
    \node at (10.5,-.41) {$b$};
    \node at (11,-.5) {$c$};
    \node at (11.5,-.41) {$d$};

    \node (131) at (14,1) {$a$};
    \node (152) at (16,2) {$b$};
    \node (134) at (14,4) {$c$};
    \node (155) at (16,5) {$d\ibullet$};

    \path (131) edge (134);
    \path (131) edge (155);
    \path (152) edge (155);
    \path (131) edge[dashed] (152);
    \path (134) edge[dashed] (155);
     \path (152) edge[dashed] (134);

  \end{tikzpicture}
  \caption{Interval ipomset of path in HDA}
  \label{fig:abiposet}
\end{figure}

Figure~\ref{fig:abiposet} shows the iposet of an accepting path
$\alpha$ on the HDA from Figure~\ref{fig:abpath} on the left. The bar
codes in the centre indicate when the events $a$, $b$, $c$ and $d$ in
the HDA start, are active and terminate relative to each other. The
dashed part of the interval for $d$ indicates that $d$ remains active
in the accept state $\tgt(\alpha)$.

The Hasse diagram on the right shows the ipomset of $\alpha$. The
precedence on events is indicated by solid arrows. The event order,
given here by the lexicographical order on events, is indicated by dashed
arrows.  As $d$ does not terminate in $\tgt(\alpha)$, it is in
$\ev(\tgt(\alpha))$, the concurrency list of $\tgt(\alpha)$, and hence
in the target interface of the ipomset of $\alpha$. We indicate this
by writing $d\ibullet$. Similarly, we would have indicated membership
in the source interface of this ipomset by adding a bullet to the left
of the event, for instance $\ibullet e$, but here $\ev(\src(\alpha))$ is
empty because no event is active in the HDA at the beginning of the
path.
\end{exa}

\section{Higher-dimensional automata}\label{s:hdas}

In this section we provide a formal definition of higher-dimensional
automata.  It differs from those in the literature and it is slightly
more general. We relate
our definition to previous ones in Appendix~\ref{ap:hda}. It is
technically convenient to model HDAs as labelled precubical sets
equipped with start and accept cells.  Precubical sets in turn can be
modelled as presheaves on a so-called labelled precube category.
Objects of this category are concurrency lists; morphisms are precube
maps, which are order and label preserving maps enriched with
information about the activity of events, their start and termination.

\subsection*{Concurrency lists}

\def\concmap{{lo-map}}
\def\cofacemap{{conclist map}}
\def\Cofacemap{{Conclist map}}

Throughout the paper, we fix an alphabet $\Sigma$ of labels, which are
meant to represent the actions of a concurrent system.

A \emph{concurrency list} or \emph{conclist} $(U,\intord,\lambda )$
consists of a finite set $U$ equipped with a strict total \emph{event
  order} $\intord$ on $U$ and a labelling function
$\lambda:U\to \Sigma$ that assigns actions to events. An
\emph{\concmap} $f:U\to V$ between conclists $U$ and $V$ is a label and
(event) order preserving map. Conclists and {\concmap}s form a
category.

We often write conclists vertically as vectors of events or actions,
which we often do not distinguish, especially in diagrams. Recall from
Section~\ref{s:overview} that the set $U$ models the concurrent local
events active in a cell of an HDA, $\intord$ can be seen as their
index order and $\lambda$ associates events with their actions.  Since
$\intord$ is strict and total, every {\concmap} is injective.

\begin{figure}
\begin{equation*}
  \begin{tikzcd}[row sep=0.18cm, column sep = small]
    &&&&c&&&&&\\
    &&&&&a&&&&\\
    &&a&& \arrow[uu, teal] 6&&d&&&\\
    &&&b&& 5\arrow[ul, dashed,
    gray]\arrow[uu, teal]&&e&&\\
    &&3\arrow[urrr, orange]\arrow[uu,
    teal, "\textcolor{black}{\lambda}"]&&a&&
    4\arrow[ul, dashed, gray]\arrow[uu,
    cyan]&&b&\\
    &&&2\arrow[ul, dashed, gray]\arrow[drrrrr, orange]\arrow[uu, teal] &&&&
    3\arrow[ul, dashed, gray]\arrow[uu, teal]&&a\\
    &&&&1\arrow[ul, dashed, gray]\arrow[drrrrr, orange, swap, "\textcolor{black}{\partial_{\{3,4,6\}}}"]\arrow[uu, teal]&&&&
    2\arrow[dashed, ul, gray]\arrow[uu, teal]&\\
    &&&&&&&&& 1\arrow[ul, dashed, gray]\arrow[uu, teal]
  \end{tikzcd}
\end{equation*}
\caption{Conclists $1\intord 2\intord 3$ and
  $1\intord 2\intord \dotsc \intord 6$ with {\concmap}
  $\partial_{\{3,4,6\}}$ and  labelling function $\lambda$ into
  $\Sigma=\{a,b,c,d,e\}$.}
\label{fi:event-ord-conc-list}
\end{figure}

By construction, each {\concmap} $f:U\to V$ order-embeds the conclist
$U$ into the conclist $V$ in a unique way.  Hence $f$ identifies the
events in the conclist $U$ with events in the conclist $V$ in a way
compatible with $\intord_U$ and $\intord_V$. It further determines a
unique set $A=V\setminus f(U)$ of elements which are inserted into $U$
to obtain $V$. Conversely, the conclist $U$ and {\concmap} $f$ are
uniquely determined by the restriction of the conclist $V$ to the
events outside of $A$.

We therefore write $\partial_{A\subseteq V}:U\hookrightarrow V$ for
{\concmap}s to emphasise this relationship or simply
$\partial_A:U\hookrightarrow V$ if the dependency on $V$ is clear.
See Figure~\ref{fi:event-ord-conc-list} for an example. The composite
of $\partial_{A\subseteq V}:U\hookrightarrow V$ and
$\partial_{B\subseteq W}:V\hookrightarrow W$ is
\begin{equation*}
  \partial_{B\subseteq W}\circ \partial_{A\subseteq V} =
  \partial_{\partial_B(A)\cup B\subseteq W}:U\hookrightarrow W,
\end{equation*}
as illustrated in Figure~\ref{fi:event-conc-comp}.

Two conclists are isomorphic, $U\cong V$, if there exists a bijective
{\concmap} $U\to V$ that is an order embedding. Isomorphism classes of
conclists can be seen as lists of actions, and {\concmap}s extended to
equivalence classes. Each such map then inserts letters from $\Sigma$
into a $\Sigma$-list. Alternatively we can see isomorphism classes of
conclists as lists of actions indexed by natural numbers. This leads
to a more standard view of HDAs, see aee Appendix~\ref{ap:hda}. In our proof of the Kleene theorem,
working with maps $\partial_A$ has notational advantages.

\begin{figure}
  \begin{equation*}
  \begin{tikzcd}
   \begin{bmatrix}\textcolor{orange}{2}\end{bmatrix}\arrow[r, "\partial_{\{\textcolor{teal}{1}\}}"] & \begin{bmatrix}\textcolor{teal}{1}\\\textcolor{orange}{2}\end{bmatrix}
 \end{tikzcd}
 \circ
 \begin{tikzcd}
   \begin{bmatrix}\textcolor{teal}{1}\\
     \textcolor{orange}{2}\end{bmatrix} \arrow[r,
   "\partial_{\{\textcolor{cyan}{3},\textcolor{cyan}{4}\}}"] &
   \begin{bmatrix}\textcolor{teal}{1}\\ \textcolor{orange}2\\ \textcolor{cyan}{3}\\
   \textcolor{cyan}{4}\end{bmatrix}
 \end{tikzcd}
 =
 \begin{tikzcd}
   \begin{bmatrix}\textcolor{orange}{2}\end{bmatrix}\arrow[r,
   "\partial_{\{\textcolor{teal}{1},\textcolor{cyan}{3},\textcolor{cyan}{4}\}}"] &
   \begin{bmatrix}\textcolor{teal}{1}\\ \textcolor{orange}{2}\\
     \textcolor{cyan}{3}\\ \textcolor{cyan}{4}\end{bmatrix}
 \end{tikzcd}
\end{equation*}
\caption{Composition of two {\concmap}s.}
\label{fi:event-conc-comp}
\end{figure}

\subsection*{{\Cofacemap}s}

Next we introduce {\cofacemap}s, which form the morphisms of our
labelled precube categories.  These are {\concmap}s in which the
information, whether events that are not in their images have
terminated or not yet started, is made explicit.

A \emph{{\cofacemap}} $d_{A,B\subseteq V}:U\to V$, or shortly
$d_{A,B}:U\to V$, is a triple $(\partial_{A\cup B\subseteq V},A,B)$
with $A,B\subseteq V$ disjoint and
$\partial_{A\cup B\subseteq V}:U\hookrightarrow V$ a {\concmap}.

Intuitively, $d_{A,B}:U\to V$ identifies the events in $U$ with events
in $V\setminus(A\cup B)$, as prescribed by $\intord_U$ and
$\intord_V$, while $A$ and $B$ are those local events in $V$ that have
not yet started and have terminated in $U$, respectively.  Compared to
the face maps in Section~\ref{s:overview}, the direction of arrows is
reversed.

The composite of the {\cofacemap}s $d_{A,B\subseteq V}:U\to V$,
$d_{C,D\subseteq W}:V\to W$ is defined as
\begin{equation*}
  d_{C,D\subseteq W}\circ d_{A,B\subseteq V} = (\partial_{C\cup
    D\subseteq W}\circ \partial_{A\cup B\subseteq V},
  \partial_{C\cup D\subseteq W}(A)\cup C, \partial_{C\cup D\subseteq W}(B)\cup D).
\end{equation*}
This formula simplifies when $V$ is a subset of $W$, which can be
guaranteed up to isomorphism of conclists.  For pairwise disjoint
$A,B,C,D\subseteq W$ and $V=W\setminus(C\cup D)$,
\begin{equation}
  \label{e:simplesqcomp}
  d_{C,D\subseteq W}\circ d_{A,B\subseteq V} = d_{A\cup C, B\cup
    D\subseteq W}.
\end{equation}

Figure~\ref{fi:dotsquare-comp} shows an example.  Further,
we write $d^0_{A\subseteq V} $ for $d_{A,\emptyset\subseteq V}$ and
$d^1_{B\subseteq V}$ for $d_{\emptyset,B\subseteq V}$, or more briefly
$d^0 _A$ and $d^1_B$.

\begin{figure}
  \centering
  \begin{tikzpicture}[x=25,y=23]
    \begin{scope}[shift={(-2,0)}]
     \node[font=\normalsize] (r0) at (0,1) {$\left[\begin{smallmatrix}b\end{smallmatrix}\right]$};
      \node (r2) at (0,-2) {$b$};
    \end{scope}
    \begin{scope}[shift={(0.4,0)}]
     \node (s0) [font=\normalsize] at (0,1) {$\left[\begin{smallmatrix}a\\b\end{smallmatrix}\right]$};
      \node (s1) at (0,-1) {$a$};
      \node[above right,color=green!50!black] at (s1) {$0$};
      \node (s2) at (0,-2) {$b$};
      \path (s1) edge[densely dotted] (s2);
    \end{scope}
    \node at (1.2,1) {$\circ$};
    \node at (1.2,-2.2) {$\circ$};
    \begin{scope}[shift={(2,0)}]
      \node (ss0) [font=\normalsize] at (0,1) {$\left[\begin{smallmatrix}a\\b\end{smallmatrix}\right]$};
      \node (ss1) at (0,-1) {$a$};
      \node (ss2) at (0,-2) {$b$};
      \path (ss1) edge[densely dotted] (ss2);
    \end{scope}
    \begin{scope}[shift={(4.4,0)}]
      \node (t0) [font=\normalsize] at (0,1)
      {$\left[\begin{smallmatrix}a\\b\\c\\ \phantom{'}a'\end{smallmatrix}\right]$};
      \node (t1) at (0,-1) {$a$};
      \node (t2) at (0,-2) {$c$};
      \node[above right, color=green!50!black] at (t2) {$0$};
      \node (t3) at (0,-3) {$b$};
      \node (t4) at (0,-4) {$a'$};
      \node[above right, color=red!50!black] at (t4) {$1$};
      \path (t1) edge[densely dotted] (t2);
      \path (t2) edge[densely dotted] (t3);
      \path (t3) edge[densely dotted] (t4);
    \end{scope}
    \node at (5.3,1) {$=$};
    \node at (5.3,-2.2) {$=$};
    \begin{scope}[shift={(6,0)}]
     \node[font=\normalsize] (u0) at (0,1) {$\left[\begin{smallmatrix}b\end{smallmatrix}\right]$};
      \node (u2) at (0,-2) {$b$};
    \end{scope}
    \begin{scope}[shift={(8.4,0)}]
      \node (v0) [font=\normalsize] at (0,1)
      {$\left[\begin{smallmatrix}a\\b\\c\\ \phantom{'}a'\end{smallmatrix}\right]$};
      \node (v1) at (0,-1) {$a$};
      \node[above right, color=green!50!black] at (v1) {$0$};
      \node (v2) at (0,-2) {$c$};
      \node[above right, color=green!50!black] at (v2) {$0$};
      \node (v3) at (0,-3) {$b$};
      \node (v4) at (0,-4) {$a'$};
      \node[above right, color=red!50!black] at (v4) {$1$};
      \path (v1) edge[densely dotted] (v2);
      \path (v2) edge[densely dotted] (v3);
      \path (v3) edge[densely dotted] (v4);
    \end{scope}
    \path (r0) edge node[above] {$d_{a,\emptyset}$} (s0);
    \path (ss0) edge node[above] {$d_{c,a'}$} (t0);
    \path (u0) edge node[above] {$d_{ac,a'}$} (v0);
    \path (r2) edge (s2);
    \path (ss1) edge (t1);
    \path (ss2) edge (t3);
    \path (u2) edge (v3);
  \end{tikzpicture}
  \caption{Composition conclist maps. 
    Annotations {\color{green!50!black}$0$} and {\color{red!50!black}$1$}
    indicate events that have not yet started ({\color{green!50!black}$0$}) or
    terminated ({\color{red!50!black}$1$}), as defined in the triple
    $(\partial_{A\cup B},A,B)$.}
  \label{fi:dotsquare-comp}
\end{figure}

\subsection*{Labelled precube categories}

Next we define the base categories of precubical sets and higher-dimensional automata modelled as presheaves. 

The \emph{full labelled precube category} $\fullsq$ has conclists as
objects and {\cofacemap}s as morphisms.  To work with equivalence
classes of conclists, we define the \emph{labelled precube category}
$\sq$ as the quotient of $\fullsq$ with respect to the isomorphism
$\cong$. Its objects are isomorphism classes of conclists, its
{\cofacemap}s equivalence classes of {\cofacemap}s in $\fullsq$.

To define the latter, note that two {\cofacemap}s $d_{A,B}:U\to V$ and
$d_{A',B'}:U'\to V'$ are equivalent in $\fullsq$ if there exists a
conclist isomorphism $\psi:V\rightarrow V'$ such that
$\psi(A)=A'$ and $\psi(B)=B'$.  As mentioned before, such $\psi$ are
unique.  This definition guarantees in particular that $U$ and $U'$
are isomorphic conclists, via unique isomorphisms
\begin{equation*}
  U\cong V\setminus(A\cup B)\cong V'\setminus(A'\cup B')\cong U'.
\end{equation*}

The category $\sq$ has countably many objects and hence it is small.
It is skeletal: isomorphisms between conclists are
unique in the presence of $\intord$, and the quotient functor
$\fullsq\to \sq$ is an equivalence of categories. We switch freely
between $\fullsq$ and $\sq$ and identify morphisms $[U] \to [V]$ on
equivalence classes of event orders with representatives
$d_{A,B}:U\to V$ on conclists. See again
Figure~\ref{fi:dotsquare-comp} for an example.

\subsection*{Precubical sets and higher-dimensional automata}

Our formalisation of precubical sets and higher-dimensional automata
differs from previous definitions~\cite{book/Grandis09,
  DBLP:journals/tcs/Glabbeek06, Goubault02-cmcim}.  One difference is
that labels are directly incorporated into the base category of the
presheaf. See Appendix~\ref{ap:hda} for a comparison.

A \emph{precubical set} $X$ (a \emph{pc-set} for short) is a presheaf
on $\sq$, hence a functor $\sq^\op\to \Set$. 
We write
\begin{itemize}
\item $X[U]$ for the value of $X$ at object $U$ of $\sq$ and call the
  elements of $X[U]$ \emph{cells};
\item
	$\Cell(X)=\bigsqcup_{U\in{\sq}}X[U]$ for the set of all cells of $X$;
      \item $\ev(x)=U$ for each $x\in X[U]$ to recover the conclist of
        concurrent events that are active within the cell $X[U]$
        ($\ev(x)$ is defined only up to isomorphism);
      \item
        $\delta_{A,B\subseteq U}=X[d_{A,B\subseteq U}]:X[U]\to
        X[U\setminus(A\cup B)]$ for the \emph{face map} associated to
        the conclist map
        $d_{A,B\subseteq U}:U\setminus (A\cup B)\to U$;
\item
	$\delta_{A\subseteq U}^0=X[d_{A\subseteq U}^0]$ and
        $\delta_{B\subseteq U}^1=X[d_{B\subseteq U}^1]$ for
         face maps attaching lower and upper faces to the cells in $X[U]$.
\end{itemize}
As before, we drop the index $U$ from face maps if convenient.

A precubical set $X$ is \emph{finite} if $\Cell(X)$ is finite. The
\emph{dimension} of a cell $x\in X[U]$ is the cardinality $|U|$ of
$U$.  The \emph{dimension} $|X|$ of the presheaf $X$ is the maximal
dimension among its cells. It is finite whenever $X$ is.

A \emph{higher-dimensional automaton} (\emph{HDA}) is a finite
precubical set $X$ equipped with a set of start cells
$X_\bot\subseteq \Cell(X)$ and a set of target cells
$X^\top\subseteq \Cell(X)$. 

\begin{rem}
  The set $\Cell(X)$ may be regarded as the set of objects of the
  category of elements of presheaf $X$, and $\ev:\Cell(X)\to\sq$ may
  be regarded as the canonical projection.  
\end{rem}

\begin{figure}[tbp]
  \centering
  \begin{tikzpicture}[x=1.2cm, y=1.2cm]
    \node[circle,draw=black,fill=black!10,inner sep=0pt,minimum size=15pt]
    (aa) at (0,0) {$v$};			
    \node[circle,draw=black,fill=black!10,inner sep=0pt,minimum size=15pt]
    (ac) at (0,4) {$x$};			
    \node[circle,draw=black,fill=black!10,inner sep=0pt,minimum size=15pt]
    (ca) at (4,0) {$w$};			
    \node[circle,draw=black,fill=black!10,inner sep=0pt,minimum size=15pt]
    (cc) at (4,4) {$y$};			
    \node[circle,draw=black,fill=red!30,inner sep=0pt,minimum size=15pt]
    (ba) at (2,0) {$e$};			
    \node[circle,draw=black,fill=red!30,inner sep=0pt,minimum size=15pt]
    (bc) at (2,4) {$f$};			
    \node[circle,draw=black,fill=green!30,inner sep=0pt,minimum size=15pt]
    (ab) at (0,2) {$g$};			
    \node[circle,draw=black,fill=green!30,inner sep=0pt,minimum size=15pt]
    (cb) at (4,2) {$h$};			
    \node[circle,draw=black,fill=yellow!60,inner sep=0pt,minimum size=15pt]
    (bb) at (2,2) {$q$};
    \node[right] at (5,4) {$X[\emptyset]=\{v,w,x,y\}$};
    \node[right] at (5,3.2) {$X[a]=\{e,f\}$};
    \node[right] at (5,2.4) {$X[b]=\{g,h\}$};
    \node[right] at (5,1.6) {$X\left[\begin{smallmatrix}a\\b\end{smallmatrix}\right]=\{q\}$};
    \path (ba) edge node[above=-0.8mm] {$\delta^0_a$} (aa);
    \path (ba) edge node[above=-0.8mm] {$\delta^1_a$} (ca);
    \path (bb) edge node[above=-0.8mm] {$\delta^0_a$} (ab);
    \path (bb) edge node[above=-0.8mm] {$\delta^1_a$} (cb);
    \path (bc) edge node[above=-0.8mm] {$\delta^0_a$} (ac);
    \path (bc) edge node[above=-0.8mm] {$\delta^1_a$} (cc);
    \path (ab) edge node[left=-0.8mm] {$\delta^0_b$} (aa);
    \path (ab) edge node[left=-0.8mm] {$\delta^1_b$} (ac);
    \path (bb) edge node[left=-0.8mm] {$\delta^0_b$} (ba);
    \path (bb) edge node[left=-0.8mm] {$\delta^1_b$} (bc);
    \path (cb) edge node[left=-0.8mm] {$\delta^0_b$} (ca);
    \path (cb) edge node[left=-0.8mm] {$\delta^1_b$} (cc);
    \path (bb) edge node[above left=-1.8mm] {$\delta^1_{ab}$} (cc);
    \path (bb) edge node[above left=-1.8mm] {$\delta^0_{ab}$} (aa);
    \path (bb) edge node[above right=-1.8mm] {$\delta_{a,b}$} (ac);
    \path (bb) edge node[above right=-1.8mm] {$\delta_{b,a}$} (ca);
    \node[below left] at (aa) {$\bot\;$};
    \node[above right] at (cb) {$\;\top$};
    \node[above right] at (cc) {$\;\top$};
    \node[right] at (5,0.8) {$X_\bot=\{v\}$};
    \node[right] at (5,0) {$X^\top=\{h,y\}$};
    \begin{scope}[shift={(8.5,1)}]
      \filldraw[color=black!10] (0,0)--(2,0)--(2,2)--(0,2)--(0,0);			
      \filldraw (0,0) circle (0.05);
      \filldraw (2,0) circle (0.05);
      \filldraw (0,2) circle (0.05);
      \filldraw (2,2) circle (0.05);
      \path (0,0) edge node[below,color=red!70!black] {$a$} (1.95,0);
      \path (0,2) edge (1.95,2);
      \path (0,0) edge node[left,color=green!70!black] {$b$} (0,1.95);
      \path (2,0) edge (2,1.95);
      \node[left] at (0,0) {$\bot$};
      \node[right] at (2,2) {$\top$};
      \node[right] at (2,1) {$\top$};
    \end{scope}
  \end{tikzpicture}
  \caption{A two-dimensional HDA $X$ on
    $\Sigma=\{{\color{red!70!black}{a}},
    {\color{green!70!black}{b}}\}$.}
  \label{fig:abcube2}
\end{figure}

\begin{exa}
  Figure \ref{fig:abcube2} shows once again the HDA and its geometric
  realisation from Figure~\ref{fig:abcube}. The first four elements of
  the column in the centre show how conclists of the base category
  are mapped to sets of cells. The empty conclist, for instance, is
  mapped to the four $0$-cells where no event is active. The conclist
  $a$ is mapped to the two red $1$-cells where $a$ is active and the
  conclist $b$ to the two green $1$-cells where $b$ is
  active. Finally, the conclist $ab$ is mapped to the yellow $2$-cell
  where both of these events are active concurrently. We omit
  braces and simplify notation as in Section~\ref{s:overview}. 
\end{exa}

A \emph{map of precubical sets} (\emph{pc-map}) is a natural
transformation $f:X\to Y$ of precubical sets $X$, $Y$ regarded as
presheaves $\sq^\op\to\Set$.  Its components are given by the
functions $(f[U]:X[U]\to Y[U])_{U\in\sq}$ that commute with face maps.
An \emph{HDA-map} is a map of precubical sets that preserves start and
accept cells: $f(X_\bot)\subseteq Y_\bot$ and
$f(X^\top)\subseteq Y^\top$.

We write $\sq\Set$ and $\HDA$ for the categories of precubical sets
and HDAs.

\subsection*{Standard cubes}\label{ss:standard-cubes}

Standard cubes form the building blocks of precubical sets.  The
\emph{standard $U$\!-cube} $\sq^U$ of the conclist $U$ is the
precubical set represented by $U$, as given by the Yoneda embedding
$\sq\to\sq\Set$. Thus, for each $V\in\sq$, $\sq^U[V]=\sq(V,U)$, the
set of all {\cofacemap}s from $V$ to $U$.  We write $[A|U|B]$ and
sometimes $[A|B]$ for a {\cofacemap} $d_{A,B}:V\to U$, regarded as
cell in $\sq^U[V]$. Further, for any {\cofacemap} $d_{A, B}: U\to W$
($A,B\subseteq W$, $U=W\setminus(A\cup B)$) we write
$\sq^{d_{A,B}}: \sq^U\to \sq^W$ for the induced pc-map given by
$\sq^{d_{A,B}}([C|U|D]) = [A\cup C|W|B\cup D]$.

\begin{exa}
  \label{e:Cubes}
  Let $U= \{a,b\}$. Then $\sq^U$ has the cells $[-|-|-]$ in the
  right-hand cube in Figure~\ref{fig:abcube}, which is reproduced
  below. Their first components list the events that have not yet
  started in this cell, the second ones the events active in $U$ and
  the third ones those that have terminated.

  \begin{equation*}
      \begin{tikzpicture}[x=.5cm, y=.5cm]
      \path[fill=black!10] (0,0) to (4,0) to (4,4) to (0,4);
    \node[circle,draw=black,fill=black,inner sep=0pt,minimum size=3pt]
    (a) at (0,0) {};			
    \node[circle,draw=black,fill=black,inner sep=0pt,minimum size=3pt]
    (b) at (0,4) {};			
    \node[circle,draw=black,fill=black,inner sep=0pt,minimum size=3pt]
    (c) at (4,0) {};			
    \node[circle,draw=black,fill=black,inner sep=0pt,minimum size=3pt]
    (d) at (4,4) {};
    \path (a) edge  node [left] {$[a | ab | \emptyset]$} (b);
    \path (a) edge  node [below] {$[b | ab | \emptyset]$} (c);
    \path (b) edge  node [above] {$[\emptyset | ab | b]$} (d);
    \path (c) edge  node [right] {$[\emptyset | ab | a]$} (d);
    \node (e) at (2,2) {$[\emptyset | ab | \emptyset]$};
    \node[below left] at (0,0) {$[ab | ab | \emptyset]\,\,$};
    \node[below right] at (4,0) {$\,\, [b | ab | a]$};
    \node[above left] at (0,4) {$[a | ab | b]\,\,$};
     \node[above right] at (4,4) {$\,\, [\emptyset | ab | ab]$};
   \end{tikzpicture}
  \end{equation*}
  As an example, $[a|ab|\emptyset]$ indicates that $a$ has not yet
  started and no element has terminated in the associated face, while
  $b$ is active.  We have omitted set braces and likewise, as usual.
\end{exa}

The following property of standard cubes is immediate from the Yoneda
lemma.
\begin{lem}
\label{l:Yoneda}
  For each pc-set $X$ and $x\in X[U]$ there is a unique pc-map
  $\iota_x:\sq^U\to X$ such that $\iota_x([\emptyset|U|\emptyset])=x$.
  Hence there is a canonical bijection $X[U]\cong \sq\Set(\sq^U,X)$.\qed
\end{lem}

This allows representing the cells of HDAs as morphism of the precube
category (hence as pc-maps); see Figure \ref{fi:Yoneda} for an
example. We use such representations frequently in Section~\ref{s:cofib} and the
subsequent ones.

\begin{figure}
      \begin{tikzpicture}[x=.5cm, y=.5cm]
      \path[fill=black!10] (0,0) to (4,0) to (4,4) to (0,4);
    \node[circle,draw=black,fill=black,inner sep=0pt,minimum size=3pt]
    (a) at (0,0) {};			
    \node[circle,draw=black,fill=black,inner sep=0pt,minimum size=3pt]
    (b) at (0,4) {};			
    \node[circle,draw=black,fill=black,inner sep=0pt,minimum size=3pt]
    (c) at (4,0) {};			
    \node[circle,draw=black,fill=black,inner sep=0pt,minimum size=3pt]
    (d) at (4,4) {};
    \path (a) edge  node [left] {$b$} (b);
    \path (a) edge  node [below] {$a$} (c);
    \path (b) edge  node [above] {$a$} (d);
    \path (c) edge  node [right] {$b$}  (d);
    \node[circle,draw=black,fill=black,inner sep=0pt,minimum size=3pt]
    (q) at (-4,0) {};			
    \node[circle,draw=black,fill=black,inner sep=0pt,minimum size=3pt]
    (r) at (-4,4) {};			
    \path (q) edge  node [left] {$b$} (r);
    \node[circle,draw=black,fill=black,inner sep=0pt,minimum size=3pt]
    (s) at (8,0) {};			
    \node[circle,draw=black,fill=black,inner sep=0pt,minimum size=3pt]
    (t) at (8,4) {};			
    \path (s) edge  node [right] {$b$} (t);
    \path (-3.2,2) edge (-1.2,2);
    \path (7.2,2) edge (5.2,2);
   \end{tikzpicture}
\caption{
	The standard cube $\sq^{ab}$ has two cells with event conclist $[b]$,
	which correspond to two pc-maps from $\sq^b$.
	\label{fi:Yoneda}
}
\end{figure}


\section{Pomsets with interfaces}
\label{s:ipomsets}

Pomsets, or partially ordered multisets, from a standard model of
non-interleaving concurrency. In a nutshell, pomsets are isomorphism
classes of finite node-labelled posets, where nodes represent events
of concurrent systems and labels represent their actions.  Associating
pomsets with executions of HDAs requires some adaptations. First, we
need to equip them with source and target interfaces, which are
subsets of their minimal and maximal elements, respectively.  Second,
we add an event order, which extends the event orders of conclists.
Third, we restrict our attention by and large to interval pomsets.
These are based on posets whose nodes can be represented as intervals
on the real line and whose (strict) order relation reflects the
precedence of intervals along the real line. Interval pomsets with
interfaces have been introduced as models for the behaviours of HDAs
in~\cite{Hdalang}. They capture in particular the precedences of
activities of concurrent events in HDAs, as explained in
Section~\ref{s:overview}. Here we recall the basic definitions.

\subsection*{Iposets and ipomsets}

A \emph{labelled poset with interfaces} (\emph{iposet})
$(P,{<},{\intord},S,T,\lambda)$ consists of the following data:
\begin{itemize}
\item $P$ is a finite set (of events);
\item the \emph{precedence} $<$ is a strict order on $P$;
\item the \emph{event order} $\intord$ is a strict order on $P$, each
  pair in $P$ must be comparable by $=$, $<$ or~$\intord$;
\item the sets $S,T\subseteq P$ form the \emph{source} and
  \emph{target interface} of $P$, elements of $S$ must be $<$-minimal
  and those of $T$ $<$-maximal;
\item $\lambda:P\to \Sigma$ is a labelling function.
\end{itemize}

We write $\epsilon$ for the empty iposet. To indicate that action $a$
is part of a source or target interface, we write $\ibullet a$ and
$a\ibullet$, respectively. Hence $\ibullet a\ibullet$ indicates that
$a$ is part of both interfaces.  See the left Hasse diagram in
Figure~\ref{fi:iposet} for an example.

The event order is not part of the standard definition of labelled
posets in concurrency theory. It is inherited from HDAs that generate
them.  It is also instrumental for coordinating the gluing of iposets
along their interfaces, as discussed in the next section.  Unlike the
event order on conclists, that on iposets need not be
linear. Conclists may be regarded as iposets with empty precedence and
empty interfaces. Conversely, interfaces of iposets with $\intord$ and
labelling restricted to their elements form conclists.

Source and target interfaces allow us to model events that are active
outside a given poset. This is particularly natural when events extend
in time and we need to cut across them to decompose concurrent
systems. Accordingly, events in a poset that do not belong to an
interface start and end their activity within that poset.  

\begin{figure}
  \centering
  \begin{tikzpicture}[x=.6cm, y=.5cm]
    \begin{scope}
    \node[circle,draw=black,inner sep=0pt,minimum size=15pt]
    (1) at (2,4) {$1$};
    \node[circle,draw=black,inner sep=0pt,minimum size=15pt]
    (2) at (0,2) {$2$};
    \node[circle,draw=black,inner sep=0pt,minimum size=15pt]
    (3) at (0,0) {$3$};
    \node[circle,draw=black,inner sep=0pt,minimum size=15pt]
    (4) at (4,2) {$4$};
    \node[circle,draw=black,inner sep=0pt,minimum size=15pt]
    (5) at (4,0) {$5$};
    \node[circle,draw=black,inner sep=0pt,minimum size=15pt]
    (6) at (8,0) {$6$};
    \path (2) edge (4);
    \path (3) edge (5);
    \path (2) edge (5);
    \path (5) edge (6);
    \path (1) edge[dashed] (2);
    \path (2) edge[dashed] (3);
    \path (1) edge[dashed] (4);
    \path (4) edge[dashed] (3);
    \path (4) edge[dashed] (5);
    \path (6) edge[dashed] (4);
    \path (6) edge[dashed, bend right] (1);
    \node[above = .2cm] at (1) {$\ibullet a\ibullet$};
    \node[above = .2cm] at (2) {$b$};
    \node[below = .2cm] at (3) {$\ibullet c$};
    \node[above = .2cm] at (4) {$c\ibullet$};
    \node[below = .2cm] at (5) {$d$};
    \node[above = .2cm] at (6) {$a$};
  \end{scope}
  \begin{scope}[shift={(12,0)}]
    \node (1) at (2,4) {$\ibullet a\ibullet$};
    \node (2) at (0,2) {$b$};
    \node (3) at (0,0) {$\ibullet c$};
    \node (4) at (4,2) {$c\ibullet $};
    \node (5) at (4,0) {$d$};
    \node (6) at (8,0) {$a$};
    \path (2) edge (4);
    \path (3) edge (5);
    \path (2) edge (5);
    \path (5) edge (6);
    \path (1) edge[dashed] (2);
    \path (2) edge[dashed] (3);
    \path (1) edge[dashed] (4);
    \path (4) edge[dashed] (3);
    \path (4) edge[dashed] (5);
    \path (6) edge[dashed] (4);
    \path (6) edge[dashed, bend right] (1);
  \end{scope}
  \end{tikzpicture}

  \caption{Hasse diagram of an iposet on the left with
    $P=\{1,2,4,5,6\}$, $<$ indicate by solid arrows, $S=\{1,3\}$,
    $T=\{1,4\}$, $\Sigma=\{a,b,c,d\}$ and
    $\lambda:1\mapsto a,2\mapsto b,3\mapsto c,4\mapsto c, 5\mapsto
    d,6\mapsto a$, the corresponding ipomset on the right.}
  \label{fi:iposet}
\end{figure}

A \emph{subsumption} of an iposet $P$ by an iposet $Q$ is a bijection
$f:P\to Q$ between the elements of $P$ and $Q$ such that
\begin{itemize}
\item $f(S_P)=S_Q$ and $f(T_P)=T_Q$;
\item $f$ is $<$-reflecting ($f(x)<_Q f(y)$ implies $x<_P y$);
\item $f$ is $\intord$-preserving on $<_P$-incomparable
  elements ($x\not <_P y$, $y\not<_P x$ and $x\intord_P y$ imply $f(x)\intord_Q f(y)$);
\item labels are respected ($\lambda_P = \lambda_Q\circ f$).
\end{itemize}

This definition adapts the standard one
\cite{DBLP:journals/fuin/Grabowski81} to event orders and interfaces.
Intuitively, $P$ has more order and less concurrency than $Q$ if
$P\to Q$ is a subsumption.  See Figure~\ref{fi:subsumption} for an
example.

\begin{figure}
  \centering
  \begin{tikzpicture}[x=.8cm, y=.5cm]
    \node[circle,draw=black,inner sep=0pt,minimum size=15pt]
    (1) at (0,2) {$1$};
    \node[circle,draw=black,inner sep=0pt,minimum size=15pt]
    (2) at (0,0) {$2$};
    \node[circle,draw=black,inner sep=0pt,minimum size=15pt]
    (3) at (2,2) {$3$};
    \node[circle,draw=black,inner sep=0pt,minimum size=15pt]
    (4) at (2,0) {$4$};
       \node[circle,draw=black,inner sep=0pt,minimum size=15pt]
    (5) at (5,2) {$5$};
    \node[circle,draw=black,inner sep=0pt,minimum size=15pt]
    (6) at (5,0) {$6$};
    \node[circle,draw=black,inner sep=0pt,minimum size=15pt]
    (7) at (7,2) {$7$};
    \node[circle,draw=black,inner sep=0pt,minimum size=15pt]
    (8) at (7,0) {$8$};
    \path (1) edge (3);
    \path (2) edge (4);
    \path (1) edge (4);
    \path (1) edge[dashed] (2);
    \path (2) edge[dashed] (3);
    \path (3) edge[dashed] (4);
    \path (5) edge (8);
    \path (6) edge[dashed] (8);
    \path (5) edge[dashed] (6);
    \path (5) edge[dashed] (7);
    \path (6) edge[dashed] (7);
    \path (7) edge[dashed] (8);
    \path (1) edge [bend left, orange] (5);
    \path (3) edge [bend left , orange] (7);
       \path (2) edge [bend right, orange] (6);
     \path (4) edge [bend right, orange] (8);
    \node[above = .2cm] at (1) {$a$};
    \node[below = .2cm] at (2) {$\ibullet b$};
    \node[above = .2cm] at (3) {$c\ibullet$};
    \node[below = .2cm] at (4) {$d$};
      \node[above = .2cm] at (5) {$a$};
    \node[below = .2cm] at (6) {$\ibullet b$};
    \node[above = .2cm] at (7) {$c\ibullet$};
    \node[below = .2cm] at (8) {$d$};
  \end{tikzpicture}

  \caption{Subsumption map, indicated in orange, between two iposets.}
  \label{fi:subsumption}
\end{figure}

An \emph{isomorphism} of iposets is a subsumption that is an order
isomorphism.  The event order makes such isomorphisms unique.  We
write $P\subsu Q$ and say that $P$ is subsumed by $Q$, or that $Q$
subsumes $P$, if there exists a subsumption $P\to Q$. We write
$P\cong Q$ if $P$ and $Q$ are isomorphic. An \emph{ipomset} is an
isomorphism class of  iposets.

Intuitively, isomorphic iposets have the same order and action
structure, while the identity of events has been forgotten.  The uniqueness of
isomorphisms allows us to switch freely between ipomsets and
iposets. In particular it makes sense to say that an ipomset is
subsumed by another. This is the case if one can choose
representatives in the two ipomsets in such a way that the subsumption
map is the identity on representatives. The Hasse diagram on the right
of Figure~\ref{fi:iposet} shows the ipomset corresponding to the
iposet on the left.

An ipomset $P$ is \emph{discrete} if $<$ is empty and hence $\intord$
total.
For each  conclist $(U,\evord_U, \lambda_U)$ and subsets $S,T\subseteq U$ 
we  define the discrete ipomset 
\begin{equation*}
	\ilo{S}{U}{T}=(U,\emptyset,\evord_U,S,T,\lambda_U).
\end{equation*}
Ipomsets $\ilo{U}{U}{U}$ are called \emph{identity} ipomsets.  We
often write discrete ipomsets as vectors and indicate interfaces by
bullets:
$\left[\begin{smallmatrix} \bullet\, a\, \phantom{\bullet}\\
    \phantom{\bullet}\, b\, \bullet\end{smallmatrix}\right]$, for
instance, stands for the ipomset $\{a\evord b\}$ with $a$ in the
source and $b$ in the target interface.

Recall that a strict partial order $<$ on $P$ is an \emph{interval
  order} if it admits an interval
representation~\cite{book/Fishburn85}: a pair
$\beInt,\enInt:P\to \mathbb{R}$ such that $\beInt(x)\le \enInt(x)$ for
all $x\in P$ and $x< y$ if and only if $\enInt(x)<\beInt(y)$ for all
$x, y\in P$.  This excludes precisely the poset $2+2$ of shape
\begin{equation*}
  \begin{tikzpicture}[x=.6cm, y=.3cm]
    \node (w) at (0,2) {$w$};
    \node (x) at (2,2) {$x$};
    \node (y) at (0,0) {$y$};
     \node (z) at (2,0) {$z$};  
     \path (w) edge (x);
     \path (y) edge (z);
  \end{tikzpicture}
\end{equation*}
as an induced subposet, so that $w<z$ or $y<x$ whenever $w<x$ and
$y<z$.

This notion extends to iposets and ipomsets: an iposet is
\emph{interval} if its precedence is an interval order. We write
$\iPoms$ and $\iiPoms$ for the sets of ipomsets and interval ipomsets,
respectively.

\subsection*{Compositions}

The standard serial and parallel compositions of pomsets
\cite{DBLP:journals/fuin/Grabowski81} can be adapted for ipomsets. The
serial composition, in particular, becomes a gluing composition, as
studied previously by
Winkowski~\cite{DBLP:journals/ipl/Winkowski77}. Yet he considered a
less general class of ipomsets, in which interfaces are formed by
all minimal and all maximal elements, respectively, and where events
with the same label must be related by precedence. 

The \emph{parallel composition} $P\parallel Q$ of labelled iposets
$P$, $Q$ is the coproduct with respect to precedences and interfaces,
while the event order is extended so that events in $P$ are prior to
those in $Q$. Formally, $P\parallel Q$
\begin{itemize}
 \item  has the disjoint union
   $P\sqcup Q$ as carrier set;
   \item $S_{P\parallel Q}=S_P\sqcup S_Q$ and 
     $T_{P\parallel Q}=T_P\sqcup T_Q$;
   \item ${<_{P\parallel Q}}={<_P}\sqcup{<_Q}$;
     \item $x\intord_{P\parallel Q} y$ iff $x\intord_P y$, $x\intord_Q y$, or
$x\in P$ and $y\in Q$;
\item $\lambda_{P\parallel Q}$ is the standard extension of $\lambda_P$ and
$\lambda_Q$ to $P\sqcup Q$.
\end{itemize}

The \emph{gluing composition} $P*Q$ of labelled iposets $P$, $Q$ is a
partial operation,  defined whenever $T_P\cong S_Q$, and
\begin{itemize}
\item its carrier set is the quotient $(P\sqcup Q)_{/x\sim f(x)}$,
  where $f: T_P\to S_Q$ denotes the unique order isomorphism between
  these interfaces;
\item $S_{P*Q}=S_P$ and $T_{P*Q}=T_Q$;
    \item $x<_{P*Q} y$ iff
  $x<_P y$, $x<_Q y$, or $x\in P\setminus T_P$ and
  $y\in Q\setminus S_Q$;
\item $\intord_{P*Q}$ is the transitive closure of ${\intord_P}$ and
  ${\intord_Q}$ on $(P\sqcup Q)_{/x\sim f(x)}$;
  \item 
  $\lambda_{P * Q}$ is the standard extension of $\lambda_P$ and
  $\lambda_Q$ to $(P\sqcup Q) _{/x\sim f(x)}$.
\end{itemize}
The structural inclusions $P\hookrightarrow P*Q\hookleftarrow Q$
preserve both the precedence and the event order.  The event order is
crucial in this definition: it allows identifying elements of $T_P$
with elements in $S_Q$ in a unique way.

For ipomsets with empty interfaces, the gluing composition becomes the
standard serial pomset composition
\cite{DBLP:journals/fuin/Grabowski81}. For ipomsets, in which
interfaces are given by minimal and maximal elements and where events
with the same label are related by precedence, we recover Winkowski's
definition~\cite{DBLP:journals/ipl/Winkowski77}. In both cases we
ignore of course the event order. 

The gluing and parallel compositions of ipomsets respect isomorphisms
and thus lift to associative, non-commutative operations on ipomsets
(commutativity of $\parallel$ is broken by the event order).  Ipomsets
form a category with identity ipomsets as objects, ipomsets as arrows
and $*$ as composition. Examples of gluing and parallel compositions
can be found in Figure~\ref{fi:compositions}.

 \begin{figure}
  \centering
  \begin{tikzpicture}[x=.4cm, y=.4cm]
    \begin{scope}
    \node (1) at (2,2) {$a$};
    \node (2) at (0,0) {$b$};
    \node (3) at (4,0) {$c\ibullet$};
    \path (2) edge (3);
    \path (1) edge[dashed] (2);
    \path (3) edge[dashed] (1);
      \node (ast) at (5.5,1) {$\ast$};
    \node (4) at (7,2) {$d$};
    \node (5) at (7,0) {$\ibullet c$};
    \path (4) edge[dashed] (5);
    \node (equals) at (8.5,1) {$=$};
    \node (6) at (10,2) {$a$};
    \node (7) at (10,0) {$b$};
    \node (8) at (14,2) {$d$};
    \node (9) at (14,0) {$c$};
    \path (6) edge (8);
    \path (7) edge (9);
    \path (7) edge (8);
    \path (8) edge[dashed] (9);
    \path (9) edge[dashed] (6);
    \path (6) edge[dashed] (7);
  \end{scope}
  
 \begin{scope}[shift={(18,0)}]
    \node (1) at (2,2) {$a$};
    \node (2) at (0,0) {$b$};
    \node (3) at (4,0) {$c\ibullet$};
    \path (2) edge (3);
    \path (1) edge[dashed] (2);
    \path (3) edge[dashed] (1);
     \node (ast) at (5.5,1) {$\parallel$};
    \node (4) at (7,2) {$d$};
    \node (5) at (7,0) {$\ibullet c$};
    \path (4) edge[dashed] (5);
    \node (equals) at (8.5,1) {$=$};
       \node (6) at (12,4) {$a$};
    \node (7) at (10,2) {$b$};
    \node (8) at (14,2) {$c\ibullet$};
    \path (7) edge (8);
    \path (6) edge[dashed] (7);
    \path (8) edge[dashed] (6);
      \node (9) at (12,0) {$d$};
    \node (10) at (12,-2) {$\ibullet c$};
    \path (9) edge[dashed] (10);
    \path(7) edge[dashed] (9);
  \end{scope}
  \end{tikzpicture}
  
  \caption{Gluing and parallel composition of ipomsets.}
  \label{fi:compositions}
\end{figure}

\begin{exa}
  Interval ipomsets are closed under gluing
  compositions~\cite{Hdalang}, but not under parallel composition: the
  parallel composition of the interval ipomset
  $a\to b$ with itself yields the ipomset
  \begin{equation*}
  \begin{tikzpicture}[x=.6cm, y=.4cm]
    \node (02) at (0,2) {$a$};
    \node (22) at (2,2) {$b$};
    \node (00) at (0,0) {$a$};
     \node (20) at (2,0) {$b$};  
     \path (00) edge (20);
     \path (02) edge (22);
     \path[dashed] (02) edge (00);
     \path[dashed] (22) edge (20);
     \path[dashed] (02) edge (20);
      \path[dashed] (22) edge (00);
  \end{tikzpicture}
\end{equation*}
which obviously contains $2+2$ as an induced subposet in its
precedence. So it does not have the interval property.
\end{exa}

The following fact is important for constructing interval ipomsets
from paths of HDAs in Section~\ref{s:languages}. 
   
   \begin{propC}[{\cite[Proposition 44]{Hdalang}}]
     \label{p:hdalang}
     Interval ipomsets are closed under gluing composition, and
     all interval ipomsets can be generated by gluing finitely many
     discrete ipomsets.
   \end{propC}

   The \emph{width} $\wid(P)$ of an ipomset $P$ is the cardinality of a
   maximal $<$-antichain; its \emph{size} is
   $\size(P)=|P|-\tfrac{1}{2}(|S|+|T|)$.

   We glue ipomsets along interfaces and hence remove half of the
   interfaces when computing $\size$, which may thus be fractional.
   All identity ipomsets have size~$0$. The following lemmas are
   immediate consequences of the definitions.
\begin{lem}
  \label{l:widsize}
  Let $P$ and $Q$ be ipomsets.  Then
  \begin{enumerate}
  \item $\wid(P\parallel Q)=\wid(P)+\wid(Q)$ and
    $\size(P\parallel Q)= \size(P)+\size(Q)$,
  \item $T_P=S_Q$ implies $\wid(P*Q)=\max(\wid(P), \wid(Q))$ and
    $\size(P*Q)=\size(P)+\size(Q)$,
   \item $P\subsu Q$ implies $\wid(P)\leq \wid(Q)$ and $\size(P)=\size(Q)$.
  \end{enumerate}
\end{lem}

\begin{lem}~
  \label{l:ElIPoms}
  \begin{enumerate}
  \item For conclists $W\subseteq V\subseteq U$\!,
  $\ilo{W}{V}{V} * \ilo{V}{U}{U}=\ilo{W}{U}{U}$.
   \item For conclists $V, W\subseteq U$ with $U=V\cup W$\!,
   $\ilo{V}{V}{V\cap W}*\ilo{V\cap
   W}{W}{W}\subsu \ilo{V}{U}{W}$.
  \end{enumerate}
\end{lem}
Lemma~\ref{l:ElIPoms} can be illustrated by the following pictures:
\begin{equation*}
\vcenter{\hbox{\begin{tikzpicture}
    \node (a) at (0,0.6) {$\phantom{\ibullet} a\ibullet$};
    \node (b) at (0,-1.8) {$\ibullet b \ibullet$};
     \path (a) edge[dashed] (b);
  \end{tikzpicture}}}
  \ast
\vcenter{\hbox{\begin{tikzpicture}
    \node (a) at (0,0.6) {$\ibullet a\ibullet$};
    \node (c) at (0,-0.6) {$\phantom{\ibullet} c \ibullet$};
    \node (b) at (0,-1.8) {$\ibullet b \ibullet$};
     \path (a) edge[dashed] (c);
     \path (c) edge[dashed] (b);
  \end{tikzpicture}}}
  =
\vcenter{\hbox{\begin{tikzpicture}
    \node (a) at (0,0.6) {$\phantom{\ibullet} a\ibullet$};
    \node (c) at (0,-0.6) {$\phantom{\ibullet} c \ibullet$};
    \node (b) at (0,-1.8) {$\ibullet b \ibullet$};
     \path (a) edge[dashed] (c);
     \path (c) edge[dashed] (b);
  \end{tikzpicture}}}
\qquad\qquad
	\vcenter{\hbox{\begin{tikzpicture}
    \node (a) at (0,0.6) {$\ibullet a\ibullet$};
    \node (b) at (0,-0.6) {$\ibullet b \phantom{\ibullet}$};
     \path (a) edge[dashed] (b);
  \end{tikzpicture}}}
  \ast
	\vcenter{\hbox{\begin{tikzpicture}
    \node (a) at (0,0.6) {$\ibullet a\ibullet$};
    \node (b) at (0,-0.6) {$\phantom{\ibullet} c \ibullet$};
     \path (a) edge[dashed] (b);
  \end{tikzpicture}}}
  =
	\vcenter{\hbox{\begin{tikzpicture}
    \node (a) at (0,0.6) {$\ibullet a\ibullet$};
    \node (b) at (-0.8,-0.6) {$\ibullet b $};
    \node (c) at (0.8,-0.6) {$ c \ibullet$};
     \path (a) edge[dashed] (b) edge[dashed] (c);
     \path (b) edge (c);
  \end{tikzpicture}}}
  \sqsubseteq
	\vcenter{\hbox{\begin{tikzpicture}
    \node (a) at (0,0.6) {$\ibullet a\ibullet$};
    \node (b) at (-0.8,-0.6) {$\ibullet b $};
    \node (c) at (0.8,-0.6) {$ c \ibullet$};
     \path (a) edge[dashed] (b);
     \path (b) edge[dashed] (c);
  \end{tikzpicture}}}
\end{equation*}

\subsection*{Ipomset languages and rational languages}

We define an \emph{interval ipomset language} (a \emph{language} for
short) as a subset $L\subseteq \iiPoms$ that is \emph{down-closed}
with respect to subsumption: if $P\subsu Q$ and $Q\in L$, then
$P\in L$. If $X$ is a set of ipomsets, then
\begin{equation*}
	X\down=\{P\in\iiPoms\mid \exists\, Q\in X:  P\subsu Q\}
\end{equation*}
      indicates the language that is its down-closure with respect to
      subsumption.

We define the \emph{rational operations} $\cup$, $*$, $\|$ and $^+$,
the Kleene plus, for languages as set union,
\begin{align*}
  L*M &= \{P*Q\mid P\in L,\; Q\in M,\; T_P=S_Q\}\down, \\
  L\parallel M &= \{P\parallel Q\mid P\in L,\; Q\in M\}\down, \\
  L^+ &= \bigcup_{n\ge 1} L^n,\quad \text{ for } L^1=L \text{ and } L^{n+1}=
  L*L^n.
\end{align*}

Down-closure is needed because parallel compositions of interval
ipomsets may not be interval ipomsets and gluing and parallel
compositions of down-closed languages may not be down-closed.

\begin{exa}
  $\{[a]\parallel [b]\}=
  \{[\begin{smallmatrix}a\\b\end{smallmatrix}]\} =
  \left\{\left[\begin{smallmatrix}a & \ibullet\\ b
        &\ibullet\end{smallmatrix}\right] *
    \left[\begin{smallmatrix}\ibullet& a\\ \ibullet&
        b\end{smallmatrix}\right] \right\}$ is not down-closed.
\end{exa}
It is routine to check that gluing and parallel composition of
languages are associative and that neither operation is commutative.
The identity of $\parallel$ is $\{\epsilon\}$, that of $*$ is the
\emph{identity language} $\Id = \{\ilo U U U\mid U\in \sq\}$ of all
identity ipomsets.

The \emph{rational languages} are then the smallest class of languages
that contains the empty language, the empty-pomset language and the
singleton pomset languages
\begin{equation}
  \label{e:singletons}
  \emptyset,\;\; \{\epsilon\},\;\; \{[a]\},\;\; \{[\ibullet\, a]\},\;\;\{[a\,
  \ibullet]\},\;\; \{[\ibullet\, a\, \ibullet]\},\;\; a\in \Sigma,
\end{equation}
and that is closed under the rational operations $\cup$, $*$,
$\|$ and  $^+$.

We define the \emph{width} of a language $L$ as the maximal width among its
elements:
\begin{equation*}
  \wid(L) = \sup\{\wid(P)\mid P\in L\}.
\end{equation*}

Lemma \ref{l:widsize} implies that all rational languages have finite
width. The identity language $\Id$, however, has infinite width and is
therefore not rational.  This explains why we consider the Kleene plus
instead of the more conventional Kleene star in the definition of
rationality: $L^\ast = \Id\cup L^+$, like $\Id$, is not rational.

\subsection*{Separated languages}

An ipomset $P$ is \emph{separated} if
$P\setminus(S_P\cup T_P)\neq \emptyset$, that is, it contains an
``interior'' element that does not belong to an interface.  A language is
\emph{separated} if all its ipomsets are separated.

\begin{lem}
  \label{l:SeparatingPower}
  If a language $L$ with $L\cap \Id=\emptyset$ has finite width and if
  $n$ is sufficiently large, then $L^n$
  is separated.
\end{lem}
\begin{proof}
  For every ipomset $Q\in L^n$ there exists an ipomset
  $P=P_1*\dotsc*P_n$ such that each $P_k\in L$
  and $Q\subsu P$. As $\size(P_k)\ge \frac 1 2$, additivity of size
  implies
  \begin{equation*}
    \size(Q)=\size(P)=\size(P_1)+\dotsc+\size(P_n)\ge \tfrac n 2.
  \end{equation*}
  Thus
  $|S_Q|,|T_Q|\leq \wid(Q)\leq \wid(P)= \max_k\wid(P_k)\leq \wid(L)$,
  as gluing compositions do not increase width. Eventually,
  \begin{equation*}
    |S_Q|+|T_Q|\leq 2\wid(L)< n\leq 2\size(Q)=2|Q|-|S_Q|-|T_Q|
  \end{equation*}
  holds for $n\ge 2\wid(L)+1$ and therefore $|S_Q|+|T_Q|<|Q|$.
\end{proof}


\section{Executions of higher-dimensional automata}
\label{s:languages}

Executions of HDAs are higher-dimensional paths that keep track of the
cells and face maps traversed~\cite{DBLP:journals/tcs/Glabbeek06}. In
this section we recall their definition. As an important stepping
stone towards a Kleene theorem, we then relate paths of HDAs with
ipomsets -- for a more general class than in~\cite{Hdalang}. We also
introduce notions of path equivalence and subsumption.  The latter
corresponds to ipomset subsumption. We end with a definition of
regular languages.

\subsection*{Paths}

A \emph{path} of length $n$ in a precubical set $X$ is a sequence
\begin{equation}
\label{e:Path}
  \alpha = (x_0,\phi_1,x_1,\phi_2,\dotsc,\phi_n,x_n),
\end{equation}
where the $x_k\in X[U_k]$ are cells and, for all $k$, either
\begin{itemize}
\item an \emph{up-step} $\phi_k=d^0_{A}\in \sq(U_{k-1},U_k)$, $A\subseteq U_k$ and
  $x_{k-1}=\delta^0_A(x_k)$ or
\item a \emph{down-step} $\phi_k=d^1_B\in \sq(U_{k},U_{k-1})$, $B\subseteq U_{k-1}$,
  $\delta^1_B(x_{k-1})=x_{k}$.
\end{itemize}
We write $x_{k-1}\arrO{A}x_k$ for the up-steps and
$x_{k-1}\arrI{B}x_k$ for the down-steps in $\alpha$, generally
assuming that $A\neq\emptyset\neq B$. We further refer to the up- or
down-steps in paths as \emph{steps} and write $\Qath{X}$ for the set
of all paths on the precubical set $X$.

\begin{exa}
  \label{ex:pathex}
  The diagram on the right of Figure \ref{fi:hda-cylinder} in the
  introduction depicts the paths
  \begin{align*}
  \alpha_1 &= (\delta_{ab}^0(x)\arrO{a}
  \delta_b^0(x)\arrI{a} \delta_{b,a}(y)),\\
  \alpha_2 &= (\delta_{ab}^0(x)\arrO{ab} x\arrI{b}
  \delta_b^1(x)\arrO{c} y\arrI{ac} \delta_{ac}^1(y)),\\
  \alpha_3 &= (\delta_{ab}^0(x)\arrO{b}
  \delta_a^0(x)\arrI{b} \delta_{ac}^0(y)\arrO{ac} y\arrI{ac}
  \delta_{ac}^1(y)).
  \end{align*}
\end{exa}

We define the \emph{source} and \emph{target} of a path $\alpha$, as
in formula (\ref{e:Path}), as $\src(\alpha)=x_0$ and
$\tgt(\alpha)=x_n$.  Each pc-map $f: X\to Y$ induces a map
$f: \Qath{X}\to \Qath{Y}$. For $\alpha$ as above it is
\begin{equation}
  f(\alpha)=(f(x_0),\phi_1,f(x_1),\phi_2,\dotsc,\phi_n,f(x_n)).\label{eq:ind-pcmap}
\end{equation}

The \emph{concatenation} of paths $\alpha=(x_0, \phi_1,\dotsc, x_n)$
and $\beta=(y_0, \psi_1,\dotsc, y_m)$ with $\tgt(\alpha)=\src(\beta)$
is defined as
$\alpha*\beta=(x_0, \phi_1,\dotsc, x_n, \psi_1,\dotsc, y_m)$, hence
again by gluing ends. This turns $\Qath{X}$ into a category with cells
of $X$ as objects and paths as morphisms, in generalisation of the
standard path categories generated by digraphs. Moreover, for
$x, y\in X$, we write
\begin{equation*}
\Qathft{X}{x}{y} = \{\alpha\in\Qath{X}\mid \src(\alpha)=x,
\tgt(\alpha)=y\}
\end{equation*}
for the homset of paths from $x$ to $y$.

\subsection*{Reachability and accessibility}

The cell $y\in X$ is \emph{reachable} from the cell $x\in X$, denoted
$x\preceq y$, if there is a path from $x$ to $y$ in $X$.  This
reachability preorder is generated by
$\delta_A^0(x)\preceq x\preceq \delta^1_B(x)$ for $x\in X$ and
$A, B\subseteq \ev(x)$.  The precubical set $X$ is \emph{acyclic} if
$\preceq$ is a partial order, or equivalently, if $\Qathft{X}{x}{x}$
contains only the constant path $(x)$ for each $x\in X$.

A path $\alpha\in\Qath{X}$ in an HDA $X$ is \emph{accepting} if
$\src(\alpha)\in X_\bot$ and $\tgt(\alpha)\in X^\top$.  A cell $x$ is
\emph{accessible} if there exists a path from a start cell to $x$, and
\emph{co-accessible} if there is a path from $x$ to an accept cell.  A
cell is \emph{essential} if it is both accessible and co-accessible.
All cells in accepting paths are essential.

\subsection*{Ipomsets of paths}

Next we introduce a map $\ev$ that computes
ipomsets of paths.

The interval ipomset $\ev(\alpha)$ of a path $\alpha\in\Qath{X}$ is
computed recursively:
\begin{itemize}
\item If $\alpha=(x)$ is a path of length $0$, then
  $\ev(\alpha)=\ilo{\ev(x)}{\ev(x)}{\ev(x)}$.
\item If $\alpha=(y\arrO{A} x)$, then
  $\ev(\alpha)=\ilo{\ev(x)\setminus A}{\ev(x)}{\ev(x)}$.
\item If $\alpha=(x\arrI{B} y)$, then
  $\ev(\alpha)=\ilo{\ev(x)}{\ev(x)}{\ev(x)\setminus B}$.
\item If $\alpha=\beta_1*\dotsm*\beta_n$ is a concatenation
  of steps $\beta_i$, then
  $\ev(\alpha)=\ev(\beta_1)*\dotsm*\ev(\beta_n)$.
\end{itemize}
Interfaces and gluings of ipomsets are essential for this
construction. The event order allows us to identify the target events
of a preceding ipomset with the events of a succeeding one.

\begin{exa}\label{ex:path2pomset}
  The ipomset of the path $\alpha_1$ in Example \ref{ex:pathex} is
  computed as
  \begin{align*}
    \ev(\alpha_1) = \ev(\delta_{ab}^0(x)\arrO{a} \delta_b^0(x)) *
                    \ev(\delta_b^0(x)\arrI{a} \delta_{b,a}(y))
    = \ilo{\emptyset}{a}{a}
    * \ilo{a}{a}{\emptyset}
    = a.
\end{align*}
Those of the other two paths in the example are
$\ev(\alpha_2) = a\parallel (b\to c) $ and $\ev(\alpha_3) = b * \left[
    \begin{smallmatrix}
      a \\ c
    \end{smallmatrix}
  \right]$. Figure~\ref{fig:abiposet} contains an additional example.
\end{exa}

Proposition~\ref{p:hdalang} guarantees the following important
structural property.

\begin{lem}\label{l:IntPath}
 For each path $\alpha \in \Qath{X}$,  $\ev(\alpha)$ is an
  interval ipomset.
\end{lem}

The following facts are immediate from the definition of $\ev$ and
induced paths maps, as well as associativity of gluing composition.

\begin{lem}
  \label{l:EvComp}
  Let $\alpha, \beta\in \Qath{X}$. Then
  $\ev(\alpha*\beta)=\ev(\alpha)*\ev(\beta)$ whenever $\tgt(\alpha)=\src(\beta)$.
\end{lem}

\begin{lem}
  \label{l:EvFunc}
 If $f:X\to Y$ is a pc-map and $\alpha\in\Qath{X}$,
 then $\ev(f(\alpha))=\ev(\alpha)$.
\end{lem}

\subsection*{Event consistency for paths}

Let $X$ be an HDA. For any path
$\alpha=(x_0,\varphi_1,\dotsc,x_n)\in\Qath{X}$, the conclists
$\ev(x_k)$ are defined only up to isomorphism. Similarly, the
$\varphi_k$ are morphisms in $\sq$ and not actual {\cofacemap}s, but
rather their equivalence classes.  The next lemma allows choosing
conclists and {\cofacemap}s as representatives in a consistent way,
and using the simple composition of {\cofacemap}s in formula
\eqref{e:simplesqcomp} in calculations.
\begin{lem}
  \label{l:ConsistentEvents}
  For every $\alpha=(x_0, \phi_1,\dotsc, x_n)\in\Qath{X}$ there exist
  conclists $U_0,\dotsc,U_n\subseteq \ev(\alpha)$ such that $\ev(x_k)=U_k$ and either
  \begin{itemize}
  \item $U_{k-1}\subseteq U_k$ and
    $\phi_k=d^0_{U_k\setminus U_{k-1}}:U_{k-1}\to U_k$ or
  \item $U_{k-1}\supseteq U_k$ and
    $\phi_k=d^1_{U_{k-1}\setminus U_{k}}:U_k\to U_{k-1}$.
  \end{itemize}    
\end{lem}
\begin{proof}
The structural inclusion
\[
    \ev(x_k)\subseteq\ev(x_0,\varphi_1,\dotsc,x_{k})*\ev(x_k)*\ev(x_k,\varphi_{k+1},\dotsc,x_{n})
\]
defines an ipomset inclusion $j_k:\ev(x_k)\subseteq \ev(\alpha)$.  So
let $U_k=j_k(\ev(x_k))$.  If $\varphi_k=d^0_A$ is an up-step, then
$U_{k-1}\subseteq U_k$ and
$x_{k-1}=\delta^0_{U_k\setminus U_{k-1}}(x_k)$; if $\varphi_k=d^1_B$
is a down-step, then $U_{k}\subseteq U_{k-1}$ and
$x_{k}=\delta^1_{U_{k-1}\setminus U_{k}}(x_{k-1})$.
\end{proof}

Note that if $x_i=x_j$ for $i\neq j$, then $U_i$ and $U_j$ are
different, but isomorphic subsets of $\ev(\alpha)$.
Henceforth we choose conclists of cells in paths as in Lemma\
\ref{l:ConsistentEvents}.  This simplifies calculations and underlines
the relevance of $\fullsq$ as a base category for precubical sets.

\subsection*{Path equivalence and subsumption}

\emph{Path equivalence} is the congruence $\simeq$ on $\Qath{X}$
generated by
\begin{enumerate}
\item
	$(z\arrO{A}y\arrO{B} x)\simeq (z\arrO{A\cup B}x)$,
\item
	$(x\arrI{A}y\arrI{B} z)\simeq (x\arrI{A\cup B}z)$,
\item
	$\gamma*\alpha*\delta \simeq\gamma*\beta*\delta$
whenever $\alpha\simeq\beta$.
\end{enumerate}
Further, \emph{path subsumption} is the
transitive relation $\subsu$ on $\Qath{X}$ generated by
\begin{enumerate}\setcounter{enumi}{3}
\item
	$(y\arrI{B}w\arrO{A}z)\subsu(y\arrO{A}x\arrI{B}z)$, for disjoint
$A,B\subseteq \ev(x)$,
\item
	$\gamma*\alpha*\delta\subsu \gamma*\beta*\delta$ whenever
	$\alpha\subsu\beta$,
\item
	$\alpha\subsu\beta$ whenever
	$\alpha\simeq\beta$.
\end{enumerate}
We say that $\beta$ \emph{subsumes} $\alpha$ if $\alpha\subsu \beta$.

Intuitively, if $\beta$ subsumes $\alpha$, then $\beta$ is more
concurrent than $\alpha$ and $\alpha$ more sequential than $\beta$.
Both $\simeq$ and $\subsu$ preserve sources and targets of paths, and
they translate to ipomsets as follows.
\begin{lem}
  If $\alpha,\beta\in\Qath{X}$, then
  \begin{enumerate}
    \item $\alpha\simeq \beta \Rightarrow
  \ev(\alpha)=\ev(\beta)$,
  \item $\alpha\subsu \beta \Rightarrow
    \ev(\alpha)\subsu\ev(\beta)$.
\end{enumerate}
\end{lem}
\begin{proof}
  We need to check items (1)--(6) from the definition of path
  equivalence and subsumption. Item (1) holds because
  \begin{align*}
 	\ev(z\arrO{A}y\arrO{B} x)
  	&=\ev(z\arrO{A}y)*\ev(y\arrO{B} x) \tag{Lemma \ref{l:EvComp}}
  	\\
  	&=\ilo{\ev(y)\setminus A}{\ev(y)}{\ev(y)}*\ilo{\ev(x)\setminus B}{\ev(x)}{\ev(x)}
  	\\
  	&=\ilo{\ev(z)}{\ev(y)}{\ev(y)}*\ilo{\ev(y)}{\ev(x)}{\ev(x)}
  	\\
  	&= \ilo{\ev(z)}{\ev(x)}{\ev(x)}	\tag{Lemma \ref{l:ElIPoms}.(1)}
  	\\
  	&= \ilo{\ev(x)\setminus(A\cup B)}{\ev(x)}{\ev(x)}
  	\\
  	&=\ev(z\arrO{A\cup B}x),
  \end{align*}
  The proof of (2) is similar and (3) follows immediately from Lemma
  \ref{l:EvComp}.  For (4), fix $x\in X$ and suppose
  $A,B\subseteq \ev(x)$ are disjoint subsets.  Let $y=\delta^0_A(x)$,
  $z=\delta^1_B(x)$, $w=\delta_{A,B}(x)$ and denote $U=\ev(x)$,
  $V=U\setminus A=\ev(y)$, $W=U\setminus B=\ev(z)$.  Then
  \begin{align*}
 	\ev(y\arrI{A}w\arrO{B} z)
  	&=\ev(y\arrI{A}w)*\ev(w\arrO{B} z) \tag{Lemma \ref{l:EvComp}}
  	\\
  	&=\ilo{\ev(y)}{\ev(y)}{\ev(y)\setminus A}*\ilo{\ev(z)\setminus B}{\ev(z)}{\ev(z)}
  	\\
  	&=\ilo{V}{V}{V\cap W}*\ilo{V\cap W}{W}{W}
  	\\
  	&\subsu \ilo{V}{U}{W}=\ilo{V}{U}{U}*\ilo{U}{U}{W}	\tag{Lemma \ref{l:ElIPoms}.(2)}
  	\\
  	&= \ilo{\ev(x)\setminus A}{\ev(x)}{\ev(x)}*\ilo{\ev(x)}{\ev(x)}{\ev(x)\setminus B }
  	\\
  	&=\ev(z\arrO{A} x \arrI{B}z).
  \end{align*}
  Finally, (5) follows again from Lemma \ref{l:EvComp} and (6) is
  straightforward.
\end{proof}

\begin{exa}
  \label{ex:pathex-subs}
  It is easy to check that the path $\alpha_3$ in Example~\ref{ex:pathex}
  is subsumed by $\alpha_2$, and so are the corresponding pomsets in
  Example~\ref{ex:path2pomset}:
  $\ev(\alpha_3) = b * \left[
    \begin{smallmatrix}
      a \\ c
    \end{smallmatrix}
  \right] \subsu a\parallel (b\to c)
  = \ev(\alpha_2)$.
\end{exa}

\subsection*{Regular languages}
An ipomset $P$ is \emph{recognised} by the HDA $X$ if $P=\ev(\alpha)$
for some accepting path $\alpha$ of $X$.  We write
\[
	\Lang(X)=\{\ev(\alpha)\mid \text{$\alpha\in\Qath{X}$ is
  accepting}\}
  \]
  for the set of interval ipomsets recognised by $X$. The language $L$
  is \emph{regular} if it is recognised by an HDA, that is,
  $L=\Lang(X)$ for some HDA $X$.
 
  Every regular language is down-closed by
  Proposition~\ref{p:LangsAreSubClosed} below and an interval ipomset
  language by Lemma~\ref{l:IntPath}.

\begin{lem}
  \label{l:widledim}
 Regular languages  have finite width.
\end{lem}
\begin{proof}
  Let $X$ be an HDA. We show that $\wid(\Lang(X))\leq \dim(X)$.  It is
  clear that $\wid(\ev(\alpha))\le \dim(X)$ for any path $\alpha$ in
  $X$.  The claim then follows by Lemma~\ref{l:widsize}.
\end{proof}

By a distant analogy with topology, we call an HDA map $f:X\to Y$ a
\emph{weak equivalence} if for every accepting path $\beta\in\Qath{Y}$
there exists an accepting path $\alpha\in\Qath{X}$ with
$f(\alpha)=\beta$ with respect to the induced $f:\Qath{X}\to \Qath{Y}$
defined in (\ref{eq:ind-pcmap}).

\begin{lem}
	Let $f:X\to Y$ be an HDA-map. Then
  \label{l:FunctorialityOfLanguages}
  \label{l:EqHDA}
  \begin{enumerate}
 \item  $\Lang(X)\subseteq \Lang(Y)$,
 \item if $f$ is a weak equivalence, then
   $\Lang(X)=\Lang(Y)$.
  \end{enumerate}
\end{lem}

\begin{proof}
  Suppose $P\in\Lang(X)$. Then there is an accepting path
  $\alpha\in\Qath{X}$ such that $\ev(\alpha)=P$.  Thus the induced
  path $f(\alpha)$ is also accepting and
  $P=\ev(\alpha)=\ev(f(\alpha))\in\Lang(Y)$ by Lemma~\ref{l:EvFunc}.
  The second claim is clear.
\end{proof}

We conclude this section with two elementary facts about regular
languages.

\begin{prop}
  \label{p:RegSingletons}
  The empty language, the empty-pomset language and the singleton
  pomset languages in \eqref{e:singletons} are regular.
\end{prop}

\begin{proof}
  These languages are recognised by the following HDAs:
  \begin{equation*}
    \hfill
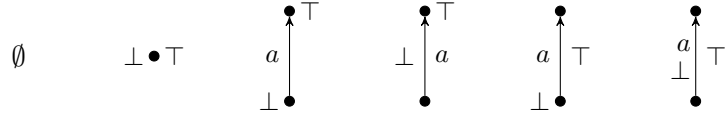
    \begin{tikzpicture}[x=1.8cm,y=1.2cm]
      \begin{scope}[shift={(0,.5)}]
        \node at (0,0) {\normalsize $\emptyset$};
      \end{scope}
      \begin{scope}[shift={(1,.5)}]
        \node[state] (00) at (0,0) {};
        \node [left] at (00) {$\bot$};
        \node [right] at (00) {$\top$};
      \end{scope}
      \begin{scope}[shift={(2,0)}]
        \node[state] (0) at (0,0) {};
        \node[left] at (0) {$\bot$};
        \node[state] (1) at (0,1) {};
        \node[right] at (1) {$\top$};
        \path (0) edge node {$a$} (1);
      \end{scope}
      \begin{scope}[shift={(3,0)}]
        \node[state] (0) at (0,0) {};
        \node[state] (1) at (0,1) {};
        \node[right] at (1) {$\top$}; 
        \path (0) edge node[right] {$a$} node[left] {$\bot$} (1);
      \end{scope}
      \begin{scope}[shift={(4,0)}]
        \node[state] (0) at (0,0) {};
        \node[left] at (0) {$\bot$};
        \node[state] (1) at (0,1) {};
        \path (0) edge node {$a$} node[right] {$\top$} (1);
            \end{scope}
      \begin{scope}[shift={(5,0)}]
        \node[state] (0) at (0,0) {};
        \node[state] (1) at (0,1) {};
         \path (0) edge node[above left=-1mm] {$a$} node[below left=-1mm]
         {$\bot$} node[right] {$\top$} (1);
      \end{scope}
    \end{tikzpicture}
    \hfill \mbox{}%
    \qedhere
  \end{equation*}
\end{proof}

\begin{prop}
  \label{p:RegUnion}
  Finite unions of regular languages are regular.
\end{prop}

\begin{proof}
  $\Lang(X\sqcup Y)=\Lang(X)\cup \Lang(Y)$, where the HDA $X\sqcup Y$
  is the coproduct of the HDAs $X$ and $Y$.
\end{proof}


\section{Kleene theorem}
\label{s:KleeneTheorem}

We can now state the Kleene theorem for HDAs, which relates them with
interval ipomset languages. In this section, we also provide a roadmap
towards its proof. Technical details and advanced concepts needed for
it are introduced in the remaining sections of this article.

\begin{thm}[Kleene theorem for HDAs]
  \label{t:kleene}
  A language is regular if and only if it is rational.
\end{thm}

The Kleene theorem follows from a series of propositions, which we
explain in the sequel. First we outline its left-to-right direction.

\begin{prop}
  \label{p:RegIsRat}
  Every regular language is rational.
\end{prop}

This proposition is obtained from a translation to the Kleene theorem
for standard finite state automata in Section \ref{s:reg2rat}. For
each HDA we construct a standard automaton with an alphabet ranging
over discrete ipomsets, and we show that it accepts the same language
as the HDA.

Proving the right-to-left direction of Theorem~\ref{t:kleene} is
harder. Our proof follows that of the classical Kleene theorem. We
inductively construct HDAs that accept the generators of rational
languages and the languages obtained by application of the rational
operations to regular languages. We have already shown
(Propositions~\ref{p:RegSingletons} and \ref{p:RegUnion}) that the
empty language, the empty-pomset language and the singleton pomset
languages are regular, and that regularity is preserved by finite
unions.  So it remains to prove that the remaining rational operations
-- parallel compositions, gluing composition and the Kleene plus --
preserve regularity as well.

\begin{prop}
  \label{p:RegParallel}
	Parallel compositions of regular languages are regular.
\end{prop}
      
In Section \ref{s:tensor} we introduce tensor products of HDAs and
show that tensor products of HDAs recognise the parallel
composition of their languages.  The proof uses an alternative
definition of languages of HDAs via track objects, introduced in
Section \ref{s:Tracks}.

The corresponding proofs for gluing compositions and the Kleene plus
are more intricate and require additional machinery. They constitute
the main technical contribution of this paper, and the tools
introduced may be of independent interest.

\begin{prop}
  \label{p:RegConcat}
  Gluing compositions of regular languages are regular.
\end{prop}

\begin{prop}
  \label{p:RegPlus}
  The Kleene plus of a regular language is regular.
\end{prop}

The ideas behind the proofs of these propositions are outlined in the
remainder of this section; the proofs themselves are developed in
Sections \ref{s:ihda} to \ref{s:Spider}.  The left-to-right direction
of the Kleene theorem then follows.

\begin{cor}
  \label{c:RatIsReg}
  Every rational language is regular. 
\end{cor}

\begin{proof}
  By Propositions \ref{p:RegSingletons}, \ref{p:RegUnion} and 
  \ref{p:RegParallel}--\ref{p:RegPlus}.
\end{proof}

It then remains to reap what we have sown.

\begin{proof}[Proof of Theorem \ref{t:kleene}]
  By Proposition~\ref{p:RegIsRat} and Corollary \ref{c:RatIsReg}.
\end{proof}

The ideas behind the proofs of Propositions \ref{p:RegConcat} and
\ref{p:RegPlus} and the intricacies encountered are similar.  We focus
on Proposition \ref{p:RegConcat} because the tools needed for proving
Proposition \ref{p:RegPlus} are more complicated.

Our goal in the proof of Proposition \ref{p:RegConcat} is the
construction, for
each pair of HDAs $X$ and $Y$, of an HDA $Z$ that recognises $\Lang(X)*\Lang(Y)$ .  For simplicity, we assume that both HDAs have
one start cell ($X_\bot=\{x_\bot\}$, $Y_\bot=\{y_\bot\}$) and one
accept cell ($X^\top=\{x^\top\}$, $Y^\top=\{y^\top\}$),
respectively. We further assume that the conclists of $x^\top$ and
$y_\bot$ agree: $\ev(x^\top)=\ev(y_\bot)=U$.  A natural candiate for
$Z$ is the HDA obtained from $X\sqcup Y$ by identifying $x^\top$ in
$X$ and $y_\bot$ in $Y$, or more formally, from a \emph{gluing
  composition} of $X$ and $Y$ defined as
\begin{equation*}
	X*Y=\colim\left( X \xleftarrow{\iota_{x^{\top}}} \sq^U \xrightarrow{\iota_{y_\bot}} Y   \right).
\end{equation*}
It is then routine to check that
$\Lang(X)*\Lang(Y)\subseteq \Lang(X*Y)$.

For standard finite automata $X$, $Y$, this construction yields indeed
$\Lang(X*Y)=\Lang(X)*\Lang(Y)$ whenever there are no transitions from
$x^\top$ into a state of $X$ and no transitions from a state of $Y$
into $y_\bot$.  Otherwise, $Z$ could allow scanning strings in
$\Lang(X)$ after having started scanning strings in $\Lang(Y)$. For
HDAs, we thus need a construction that brings $X$ and $Y$ into a
similar shape to prevent such backdoor scanning. A second complication
that is particular to HDAs is that gluing compositions $X*Y$ not only
identify $x^\top$ and $y_\bot$. Paths that do not cross the ``gluing''
cell may therefore appear and contribute to the language of $X*Y$,
though their prefixes or suffixes are not in the language of $X$ or
$Y$, see Figure \ref{fi:GluingExample}.

 \begin{figure}
  \centering
\begin{tikzpicture}
    \begin{scope}[x=1cm, y=1cm, shift={(0,0)}]
        \path[fill=black!10] (0,0) to (2,0) to (2,2) to (0,2) to (0,0);
        \node[state] (00) at (0,0) {};
        \node[state] (20) at (2,0) {};
        \node[state] (02) at (0,2) {};
        \node[state] (22) at (2,2) {};    	
        \path (00) edge node[below] {$a$} (20);
        \path (02) edge  (22);
        \path (00) edge node[left] {$c$} (02);
        \path (20) edge (22);
        \node[below left] at (00) {$\bot$};
        \node[right] at (2,1) {$\top$};
		\node at (-0.4, 2.4) {$X$};
	\end{scope}
    \begin{scope}[x=1cm, y=1cm, shift={(4,0)}]
        \path[fill=black!10] (0,0) to (2,0) to (2,2) to (0,2) to (0,0);
        \node[state] (00) at (0,0) {};
        \node[state] (20) at (2,0) {};
        \node[state] (02) at (0,2) {};
        \node[state] (22) at (2,2) {};    	
        \path (00) edge node[below] {$b$} (20);
        \path (02) edge  (22);
        \path (00) edge  (02);
        \path (20) edge node[right] {$c$} (22);
        \node[below right] at (20) {$\top$};
        \node[left] at (0,1) {$\bot$};
		\node at (-0.4, 2.4) {$Y$};
	\end{scope}
      \begin{scope}[x=1cm, y=1cm, shift={(8,0)}]
        \path[fill=black!10] (0,0) to (4,0) to (4,2) to (0,2);
        \node[state] (00) at (0,0) {};
        \node[state] (20) at (2,0) {};
        \node[state] (40) at (4,0) {};
        \node[state] (02) at (0,2) {};
        \node[state] (22) at (2,2) {};
        \node[state] (42) at (4,2) {};
        \path (00) edge node[below] {$a$} (20);
	    \path (20) edge node[below] {$b$} (40);
        \path (02) edge  (22);
	    \path (22) edge  (42);
        \path (00) edge node[left] {$c$} (02);
        \path (20) edge (22);
        \path (40) edge node[right] {$c$} (42);
        \node[below left] at (00) {$\bot$};
        \node[below right] at (40) {$\top$};
		\node at (-0.4, 2.4) {$X*Y$};
       \end{scope}
       \node at (3,1) {$*$};
       \node at (7,1) {$=$};
\end{tikzpicture}  
\caption{The language of gluings of HDAs need not be the gluing
  composition of their languages: $\Lang(Y)=\emptyset$, but
  $ab\in \Lang(X*Y) \neq \Lang(X)*\Lang(Y)= \emptyset$.}
  \label{fi:GluingExample}
\end{figure}

We introduce two tools to deal with this situation.  First, we
introduce higher-dimensional automata with interfaces (iHDAs) in
Section \ref{s:ihda} as an alternative to HDAs, which allows us to
mark events that cannot terminate and those that cannot be
unstarted in an HDA, and to trace such elements across cells. HDAs and
iHDAs are related to each other via a pair of language-preserving
functors $\Res:\HDA\to\IHDA$ and $\Cl:\IHDA\to\HDA$, introduced in
Section \ref{s:hda-ihda}, which we call \emph{resolution} and
\emph{closure}. In particular, we use these functors to show that HDAs
and iHDAs recognise the same class of regular languages.  This allows
us to work with both kinds of automata, depending on the context, but
using iHDAs guarantees better properties of some of our gluing
constructions.

Second, we introduce a construction that removes transitions into
start cells or out of accept cells -- but generally not both -- and which
separates start or accept cells, so that the sets of their faces are
disjoint.  The resulting iHDAs are called (start or accept) proper. To
enable this construction, we introduce cylinders in Section
\ref{s:cofib}. It is once again important for gluing HDAs in a
principled way.

In Sections \ref{s:Toolbox} and \ref{s:SeqComp} we use these tools to
prove Proposition \ref{p:RegConcat}. Finally, in Section
\ref{s:Spider}, we prove Proposition \ref{p:RegPlus} while dealing
with the additional issue that iHDAs are generally not both start and
accept proper.


\section{Regular languages are rational}
\label{s:reg2rat}

In this section we construct for any finite HDA a finite
automaton that recognises essentially the same language, in a sense
explained below.

Let $X$ be an HDA with $\dim(X)=n$.  We define the automaton
$G(X)=(\Omega, Q, I, E, F)$ by the following data:
\begin{itemize}
\item The \emph{input alphabet} $\Omega$ consists of the set of
  discrete ipomsets with at most $n$ elements.
\item The \emph{set of states}
  $Q=\Cell(X)\cup\{x_\bot\mid x \in X_\bot\}$.  The states of $G(X)$ are
  the cells of $X$ with an extra copy of every start cell added.
\item The \emph{start states} $I=\{x_\bot\mid x\in X_\bot\}$ and the
  \emph{accept states} $F=\{x\mid x\in X^\top\}$.
\item The \emph{set of transitions} $E$ is given by:
\begin{itemize}
\item For every $x\in X[U]$ and $A\subseteq U$ there is a transition
  $d^0_{A}:\delta^0_A(x)\to x$ labelled with 
  $\ilo{(U\setminus A)}{U}{U}$.
\item For every $x\in X[U]$ and $B\subseteq U$ there is a transition
  $d^1_B:x\to \delta^1_B(x)$ labelled with
  $\ilo{U}{U}{(U\setminus B)}$.
\item For every $x\in X_\bot$, $x\in X[U]$, there is a transition
  $\varrho_{x}:x_\bot\to x$ labelled with $\ilo{U}{U}{U}$.
\end{itemize} 
\end{itemize}

We write $\Lang(G(X))$ for the language of words over $\Omega$
recognised by the automaton $G(X)$,
and $\Qath{G(X)}$ for the set of paths in $G(X)$.

\begin{exa}
  Figure~\ref{fig:GX} shows an example for the construction of the
  standard automaton $G(X)$ from a simple HDA $X$.  
\end{exa}

\begin{figure}[tbp]
  \centering
  \begin{tikzpicture}[x=1cm, y=1cm]
    \node[circle,draw=black,fill=black!10,inner sep=0pt,minimum size=15pt]
    (aa) at (0,0) {$v$};			
    \node[circle,draw=black,fill=black!10,inner sep=0pt,minimum size=15pt]
    (za) at (-2,0) {$v_\bot$};			
    \node[circle,draw=black,fill=black!10,inner sep=0pt,minimum size=15pt]
    (ac) at (0,4) {$x$};			
    \node[circle,draw=black,fill=black!10,inner sep=0pt,minimum size=15pt]
    (ca) at (4,0) {$w$};			
    \node[circle,draw=black,fill=black!10,inner sep=0pt,minimum size=15pt]
    (cc) at (4,4) {$y$};			
    \node[circle,draw=black,fill=black!10,inner sep=0pt,minimum size=15pt]
    (ba) at (2,0) {$e$};			
    \node[circle,draw=black,fill=black!10,inner sep=0pt,minimum size=15pt]
    (bc) at (2,4) {$f$};			
    \node[circle,draw=black,fill=black!10,inner sep=0pt,minimum size=15pt]
    (ab) at (0,2) {$g$};			
    \node[circle,draw=black,fill=black!10,inner sep=0pt,minimum size=15pt]
    (zb) at (-2,2) {$g_\bot$};			
    \node[circle,draw=black,fill=black!10,inner sep=0pt,minimum size=15pt]
    (cb) at (4,2) {$h$};			
    \node[circle,draw=black,fill=black!10,inner sep=0pt,minimum size=15pt]
    (bb) at (2,2) {$q$};
    \path (aa) edge node[above=-0.8mm] {$\left[\begin{smallmatrix} a\ibullet\end{smallmatrix}\right]$} (ba);
    \path (ba) edge node[above=-0.8mm] {$\left[\begin{smallmatrix} \ibullet a\end{smallmatrix}\right]$} (ca);
    \path (ab) edge node[above=-0.8mm] {$\left[\begin{smallmatrix} \phantom{\ibullet}a\ibullet\\ \ibullet b\ibullet\end{smallmatrix}\right]$} (bb);
    \path (bb) edge node[above=-0.8mm] {$\left[\begin{smallmatrix} \ibullet a\phantom{\ibullet}\\ \ibullet b\ibullet\end{smallmatrix}\right]$} (cb);
    \path (ac) edge node[above=-0.8mm] {$\left[\begin{smallmatrix} a\ibullet\end{smallmatrix}\right]$} (bc);
    \path (bc) edge node[above=-0.8mm] {$\left[\begin{smallmatrix} \ibullet a\end{smallmatrix}\right]$} (cc);
    \path (aa) edge node[left=-0.8mm] {$\left[\begin{smallmatrix} b\ibullet\end{smallmatrix}\right]$} (ab);
    \path (ab) edge node[left=-0.8mm] {$\left[\begin{smallmatrix} \ibullet b\end{smallmatrix}\right]$} (ac);
    \path (ba) edge node[right=-0.8mm] {$\left[\begin{smallmatrix} \ibullet a\ibullet\\ \phantom{\ibullet} b\ibullet\end{smallmatrix}\right]$} (bb);
    \path (bb) edge node[left=-0.8mm] {$\left[\begin{smallmatrix} \ibullet a\ibullet\\ \ibullet b\phantom{\ibullet}\end{smallmatrix}\right]$} (bc);
    \path (ca) edge node[right=-0.8mm] {$\left[\begin{smallmatrix} b\ibullet\end{smallmatrix}\right]$} (cb);
    \path (cb) edge node[right=-0.8mm] {$\left[\begin{smallmatrix} \ibullet b\end{smallmatrix}\right]$} (cc);
    \path (bb) edge node[above left=-0.8mm] {$\left[\begin{smallmatrix} \bullet a\\ \ibullet b\end{smallmatrix}\right]$} (cc);
    \path (aa) edge node[above left=-0.8mm] {$\left[\begin{smallmatrix} a\bullet \\ b\ibullet \end{smallmatrix}\right]$} (bb);
    \path (zb) edge node[above=-0.8mm] {$\left[\begin{smallmatrix}\ibullet b\ibullet\end{smallmatrix}\right]$} (ab);
    \path (za) edge node[above=-0.8mm] {$\varepsilon$} (aa);
    \node[below left] at (za) {$\bot\;$};
    \node[below left] at (zb) {$\bot\;$};
    \node[above right] at (cb) {$\;\top$};
    \node[above right] at (cc) {$\;\top$};
    \node at (-2, 4) {$G(X)$};
    \begin{scope}[shift={(-8,0.5)}]
      \filldraw[color=black!10] (0,0)--(3,0)--(3,3)--(0,3)--(0,0);			
      \filldraw (0,0) circle (0.05);
      \filldraw (3,0) circle (0.05);
      \filldraw (0,3) circle (0.05);
      \filldraw (3,3) circle (0.05);
      \path (0,0) edge node[below] {$a$} (2.95,0);
      \path (0,3) edge (2.95,3);
      \path (0,0) edge node[above left] {$b$} (0,2.95);
      \path (3,0) edge (3,2.95);
      \node[left] at (0,0) {$\bot$};
      \node[below left] at (0,1.5) {$\bot$};
      \node[right] at (3,3) {$\top$};
      \node[right] at (3,1.5) {$\top$};
	    \node at (-0.4, 3.4) {$X$};
    \end{scope}
  \end{tikzpicture}
  \caption{An HDA $X$ and the corresponding standard finite automaton $G(X)$.}
  \label{fig:GX}
\end{figure}

\begin{prop}
  \label{p:HDAtoFSA}
  $\Lang(X)=\{P_1*P_2*\dotsm*P_n\mid P_1P_2\dotsm P_n\in\Lang(G(X)), n\ge 1\}$.
\end{prop}

\begin{proof}
  There is a one-to-one correspondence between the accepting paths in
  $X$ and $G(X)$:
  \[
    \Qath{X}\ni \alpha=
    (x_0,\varphi_1,x_1,\dotsc,x_n)
    \mapsto
    \left(
      (x_0)_\bot\xrightarrow{\varrho_x}
      x_0 \xrightarrow{\varphi_1}
      x_1\xrightarrow{\varphi_2}
      \dotsm
      \xrightarrow{\varphi_n}
      x_n
    \right)
    =\omega
    \in \Qath{G(X)}.
  \]
  Suppose $\beta_i=\ev(x_{i-1},\varphi_i,x_i)$, which is a discrete
  ipomset.  If
  \[
    Q=\ev(\alpha)=\ev(\beta_1)*\dotsm*\ev(\beta_n)\in\Lang(X),
  \]
  then $\ev(x_0) \ev(\beta_1)\dotsm \ev(\beta_n)\in\Lang(G(X))$.  This
  shows the inclusion $\subseteq$.
  
  If $P_1P_2\dotsm P_n\in\Lang(G(X))$ is recognised by a path
  $\omega$, then $P_1$ is an identity pomset.  If the corresponding
  path $\alpha$ in $X$ is not constant, then it recognises
  $P_2*\dots* P_n=P_1*\dotsm*P_n$. If $\alpha$ is constant, then it
  recognises $P_1$ (and $n=1$).  This shows  $\supseteq$.
\end{proof}
      
We have added copies of start cells in the definition above to avoid
states in $G(X)$ that are both start and accept states.  Otherwise,
constant paths in $G(X)$ could recognise the empty word while their
counterparts in $X$ would recognise non-empty identity ipomsets.

\begin{proof}[Proof of Proposition \ref{p:RegIsRat}]
  Let $X$ be an HDA of dimension $n$. Then $\Lang(G(X))$ is a regular
  language over $\Omega$ that does not contain the empty word.  The
  Kleene theorem for finite state automata thus guarantees that
  $\Lang(G(X))$ can be represented by a regular expression
  $w(U_1,\dotsc,U_n)$ with operations $\cup$, $*$ and $(-)^+$ and
  $U_i\in\Omega$.
 
  Each $U_i$ can be presented as a parallel composition
  $U_i=e^1_i\parallel \dotsm \parallel \smash[t]{e_i^{k(i)}}$ of
  singleton ipomsets.  Using Proposition \ref{p:HDAtoFSA}, we conclude
  that $\Lang(X)$ is represented by
  \[
    \smash[t]{w(e_1^1\parallel\dotsm\parallel
      e_1^{k(1)},\dotsc,e_n^1\parallel\dotsm\parallel e_n^{k(n)})}
  \]
  and therefore rational.
\end{proof}

\begin{rem}
  The proof above, in combination with the other direction of the
  Kleene theorem (Corollary \ref{c:RatIsReg}), implies that any
  regular expression can be normalised so that parallel compositions
  appear below all other operators in parse trees. For example, 
  \[
  	\{a\}^+\parallel \{b\}
  	=
  	(\{a\}\parallel\{b\})
  	\cup
  	(\{a\ibullet\}\parallel\{b\ibullet\})
  	*
  	\big(
  	(\{\ibullet a\}\parallel\{\ibullet b\ibullet\})
  	*
  	(\{a\ibullet\}\parallel\{\ibullet b\ibullet\})
  	\big)^+
  	*
  	(\{\ibullet a\}\parallel\{\ibullet b\}).
  \]
\end{rem}


\section{Track objects and tracks}
\label{s:Tracks}

Track objects and tracks on HDAs have been introduced in
\cite{Hdalang}.  They provide an alternative description of the
executions and languages of HDAs, which is more abstract, and
sometimes more convenient.  Like cells of an HDA can be represented as
pc-maps from standard cubes (see Lemma \ref{l:Yoneda}), paths can be
represented as maps from track objects. Here we extend results on
tracks, which we proved for the subclass of event consistent HDAs
in~\cite{Hdalang}, to general HDAs, as they are needed in our
constructions.

The \emph{track object} $\sq^P$ of an ipomset $P$ 
is an HDA defined as follows.
\begin{itemize}
\item $\sq^P[U]$ is the set of functions $c:P\to\{0,\exec ,1\}$ such that
$c^{-1}(\exec )\cong U$ and for all $p,q\in P$,
\begin{equation}
\label{e:TrackInequalities}
		p<q
		\implies 
		(c(p),c(q))\in\{(0,0),(\exec ,0),(1,0),(1,\exec ),(1,1)\}.
\end{equation}
\item For $A,B\subseteq U\conceq c^{-1}(\exec )$ with $A\cap
  B=\emptyset$, 
\[
	\delta_{A,B}(c)(p)
	=
	\begin{cases}
		0 &\text{if $c(p)=0$ or $p\in A$,}\\
		\exec  & \text{if $p\in U\setminus (A\cup B)$,}	\\
		1 & \text{if $c(p)=1$ or $p\in B$.}
	\end{cases}
\]
\item $\sq^P$ has one source cell $c^P_\bot$ and one target cell $c^\top_P$:
\[
	c_\bot^P(p)=
	\begin{cases}
		\exec  & \text{for $p\in S_P$},\\
		0 & \text{for $p\not\in S_P$},
	\end{cases}
	\qquad\qquad
	c^\top_P(p)=
	\begin{cases}
		\exec  & \text{for $p\in T_P$},\\
		1 & \text{for $p\not\in T_P$}.
	\end{cases}
      \]
\end{itemize}
A cell $c$ of $\sq^P$ can be regarded as a temporary snapshot of an
execution of events in $P$: for $p\in P$, $c(p)$ is $0$ if $p$ has not
yet started, $\exec$ if $p$ is currently active and $1$ if $p$ has
terminated.  This clearly enforces condition
(\ref{e:TrackInequalities}).

See \cite[Example 61]{Hdalang} for the construction of a track object
of a particular ipomset.

\begin{lemC}[{\cite[Proposition 92]{Hdalang}}]
  \label{l:TrackSubsumption}
  $\Lang(\sq^P)=\{P\}{\downarrow}$.
\end{lemC}

The following proposition relates paths with track objects.

\begin{prop}
	\label{p:TracksEqToPaths}
	Let $X$ be a precubical set, $x$ and $y$ cells of $X$ and $P$
        an ipomset.  The following conditions are equivalent:
	\begin{enumerate}
	\item
		There exists a path $\alpha\in \Qathft{X}{x}{y}$ such that $\ev(\alpha)=P$.
	\item
		There exists a pc-map $f:\sq^{P}\to X$
		such that $f(c_\bot^P)=x$ and $f(c^\top_P)=y$.
	\end{enumerate}
	Thus, for each HDA $X$, 
	\[
		\Lang(X)=\{P\in\iiPoms\mid \HDA(\sq^P,X)\neq\emptyset\}.
	\]
\end{prop}
\begin{proof}

\noindent
(2)$\Rightarrow$(1) There exists a path
$\beta\in\Qathft{\sq^P}{c_\bot^P}{c^\top_P}$ such that $\ev(\beta)=P$
(see~\cite[Proposition 67]{Hdalang} for further information).  So
$\alpha=f(\beta)$ satisfies the conditions required.
	
\noindent
(1)$\Rightarrow$(2) By induction on the length $n$ of $\alpha$.
First, suppose $n=0,1$. We abbreviate $U=\ev(x)$.  If $\alpha=(x)$,
then $\sq^{P}=\sq^U$, $c_\bot^P=c^\top_P=[\emptyset|U|\emptyset]$ and
then $f=\iota_x$ satisfies (b).  If $\alpha=(x\arrI{B} \delta^1_B(x))$
is a down-step, then $P=\ilo U U {U\setminus B}$, $\sq^P=\sq^U$,
$c_\bot^P=[\emptyset|U|\emptyset]$, $c^\top_P=[\emptyset|U|B]$, and,
as before, $f=\iota_x$ satisfies (b).  The proof for up-steps is
symmetric.
	
If $n>1$, then $\alpha=\beta*\gamma$ for paths $\beta$, $\gamma$ of
length $<n$.  We write $z=\tgt(\beta)=\src(\gamma)$, $Q=\ev(\beta)$,
$R=\ev(\gamma)$ for short.  By the inductive hypothesis, there are
precubical maps $g:\sq^Q\to X$ and $h:\sq^R\to X$ such that
$g(c_\bot^Q)=x$, $g(c^\top_Q)=h(c_\bot^R)=z$ and $h(c^\top_R)=y$.
Let $U=\ev(z)\conceq T_Q\conceq S_R$. By \cite[Lemma 65]{Hdalang},
there is a pushout diagram
	\begin{equation*}
\begin{tikzpicture}[x=1.6cm, y=1.6cm]
	\node(Q) at (0,1) {$\sq^{Q}$};
	\node(P) at (1,1) {$\sq^P$};
	\node(U) at (0,0) {$\sq^U$};
	\node(R) at (1,0) {$\sq^R$};
	\path (U) edge node[left=-0.5mm] {$\iota_{c^\top_Q}$} (Q);
	\path (Q) edge node[above=-0.5mm] {$j_Q$} (P);
	\path (U) edge node[above=-0.5mm] {$\iota_{c_\bot^R}$} (R);
	\path (R) edge node[right=-0.5mm] {$j_R$}  (P);
\end{tikzpicture}
\end{equation*}
such that
$j_Q(c_\bot^Q)=c_\bot^P$ and $j_R(c^\top_R)=c^\top_P$.  Since
$g\circ \iota_{c^\top_Q}=\iota_z=h\circ \iota_{c_\bot^R}$, by the
universal property of pushouts, the maps $g$ and $h$ glue to a map
$f:\sq^P\to X$.  Moreover, we have
$f(c_\bot^P)=f(j_Q(c_\bot^P))=g(c_\bot^Q)=x$ and
$f(c^\top_P)=f(j_R(c^\top_R))=h(c^\top_R)=y$.
\end{proof}

\begin{prop}
  \label{p:LangsAreSubClosed}
  \label{p:LangXIsALanguage}
  Languages of HDAs are down-closed with respect to subsumption.
\end{prop}

\begin{proof}
  Let $X$ be an HDA.  If $P\subsu Q$ and $Q\in\Lang(X)$ then there is
  an HDA-map $\sq^Q\to X$ by Proposition~\ref{p:TracksEqToPaths} and
  $P\in\Lang(\sq^Q)$ by Lemma~\ref{l:TrackSubsumption}.  Thus
  $P\in\Lang(X)$ by Lemma~\ref{l:FunctorialityOfLanguages}.
\end{proof}

\section{Tensor product of higher-dimensional automata}
\label{s:tensor}

The \emph{tensor product} of HDAs $X$ and $Y$ is the HDA $X\otimes Y$
defined, for $U,V,W\in\sq$, $x\in X[V]$, $y\in Y[W]$ and $A,B\subseteq U$ as
\begin{align*}
  (X\otimes Y)[U]&=\bigcup_{V\parallel W=U} X[V]\times Y[W], \\
  \delta_{A,B}(x,y)&= (\delta_{A\cap V,B\cap V}(x),\delta_{A\cap
                     W,B\cap W}(y)),\\
  (X\otimes Y)_\bot&=X_\bot\times Y_\bot,\\
  (X\otimes Y)^\top&=X^\top\times Y^\top.
\end{align*}

See \cite[Example 107]{Hdalang} for an example. The following
proposition is shown for event-consistent HDAs in \cite[Theorem
108]{Hdalang}. We need a proof without this restriction.

\begin{prop}
  \label{p:TensorParallel}
  Let $X$ and $Y$ be HDAs. Then
  $\Lang(X\otimes Y)=\Lang(X)\parallel\Lang(Y)$.
\end{prop}

\begin{proof}
  For $\Lang(X)\parallel\Lang(Y)\subseteq \Lang(X\otimes Y)$ the
  argument in~\cite[Theorem 108]{Hdalang} works: if
  $P\in \Lang(X)\parallel \Lang(Y)$, then, by definition, there are
  $Q\in\Lang(X)$ and $R\in\Lang(Y)$ such that $P\subsu Q\parallel R$.
  By Proposition~\ref{p:TracksEqToPaths}, there are HDA-maps
  $\boldsymbol\alpha:\sq^Q\to X$ and $\boldsymbol\beta:\sq^R\to Y$.
  Their composition
  \[
    \sq^P\to \sq^{Q||R} \cong \sq^{Q}\otimes \sq^R
    \xrightarrow{\boldsymbol\alpha\otimes\boldsymbol\beta} X\otimes Y
  \]
  shows that $P\in \Lang(X\otimes Y)$.  Finally, the isomorphism
  $\sq^{Q||R} \cong \sq^{Q}\otimes \sq^R$ is shown in~\cite[Lemma
  105]{Hdalang}.
  
  The proof of the converse direction in \cite{Hdalang} depends on
  event consistency, so we need another one.  Suppose
  $\alpha=((x_0,y_0),\varphi_1,\dots,(x_n,y_n))\in\Qath{X\otimes Y}$,
  and $x_k\in X[U_k]$ as well as $y_k\in Y[V_k]$, for $k=0,\dots,n$.
  For any $k$,
  \begin{itemize}
  \item if
    $\varphi_k=d^0_{A}\in\sq(U_{k-1}\parallel V_{k-1},U_k\parallel
    V_k)$, then we put $\psi_k=d^0_{A\cap U_k}\in\sq(U_{k-1},U_k)$ and
    $\omega_k=d^0_{A\cap V_k}\in\sq(V_{k-1},V_k)$.
  \item If $\varphi_k=d^1_{B}$, we put $\psi_k=d^1_{B\cap U_{k-1}}$ and
    $\omega_k=d^1_{B\cap V_{k-1}}$.
  \end{itemize}
  It is then routine to check that
  $\beta=(x_0,\psi_0,\dots,x_n)\in\Qath{X}$ and
  $\gamma=(y_0,\omega_0,\dots,y_n)\in\Qath{Y}$.  We write
  $\pi_X(\alpha)=\beta$, $\pi_Y(\alpha)=\gamma$.
 
  We prove that $\ev(\alpha)\subsu\ev(\beta)\parallel\ev(\gamma)$ by
  induction on $n$.
  If $n=1$ and $\alpha$ is an up-step, then
  \begin{equation*}
    \ev(\alpha)=\ilo{(U_0\parallel V_0)}{(U_1\parallel V_1)}{(U_1\parallel V_1)}=
    \ilo{U_0}{(U_1)}{U_1}\parallel \ilo{V_0}{(V_1)}{V_1}
    =\ev(\beta)\parallel \ev(\gamma).
  \end{equation*}
  For down-step the same formula holds by symmetry.
  The case $n=0$ is similar.

  If $n>1$, then $\alpha$ can be decomposed into a non-trivial
  composition $\alpha=\alpha'*\alpha''$.  Let $\beta'=\pi_X(\alpha')$,
  $\gamma'=\pi_Y(\alpha')$, $\beta''=\pi_X(\alpha'')$,
  $\gamma''=\pi_Y(\alpha'')$.  Using the inductive hypothesis and the
  weak interchange law
  $(P\parallel P')*(Q\parallel Q')\subsu (P*Q)\parallel (P'*Q')$ of
  ipomsets \cite{DBLP:journals/iandc/FahrenbergJSZ22}, 
  \begin{align*}
    \ev(\alpha) &= \ev(\alpha')*\ev(\alpha'')\\
    &\subsu
    (\ev(\beta')\parallel\ev(\beta''))*(\ev(\gamma')\parallel\ev(\gamma''))\\
    &\subsu
    (\ev(\beta')*\ev(\beta''))\parallel (\ev(\gamma')*\ev(\gamma''))\\
    &=
    \ev(\beta'*\beta'')\parallel \ev(\gamma'*\gamma'')\\
    &=\ev(\beta)\parallel\ev(\gamma).
  \end{align*}

  Now let $P\in \Lang(X\otimes Y)$ and let
  $\alpha\in\Qath{X\otimes Y}$ be an accepting path such that
  $\ev(\alpha)$.  Then both $\beta=\pi_X(\alpha)\in\Qath{X}$ and
  $\gamma=\pi_Y(\alpha)\in\Qath{Y}$ are accepting, and
  \[
    P=\ev(\alpha)\subsu \ev(\beta)\parallel
    \ev(\gamma).
  \]
 As $\ev(\alpha)\in\Lang(X)$ and $\ev(\beta) \in \Lang(Y)$,
  down-closure yields $P\in \Lang(X)\parallel \Lang(Y)$.
\end{proof}

It follows that parallel compositions of regular languages are
regular.

\begin{proof}[Proof of Proposition \ref{p:RegParallel}]
  Let $L$ and $M$ be regular languages. Then $L=\Lang(X)$ and
  $M=\Lang(Y)$ for some HDA $X$ and $Y$. Thus $L\parallel M =
  \Lang(X)\parallel \Lang(Y)= \Lang(X\otimes Y)$ is recognised by
  $X\otimes Y$ by Proposition \ref{p:TensorParallel} and therefore
  regular.
\end{proof}


\section{Higher-dimensional automata with interfaces}
\label{s:ihda}

In this section we introduce higher-dimensional automata with
interfaces (iHDAs).  The main difference to HDAs is that the elements
in the lists of concurrent events that are assigned to cells are now
equipped with interfaces. Every event of an iHDA can thus be labelled
as a source event or a target event, or as both.  Target events may
not be terminated while source events cannot be ``unstarted''.
Moreover, start cells must only contain source events, and target
cells only target events.  The advantage of this variant is that
source and target events of an iHDA can be traced along its execution,
as shown for instance in Figure~\ref{fig:iHDA} below. 

\subsection*{Concurrency lists with interfaces}

A \emph{concurrency list with interfaces}
(\emph{iconclist}) is a triple $(S,U,T)$ of a conclist $U$, a \emph{source
  interface} $S\subseteq U$ and a \emph{target interface}
$T\subseteq U$.

We also write $\ilo{S}{U}{T}$ or just $U$ for an iconclist $(S,U,T)$.
In the latter case, we write $S_U$ and $T_U$ for the interfaces of
$U$.  Conclists may be regarded as iconclists with empty interfaces and
iconclists as discrete ipomsets.

An \emph{ilo-map} $f:U\to V$ is an lo-map that also
satisfies $S_U=f^{-1}(S_V)$ and $T_U=f^{-1}(T_V)$.
An \emph{iconclist isomorphism} is an invertible ilo-map.  We write $U\conceq V$
if iconclists $U$ and $V$ are isomorphic.

As for conclists, there is at most one isomorphism between iconclists.
Isomorphism classes of iconclists can be modelled as words over the
extended alphabet
$\Sigma_{\ibullet}=\{a, \ibullet a, a\ibullet, \ibullet a\ibullet\mid
a\in\Sigma\}$, where $\ibullet a$ indicates membership in a source
interface and so on. As in Figure \ref{fi:hda-cylinder}, we represent
such words as column vectors.

Let $U=(S_U,U,T_U)$ and $V=(S_V,V,T_V)$ be iconclists.
An \emph{iconclist map} from $U$ to $V$ 
is a conclist map $d_{A,B}:U\to V$ such that
\begin{itemize}
\item source and target events are preserved: $d_{A,B}^{-1}(S_V)=S_U$
  and $d_{A,B}^{-1}(T_V)=T_U$,
\item source events cannot be unstarted and target events not be
  terminated: $A\cap S_V=\emptyset=B\cap T_V$.
  \end{itemize}
See Figure \ref{fi:iconclistMap} for an example.
Compositions of iconclist maps are defined as for conclist maps.
\begin{figure}
\begin{tikzpicture}
      \node (ss1) at (-3,-2) {$\ibullet a\phantom{\ibullet}$};
      \node (ss2) at (-3,-3) {$\ibullet b\ibullet$};
      \path (ss1) edge[densely dotted] (ss2);
      \node (t1) at (0,-1) {$\ibullet a\phantom{\ibullet}$};
      \node (t2) at (0,-2) {$\phantom{\ibullet}c \ibullet$};
      \node[above right, color=green!50!black] at (t2) {$0$};
      \node (t3) at (0,-3) {$\ibullet b\ibullet$};
      \node (t4) at (0,-4) {$\ibullet e\phantom{\ibullet}$};
      \node[above right, color=red!50!black] at (t4) {$1$};
      \path (t1) edge[densely dotted] (t2);
      \path (t2) edge[densely dotted] (t3);
      \path (t3) edge[densely dotted] (t4);
      \path (ss1) edge (t1);
      \path (ss2) edge (t3);
\end{tikzpicture}
  \caption{An example of an iconclist map.
    Annotations {\color{green!50!black}$0$} and {\color{red!50!black}$1$}
    indicate events that have not yet started ({\color{green!50!black}$0$}) or
    terminated ({\color{red!50!black}$1$}), as in Figure \ref{fi:dotsquare-comp}.
    Bullets indicate source and target interfaces.
    Note that $c\ibullet$ cannot be marked by {\color{red!50!black}$1$} since it is in the target interface; similarly, $\ibullet e$ cannot be marked by {\color{green!50!black}$0$}. 
    No marking is possible for $\ibullet b \ibullet$ and thus it must be in the image.
  \label{fi:iconclistMap}
  }
\end{figure}

The \emph{full labelled precube category with interfaces}, $\fullisq$,
has iconclists as objects and iconclist maps as morphisms.
Every pair of isomorphic iconclists admits exactly one isomorphism
between them.
Thus, we define the \emph{labelled precube category with interfaces} $\isq$ as the
quotient of $\fullisq$ by isomorphisms.
The quotient functor $\fullisq\to\isq$
is an equivalence of categories.
The category $\isq$ is skeletal, and its objects are words on
$\Sigma_\bullet=\{a,\sbt{a},a\sbt,\sbt a\sbt\mid a\in\Sigma\}$.

We can assign an iconclist with empty interfaces to any conclist using
the inclusion functors
$\fullsq\ni U\mapsto\ilo{\emptyset}{U}{\emptyset}\in\fullisq$ and
$\sq\to\isq$.  Conversely, there are forgetful functors $\isq\to\sq$
and $\fullisq\to\fullsq$ that ignore interfaces and assign the
underlying conclist to each iconclist.

The involutive \emph{reversal} functor on $\fullisq$ and $\isq$ maps
$\ilo{S}{U}{T}$ to $\ilo{T}{U}{S}$ and $d_{A,B}$ to $d_{B,A}$.  It
swaps events that have not yet started and those that have terminated.

\subsection*{Precubical sets with interfaces and iHDAs}

A \emph{precubical set with interfaces} (\emph{ipc-set}) is a presheaf
on $\isq$.  We write $X[U]$ for the value of $X$ on object $U$ of
$\isq$, and $\delta_{A,B}=X[d_{A,B}]:X[U]\to X[U\setminus(A\cup B)]$
for the \emph{face map} associated to the coface map
$d_{A,B}:U\setminus (A\cup B)\to U$.  Elements of $X[U]$ are
\emph{cells} of $X$.  We write $\Cell(X)=\bigsqcup_{U\in\isq}X[U]$ for
the set of cells of $X$.  An ipc-set $Y$ is an \emph{ipc-subset} of
$X$ if $Y[U]\subseteq X[U]$ for all $U\in\isq$ and the face maps of
$Y$ are the restrictions of face maps of $X$.

If $x\in X[\ilo S U T]$, we write $\Iev(x)= \ilo{S}{U}{T}\in\isq$ and
$\ev(x)=U\in\sq$.  As before, we also write $\delta_A^0$ for
$X[d_A^0]$ and $\delta_B^1$ for $X[d_B^1]$.  We may view a precubical
set as an ipc-set $X$ such that $X[\ilo{S}{U}{T}]=\emptyset$ whenever
$S\ne\emptyset$ or $T\ne\emptyset$.
As for pc-sets, 
$\Cell(X)$ may be regarded as the category of elements of $X$,
with the projection functor $\Iev:\Cell(X)\to\isq$.
We often view $X$ as a set of cells, that is, objects of $\Cell(X)$.

A \emph{higher-dimensional automaton with interfaces} (\emph{iHDA})
is a finite ipc-set $X$ with subsets $X_\bot$ of \emph{start cells} and
$X^\top$ of \emph{accept cells}.
These are required to satisfy $S=U$ for all $x\in X_\bot$ with $\Iev(x) = \ilo{S}{U}{T}$,
and $T=U$ for all $x\in X^\top$ with $\Iev(x)= \ilo{S}{U}{T}$.
Neither $X_\bot$ nor $X^\top$ is necessarily an ipc-subset.

HDAs are not simply special cases of iHDAs
due to the above requirements on interfaces of start and accept cells.
See Figure~\ref{fig:iHDA} for examples. 

An \emph{ipc-map} is a natural transformation $f:X\to Y$ of ipc-sets
$X$, $Y$, an \emph{iHDA-map} must preserve start and accept cells as
well: $f(X_\bot)\subseteq Y_\bot$ and $f(X^\top)\subseteq Y^\top$.  We
write $\isq\Set$ and $\IHDA$ for the resulting categories of ipc-sets
and iHDAs.

The reversal on $\isq$ translates to ipc-sets and iHDAs.  It maps
$\delta_{A,B}$ to $\delta_{B,A}$ and exchanges start and accept cells
if present. 

\subsection*{Standard icubes}

\begin{figure}[tbp]
  \centering
  \label{fi:iCubes}
    \begin{tikzpicture}[x=0.5cm, y=0.5cm]
      \begin{scope}[shift={(3,2)}]
        \path[fill=black!10] (0,0) to (4,0) to (4,4) to (0,4);
        \node[state, fill=white] (00) at (0,0) {};
        \node[state] (10) at (4,0) {};
        \node[state, fill=white] (01) at (0,4) {};
        \node[state, fill=white] (11) at (4,4) {};
        \path (00) edge (10);
        \path[-, densely dashed] (00) edge (01);
        \path[-, densely dashed] (01) edge (11);
        \path (10) edge (11);
        \node[below right] at (4,0) {$[b|U|a]$};
        \node[below] at (2,0) {$[b|U|\emptyset]$};
        \node[right] at (4,2) {$[\emptyset|U|a]$};
        \node at (2,2) {$[\emptyset|U|\emptyset]$};
      \end{scope}
 \begin{scope}[shift={(14,0)}]
    \path[fill=black!10] (0,0) to (4,0) to (4,4) to (0,4);
    \path[fill=black!10] (4,0) to (8,0) to (8,4) to (4,4);
    \path[fill=black!10] (4,4) to (8,4) to (8,8) to (4,8);
    \node[state, fill=white] (00) at (0,0) {};
    \node[state, fill=white] (04) at (0,4) {};           
    \node[state] (40) at (4,0) {};           
    \node[state] (44) at (4,4) {};
    \node[state, fill=white] (80) at (8,0) {};
    \node[state, fill=white] (84) at (8,4) {};
    \node[state, fill=white] (88) at (8,8) {};
    \node[state, fill=white] (48) at (4,8) {};
    \path[-, densely dashed] (00) edge  (04);
    \path (00) edge[-]  node [below] {$[\ibullet a]$} (40);
    \path (04) edge  node [above] {$[\ibullet a]$} (44);
    \path (40) edge node [right] {$[b]$} (44);
    \path (40) edge node [below] {$[c\ibullet]$} (80);
    \path (44) edge node [above] {$[c\ibullet]$} (84);
    \path[-, densely dashed] (48) edge  (88);
    \path[-, densely dashed] (80) edge (84);
    \path (44) edge node [left] {$[d\ibullet]$} (48);
      \path[-, densely dashed] (84) edge (88);
    \node (22) at (2,2) {$\left[\begin{smallmatrix} \ibullet a \\ \phantom{\ibullet}b \end{smallmatrix}\right]$};
    \node (62) at (6,2) {$\left[\begin{smallmatrix} b\phantom{\ibullet} \\ c\bullet \end{smallmatrix}\right]$};
    \node (66) at (6,6) {$\left[\begin{smallmatrix} c\ibullet \\ d\bullet \end{smallmatrix}\right]$};
    \node[above=-0.5mm] at (2,0) {$\bot$};
    \node[above right=0.5mm] at (6,6) {$\top$};
  \end{scope}
    \end{tikzpicture}
     \caption{Left: the standard icube $\sq^U$ for $U={\left[\begin{smallmatrix} \ibullet\, a\,     \phantom{\sbt}\\
      \phantom{\sbt}\, b\, \sbt\end{smallmatrix}\right]}$ with names of cells.
      Right: an example of an iHDA with iconclists associated to
      particular cells.
      The presence of interfaces causes that some faces are ``missing''.
      Those are indicated by dashed lines or circles.
      }
  \label{fig:iHDA}
\end{figure}

The \emph{standard icube} $\isq^U$ is the presheaf represented by an iconclist $U\in\isq$,
that is, $\isq^U[W]=\isq(W,U)$.
Every morphism $d_{A,B}\in\isq(U,V)$
defines an ipc-map
$
	\isq^{d_{A,B}}:\isq^U\to\isq^V,
$
which gives a functor $\isq\ni U\mapsto \isq^U\in\isq\Set$.

An example of a standard icube is given on the left in
Figure~\ref{fig:iHDA}. 

The analogue of Lemma \ref{l:Yoneda} holds for iHDA as well.
For every iHDA $X$ and cell $x\in X[U]$
there exists a unique iHDA-map $\iota_x:\isq^U\to X$
such that $\iota_x(\id_U)=x$.
(Note that $\id_U=[\emptyset|U|\emptyset]\in \isq(U,U)\cong \isq^U[U]$ is the top cell of $\isq^U$).

\def\cC{\mathcal{C}}
\def\cD{\mathcal{D}}
\def\upstep{\mathbf{S}}
\def\downstep{\mathbf{T}}
\def\isostep{\mathbf{I}}

\subsection*{Executions of iHDAs}

Paths in iHDAs, their ipomsets and languages are defined by analogy to
HDAs in Section \ref{s:languages}.  A path in an iHDA $X$ is a
sequence $(x_0,\varphi_1,\dotsc,x_n)$ such that every step
$(x_{k-1},\varphi_k,x_k)$ is either an up-step
$\delta^0_A(x)\arrO{A} x=(\delta^0_A(x),d^0_A,x)$ or a down-step
$x \arrI{B} \delta^1_B(x)=(x,d^B_1,\delta^1_B(x))$.  The ipomset of a
path is a gluing composition of ipomsets of consecutive steps, which
are $\ilo{(U\setminus A)}{U}{U}$ and $\ilo{U}{U}{(U\setminus B)}$ for
up-steps and down-steps as specified above.

The set of ipomsets of accepting paths forms the language $\Lang(X)$
of $X$.  We will see in the next section that $\Lang(X)$ is
down-closed, that is, an interval ipomset language.  Further, any iHDA
can be translated to an HDA that recognises the same language, and
vice versa.

As for HDA maps, we call an iHDA map $f:X\to Y$ a \emph{weak
  equivalence} if for every accepting path $\beta\in\Qath{Y}$ there is
an accepting path $\alpha\in\Qath{X}$ such that $f(\alpha)=\beta$.
The analogue of Lemma \ref{l:FunctorialityOfLanguages} holds for iHDA:
for every iHDA map $f:X\to Y$ we have $\Lang(X)\subseteq \Lang(Y)$,
and $\Lang(X)=\Lang(Y)$ if $f$ is a weak equivalence.


\section{Higher-dimensional automata with and without interfaces}
\label{s:hda-ihda}

In this section we discuss the relationship between HDAs and iHDAs.
We show that there are two functors that translate between them: the
resolution $\Res:\HDA\to\IHDA$ and the closure $\Cl:\IHDA\to\HDA$,
both of which preserve languages.  Finite HDAs and finite iHDAs
therefore recognise the same class of regular languages.

\subsection*{Resolution}

The \emph{resolution} of a precubical set $X$ is the ipc-set
$\Res(X)=X\circ \sfF$, where $\sfF:\isq\to\sq$ is the forgetful
functor.

For $\ilo{S}{U}{T}\in\isq$ and $A,B\subseteq U$, the definition of
$\Res$ expands to
\begin{align*}
  \Res(X)[\ilo{S}{U}{T}] &=\{(x;S,T)\mid x\in X[U]\},\\
  \delta_{A,B}((x;S,T))&=
  (\delta_{A,B}(x); S\setminus B, T\setminus A).
\end{align*}
The functor $F:\isq \to \sq$ forgets interfaces, but the composition
$\isq\stackrel{F}{\to}\sq\stackrel{X}{\to}\Set$ produces copies of
cells of $X$ equipped with all possible combinations of interfaces.
Notation such as $(x;S,T)$ indicates that $\Res(X)[\ilo{S}U{T}]$ is
essentially the same set as $X[U]$, but each $x\in X[U]$ is tagged
with $S$ and $T$.  For every HDA $X$ we define
$(x;S,T)\in\Res(X)[\ilo{S}{U}{T}]$ to be a start cell if $x\in X_\bot$
and $S=U$, and an accept cell if $x\in X^\top$ and $T=U$.

This extends $\Res$ to a functor $\HDA\to \IHDA$. Every cell
$x\in X[U]$ produces $4^{|U|}$ cells in $\Res(X)$. Thus $\Res(X)$ is
finite whenever $X$ is.

\begin{exa}
For an HDA $X$ with $x\in X[a]$ and $v,w\in X[\emptyset]$,
\begin{equation*}
  \Res\left(
  \vcenter{\hbox{
      \begin{tikzpicture}[x=3cm, y=1.2cm]
        \node[state] (0) at (0,0) {};
        \node[state] (1) at (1,1) {};
        \path (0) edge node[above=-0.1mm] {$x$} (1);
        \node at (0) [left] {$v$};
        \node at (1) [right] {$w$};
      \end{tikzpicture}
    }}
  \right)
  \quad = \quad
  \vcenter{\hbox{
      \begin{tikzpicture}[x=3cm, y=1.2cm]
  	\node[state] (0) at (0,0) {};
  	\node[state] (1) at (1,1) {};
  	\node[state, fill=white] (0a) at (0,1) {};
  	\node[state, fill=white] (1a) at (1,0) {};
  	\node[state, fill=white] (0b) at (0,-1) {};
  	\node[state, fill=white] (1b) at (1,-1) {};
        \path (0) edge node[above left=-1.6mm] {\scriptsize{$(x;\emptyset,\emptyset)$}} (1);
        \path (0a) edge node[above=-0.4mm] {\scriptsize{$(x;a,\emptyset)$}}  (1);
        \path (0) edge  node[below=-0.4mm] {\scriptsize{$(x;\emptyset,a)$}}   (1a);
        \path (0b) edge  node[below=-0.4mm] {\scriptsize{$(x;a,a)$}}   (1b);
        \node at (0) [below left=-0.4mm] {\scriptsize{$(v;\emptyset,\emptyset)$}};
        \node at (1) [above right=-0.4mm] {\scriptsize{$(w;\emptyset,\emptyset)$}};
      \end{tikzpicture}	 
    }}
\end{equation*}
\end{exa}

\begin{prop}
  \label{l:ResPreservesAcceptingTracks}
  If $X$ is an HDA, then $\Lang(\Res(X))=\Lang(X)$.
\end{prop}

\begin{proof}
  If
  $((x_0;S_0,T_0),\phi_1,(x_1;S_1,T_1),\phi_2,\dotsc,\phi_n,(x_n;S_n,T_n))$
  is an accepting path in $\Res(X)$, then
  $(x_0,\phi_1,x_1,\phi_2\dots,\phi_n,x_n)$ is an accepting path in
  $X$ with the same event ipomset.  Conversely, if
  $\alpha=(x_0,\phi_1,x_1,\dots,\phi_n,x_n)$, $x_k\in X[U_k]$ is an
  accepting path in $X$, we define $S_k$ and $T_k$ recursively (recall
  that, by Lemma~\ref{l:ConsistentEvents}, the $U_k$ may be taken as
  subsets of $\ev(\alpha)$):
  \begin{itemize}
  \item $S_0=U_0$, $T_n=U_n$,
  \item if $\varphi_k=d^0_{U_k\setminus U_{k-1}}$, then $S_k=S_{k-1}$ and
    $T_{k-1}=T_k\cap U_{k-1}$,
  \item if $\varphi_k=d^1_{U_{k-1}\setminus U_k}$, then
    $S_k=S_{k-1}\cap U_k$ and $T_{k-1}=T_{k}$.
  \end{itemize}
  This yields an accepting path $((x_k;S_k,T_k),\varphi_k)$ in
  $\Res(X)$ with the same event ipomset as $\alpha$ in $X$.
\end{proof}

\subsection*{Closure}

The closure is the left adjoint to resolution, though we neither need
nor prove this fact in this article.  Instead, we give an explicit
definition.

The \emph{closure} of the ipc-set $X$ is the pc-set $\Cl(X)$ defined,
for all $U\in\sq$, as
\begin{equation*}
  \Cl(X)[U]=\{ [x;A,B]\mid \exists\, \ilo{S}{V}{T}\in\isq: x\in X[V], 
  A\subseteq S, B\subseteq T, A\cap B=\emptyset, U=V\setminus(A\cup B)
  \}.
\end{equation*}

We write $[x;A,B]$ instead of $(x;A,B)$ to distinguish the provenance
of these elements. Face maps are given by
\begin{equation*}
  \delta_{C,D}([x;A,B]) =
  [\delta_{C\setminus S,D\setminus T}(x); A\cup (C\cap S), B\cup (D\cap T)],
\end{equation*}
where $U$ and $\ilo{S}{V}{T}$ are as above and the $C,D\subseteq U$
satisfy $C\cap D=\emptyset$.  An ipc-map $f:X\to Y$ induces a pc-map
$\Cl(f):\Cl(X)\to\Cl(Y)$ such that $\Cl(f)[U]([x;A,B])=[f(x);A,B]$.
This makes $\Cl:\isq\Set\to\sq\Set$ a functor.

Intuitively, $\Cl(X)$ fills in the missing cells of the ipc-set $X$.
The function $\delta_{C\setminus S_V,D\setminus T_V}$ takes as much of
the face map as possible, while the remaining events that should be
unstarted or terminated are added to $A$ and $B$, respectively.  See
Figure \ref{fi:iClosure} for an example.

For an iHDA $X$ we define 
\[
	\Cl(X)_\bot = \{[x;\emptyset,\emptyset]\mid x\in X_\bot\}
	\qquad \text{and}\qquad
	\Cl(X)^\top = \{[x;\emptyset,\emptyset]\mid x\in X^\top\}.
\]	
This extends $\Cl$ to a functor $\IHDA\to\HDA$, which is the left
adjoint of the extension of the functor $\Res$.

\begin{figure}[tbp]
  \centering
\begin{tikzpicture}[x=0.6cm, y=0.5cm]
 \begin{scope}[shift={(0,0)}]
    \path[fill=black!10] (0,0) to (4,0) to (4,4) to (0,4);
    \path[fill=black!10] (4,0) to (8,0) to (8,4) to (4,4);
    \path[fill=black!10] (4,4) to (8,4) to (8,8) to (4,8);
    \node[state, fill=white] (00) at (0,0) {};
    \node[state, fill=white] (04) at (0,4) {};           
    \node[state, fill=white] (40) at (4,0) {};           
    \node[state] (44) at (4,4) {};
    \node[state, fill=white] (80) at (8,0) {};
    \node[state, fill=white] (84) at (8,4) {};
    \node[state, fill=white] (88) at (8,8) {};
    \node[state] (48) at (4,8) {};
    \path[-, densely dashed] (00) edge  (04);
    \path[-, densely dashed] (00) edge (40);
    \path (04) edge  node [above] {$a$} (44);
    \path (40) edge node [right] {$b$} (44);
    \path[-, densely dashed] (40) edge (80);
    \path (44) edge node [above] {$c$} (84);
    \path (48) edge node [above] {$c'$} node[below] {$\top$} (88);
    \path[-, densely dashed] (80) edge (84);
    \path (44) edge node [left] {$d$} (48);
      \path[-, densely dashed] (84) edge (88);
    \node (22) at (2,2) {$x$};
    \node (62) at (6,2) {$y$};
    \node (66) at (6,6) {$z$};
    \node[above left] at (48) {$q$};
    \node[below left=-0.5mm] at (2,2) {$\bot$};
  \end{scope}
 \begin{scope}[shift={(14,0)}]
    \path[fill=black!10] (0,0) to (4,0) to (4,4) to (0,4);
    \path[fill=black!10] (4,0) to (8,0) to (8,4) to (4,4);
    \path[fill=black!10] (4,4) to (8,4) to (8,8) to (4,8);
    \node[state] (00) at (0,0) {};
    \node[state] (04) at (0,4) {};           
    \node[state] (40) at (4,0) {};           
    \node[state] (44) at (4,4) {};
    \node[state] (80) at (8,0) {};
    \node[state] (84) at (8,4) {};
    \node[state] (88) at (8,8) {};
    \node[state] (48) at (4,8) {};
    \path (00) edge node [left] {$[x;a,\emptyset]$}  (04);
    \path (00) edge node [below] {$[x;b,\emptyset]$} (40);
    \path (04) edge node [above] {$[a;\emptyset,\emptyset]$} (44);
    \path (40) edge (44);
    \path (40) edge node [below] {$[y;b,\emptyset]$}  (80);
    \path (44) edge node [above] {$[c;\emptyset,\emptyset]$} (84);
    \path (48) edge node [above] {$[c';\emptyset,\emptyset]$} node[below] {$\top$} (88);
    \path (80) edge node [right] {$[y;\emptyset,c]$} (84);
    \path (44) edge (48);
     \path (84) edge node [right] {$[z;\emptyset,c]$} (88);
    \node (22) at (2,2) {$[x;\emptyset,\emptyset]$};
    \node (62) at (6,2) {$[y;\emptyset,\emptyset]$};
    \node (66) at (6,6) {$[z;\emptyset,\emptyset]$};
    \node[above left] at (48) {$[q;\emptyset,\emptyset]$};
    \node[above left] at (04) {$[a;a,\emptyset]$};
    \node[below left] at (00) {$[x;ab,\emptyset]$};
    \node[below right] at (80) {$[y;b,c]$};
    \node[right] at (84) {$[c;\emptyset,c]$};
    \node[above right] at (88) {$[c';\emptyset,c]$};
    \node[below left=-0.2mm] at (2,2) {$\bot$};
  \end{scope}
\end{tikzpicture}
     \caption{An example of an iHDA (left) and its closure (right).
     Not all cells are captioned. 
     The iconclists of cells on the left are 
     $\Iev(x)=\left[\begin{smallmatrix} \ibullet a \\ \ibullet b \end{smallmatrix}\right]$,
     $\Iev(y)=\left[\begin{smallmatrix} \ibullet b\phantom{\ibullet} \\ \phantom{\ibullet} c \ibullet \end{smallmatrix}\right]$,
     $\Iev(z)=\left[\begin{smallmatrix}  c\ibullet \\ d\phantom{\ibullet} \end{smallmatrix}\right]$.
  \label{fi:iClosure}
}
\end{figure}

\begin{lem}
  \label{l:ClosureOfCube}
  If $\ilo{S}{U}{T}$ is an iconclist, then
  $\Cl(\isq^{\ilo S U T})\cong \sq^U$.
\end{lem}

\begin{proof}
  The isomorphism maps a cell $[A|B]\in \sq^U[U\setminus(A\cup B)]$ into
  $
  	[[(A\setminus S)|(B\setminus T)];A\cap S, B\cap T]
  	$
  	in the set $\Cl(\isq^{\ilo S U T})[U\setminus(A\cup B)]$.
\end{proof}

\begin{prop}
  \label{l:ClPreservesLanguage}
  If $X$ is an iHDA, then $\Lang(\Cl(X))=\Lang(X)$.
\end{prop}

\begin{proof}
  If $(x_0,\phi_1,\dotsc,\phi_n,x_n)$ is an accepting path in $X$,
  then
  \[([x_0;\emptyset,\emptyset],\phi_1,\dotsc,\phi_n,[x_n;\emptyset,
    \emptyset])\] is an accepting path in $\Cl(X)$.  Conversely,
  let $\alpha =([x_0;A_0,B_0],\phi_1,\dotsc,\phi_n,[x_n;A_n, B_n])$ be
  a path in $\Cl(X)$. Then, for all $\phi_k=d^0_{C_k}$, we have
  $A_{k-1}\supseteq A_k$ and $B_{k-1}=B_k$, and for all
  $\phi_k=d^1_{D_k}$, $A_{k-1}= A_k$ and $B_{k-1}\subseteq B_k$.
  Hence $A_0\supseteq \dotsm\supseteq A_n$ and
  $B_0\subseteq \dotsm\subseteq B_n$.  If $\alpha$ is accepting, then
  $A_0=B_n=\emptyset$ and thus $A_k=B_k=\emptyset$ for all $k$.
\end{proof}

\begin{prop}
  \label{p:RegIsIReg}
  HDAs and iHDAs recognise the same class of languages: that of
  regular languages.
\end{prop}

\begin{proof}
  Resolution and closure preserves finiteness of automata. The result
  then follows from Propositions \ref{l:ResPreservesAcceptingTracks}
  and \ref{l:ClPreservesLanguage}.
\end{proof}

We conclude with an easy technical lemma that is needed later on.

\begin{lem}
  \label{l:FacesInClosures}
  Let $X$ be an iHDA, $x\in X$ and  $[y;A,B]\in\Cl(X)$.  Then
  $[y;A,B]$ is a face of $[x;\emptyset,\emptyset]$ in $\Cl(X)$ if and
  only if $y$ is a face of $x$ in $X$.
\end{lem}

\begin{proof}
	If $y=\delta_{C,D}(x)$, then 
	\[
		\delta_{A,B}(\delta_{C,D}([x;\emptyset,\emptyset]))
		=
		\delta_{A,B}([y;\emptyset,\emptyset])
		=[y;A,B], 
	\]
	and if $[y;A,B]=\delta_{C,D}([x;\emptyset,\emptyset])$, then
        $y=\delta_{C\setminus S,D\setminus T}(x)$, where
        $S=S_{\Iev(x)}$, $T=T_{\Iev(x)}$.
\end{proof}

\subsection*{Simple languages}

An iHDA $X$ is
\emph{start simple} if it has exactly one start cell,
\emph{accept  simple} if it has exactly one accept cell,
and \emph{simple} if it is both start and accept simple.
A regular language is \emph{simple} if it is
recognised by a simple iHDA. 

\begin{exa}
  HDAs with one start and one accept cell recognise a larger class of
  languages that simple iHDA.  The HDA $X$ with a single $0$-cell $x$,
  a $1$-loop labelled $a$ on $x$, and $X_\bot=X^\top=\{a\}$ is simple
  and recognises the language of all ipomsets
  $[\ibullet a\dotsm a\ibullet]$, but no simple iHDA does.  This is
  because $[\ibullet a \ibullet]$ may only be an ipomset of a constant
  path $(x)$ such that $x\in X[\ibullet a \ibullet]$ and
  $x\in X_\bot\cap X^\top$.  Yet the event $a$ cannot be terminated in
  any path starting at $x$, so no such path may recognise
  $[\ibullet a a\ibullet]$.
\end{exa}

\begin{lem}
  \label{l:RegIsUnionOfSimple}
  Every regular language is a finite union of simple regular languages.
\end{lem}

\begin{proof}
  Proposition~\ref{p:RegIsIReg} allows us to work with a suitable iHDA
  $X$.  Let $X_\bot=\{x_\bot^i\}_{i=1}^m$ and
  $X^\top=\{x^\top_j\}_{j=1}^n$.  For each pair
  $(i,j)$, let $X_i^j$ be the iHDA with the same underlying ipc-set as
  $X$ and $(X_i^j)_\bot=\{x_\bot^i\}$, $(X_i^j)^\top=\{x^\top_j\}$.
  Then $\Lang(X)=\bigcup_{i,j}\Lang(X_i^j)$.
\end{proof}

Intuitively, we switch off all start and accept states but one of each
in this proof.

\subsection*{Normal form of paths on closures of iHDAs}

The next lemma gives a normal form for paths on HDAs that are closures
of iHDAs.

\begin{lem}
  \label{l:Red}
  Let $X\in\IHDA$, $\alpha\in\Qath{\Cl(X)}$.  Then $\alpha$ is
  subsumed by a path of the form
  \[
    ([x;C\cup A,D] \arrO{A} [x;C,D]) *\beta* ([y;C,D] \arrI{B}
    [y;C,D\cup B]),
  \]
  where $\beta=([x_0;C,D],\phi_0,\dots,[x_n;C,D])$.
\end{lem}
See Figure \ref{fi:NormalForm} for an example.
Note that the restriction of $\alpha=([x_k;A_k,B_k],\varphi_k)\in\Qath{\Cl(X)}$
to the first coordinate gives a path $\alpha'=(x_k,\varphi'_k)\in\Qath{X}$
with, possibly, trivial steps (\ie, $\varphi_k'=d^0_\emptyset$ or $\varphi_k'=d^1_\emptyset$).
In Lemma~\ref{l:Red} we regard $A,B,C,D$ as subsets of $\ev(\alpha')$.

\begin{proof}
Without loss of generality we assume that 
$\alpha$ is a \emph{dense} path, that is, all its steps either start ($d^0_{\{a\}}$) or terminate ($d^1_{\{a\}}$) a single event $a$.
Every single step $\sigma$ in $\alpha$ falls into one of three categories:
\begin{itemize}
\item []($+$)\ 
	$\sigma=[x;A,B]\arrI{b}[x;A,B\cup\{b\}]$ for $b\in T_{\ev(x)}$,
\item[]($-$)\
	$\sigma=[x;A\cup\{a\},B]\arrO{a}[x;A,B]$ for $a\in S_{\ev(x)}$,
\item[]  ($0$)\
	neither of the above, so
	$\sigma=[x;A,B]\arrI{b}[\delta^1_b(x);A,B]$
	or $\sigma=[\delta^0_a(x);A,B]\arrO{a}[x;A,B]$.
\end{itemize}
We show that the steps of $\alpha$ can be rearranged so that all
($0$)-steps are preceded by ($-$)-steps and succeeded by
($+$)-steps. Let $\ilo{S}{U}{T}=\Iev(x)$.  For $b\in T$,
$c\in U\setminus T$,
\begin{multline*}
	\left( [x;A,B]\arrI{b}[x;A,B\cup \{b\}]\arrI{c}[\delta^1_c(x);A,B\cup \{b\}]\right)
	\\
	\simeq
	\left([x;A,B]\arrI{c}[\delta^1_c(x);A,B]\arrI{b}[\delta^1_c(x);A,B\cup \{b\}]\right).
\end{multline*}
For $a\in U\setminus S$, $b\in T$,
\begin{multline*}
	\left([\delta^0_a(x);A,B]\arrI{b}[\delta^0_a(x);A,B\cup\{b\}]\arrO{a}[x;A,B\cup \{b\}]\right)
	\\
	\subsu
	\left( [\delta^0_a(x);A,B]\arrO{a}[x;A,B]\arrI{b}[x;A,B\cup \{b\}]\right).
\end{multline*}
Thus every ($+$)-step followed by a ($0$)-step, can be swapped,
possibly passing to a subsuming path. Likewise, we can swap every
($0$)-step followed by a ($-$)-step.  Further, for $a\in S$, $b\in T$,
\begin{multline*}
	\left([x;A\cup\{a\},B]\arrI{b}[x;A\cup\{a\},B\cup\{b\}]\arrO{a}[x;A,B\cup \{b\}]\right)
	\\
	\subsu
	\left( [x;A\cup\{a\},B]\arrO{a}[x;A,B]\arrI{b}[x;A,B\cup \{b\}]\right),
\end{multline*}
so every ($+$)-step followed by a ($-$)-step can be swapped, too.
Finally, we can concatenate all ($-$)-steps, ($0$)-steps and ($+$)
steps to obtain the conclusion.
\end{proof}

\begin{figure}[tbp]
  \centering
\begin{tikzpicture}[x=0.6cm, y=0.5cm]
 \begin{scope}[shift={(0,0)}]
    \path[fill=black!05] (0,0) to (4,0) to (4,4) to (0,4);
    \path[fill=black!05] (4,0) to (8,0) to (8,4) to (4,4);
    \path[fill=black!05] (4,4) to (8,4) to (8,8) to (4,8);
    \node[state] (00) at (0,0) {};
    \node[state] (04) at (0,4) {};           
    \node[state] (40) at (4,0) {};           
    \node[state] (44) at (4,4) {};
    \node[state] (80) at (8,0) {};
    \node[state] (84) at (8,4) {};
    \node[state] (88) at (8,8) {};
    \node[state] (48) at (4,8) {};
    \path (00) edge (04);
    \path (00) edge node [below] {$[x;b,\emptyset]$} (40);
    \path (04) edge(44);
    \path (40) edge (44);
    \path (40) edge node [below] {$[y;b,\emptyset]$}  (80);
    \path (44) edge (84);
    \path (48) edge (88);
    \path (80) edge node [right] {$[y;\emptyset,c]$} (84);
    \path (44) edge (48);
    \path (84) edge node [right] {$[z;\emptyset,c]$} (88);
    \node (62)[above] at (6,2) {$[y;\emptyset,\emptyset]$};
    \node[below left] at (00) {$[x;ab,\emptyset]$};
    \node[below] at (40) {$[b;b,\emptyset]$};
    \node[right] at (84) {$[c;\emptyset,c]$};
    \node[above right] at (88) {$[c';\emptyset,c]$};
    \path[very thick, orange] (00) edge (2,0);
    \path[very thick, orange] (2,0) edge (40);
    \path[very thick, orange] (40) edge (6,0);
    \path[very thick, orange] (6,0) edge (6,2);
    \path[very thick, orange] (6,2) edge (8,2);
    \path[very thick, orange] (8,2) edge (84);
    \path[very thick, orange] (84) edge (8,6);
    \path[very thick, orange] (8,6) edge (88);
    \node[orange] at (1,1) {$\alpha$};
  \end{scope}
 \begin{scope}[shift={(12,0)}]
    \path[fill=black!5] (0,0) to (4,0) to (4,4) to (0,4);
    \path[fill=black!5] (4,0) to (8,0) to (8,4) to (4,4);
    \path[fill=black!5] (4,4) to (8,4) to (8,8) to (4,8);
    \node[state] (00) at (0,0) {};
    \node[state] (04) at (0,4) {};           
    \node[state] (40) at (4,0) {};           
    \node[state] (44) at (4,4) {};
    \node[state] (80) at (8,0) {};
    \node[state] (84) at (8,4) {};
    \node[state] (88) at (8,8) {};
    \node[state] (48) at (4,8) {};
    \path (00) edge (04);
    \path (00) edge (40);
    \path (04) edge (44);
    \path (40) edge (44);
    \path (40) edge (80);
    \path (44) edge (84);
    \path (48) edge node [above] {$[c';\emptyset,\emptyset]$} (88);
    \path (80) edge (84);
    \path (44) edge (48);
    \path (84) edge (88);
    \node (22)[above] at (2,2) {$[x;\emptyset,\emptyset]$};
    \node (62)[below] at (6,2) {$[y;\emptyset,\emptyset]$};
    \node (66)[left] at (6,6) {$[z;\emptyset,\emptyset]$};
    \node[below left] at (00) {$[x;ab,\emptyset]$};
    \node[above right] at (88) {$[c';\emptyset,c]$};
    \node[below left] at (4,2) {$[b;\emptyset,\emptyset]$};
    \node[above right] at (6,4) {$[c;\emptyset,\emptyset]$};
    \path[very thick, purple] (00) edge (2,2);
    \path[very thick, cyan] (2,2) edge (4,2);
    \path[very thick, cyan] (4,2) edge (6,2);
    \path[very thick, cyan] (6,2) edge (6,4);
    \path[very thick, cyan] (6,4) edge (6,6);
    \path[very thick, cyan] (6,4) edge (6,8);
    \path[very thick, purple] (6,8) edge (88);
    \node[cyan] at (5,3) {$\beta$};
  \end{scope}
\end{tikzpicture}
     \caption{An illustration of Lemma \ref{l:Red}
     using the iHDA from Figure \ref{fi:iClosure}.
     The orange path $\alpha$ on the left
     is subsumed by $([x;ab,\emptyset]\arrO{ab}[x;\emptyset,\emptyset])*\beta*([c';\emptyset,\emptyset]\arrI{c}[c';\emptyset,c])$ on the right.
  \label{fi:NormalForm}
     }
\end{figure}


\section{Cylinders}
\label{s:cofib}

In this section we introduce cylinders for ipc-sets.  This
construction is motivated by the double mapping cylinder from
topology
and may be regarded as a significant generalization of resolving $\varepsilon$-transitions.
For a pair of ipc-maps $f:Y\to X$ and $g:Z\to X$, we
construct an ipc-set $\Cyl(f,g)$, which is equivalent to $X$ in a
sense explained below. This allows us to replace $f$ and
$g$ by injections $\tilde{f}:Y\to\Cyl(f,g)$ and
$\tilde{g}:Z\to \Cyl(f,g)$ whose images are initial and final in
$\Cyl(f,g)$, respectively, in the sense of the following definition. 
Cylinders are used later as tools
 that separate initial and accept states in iHDAs, that is, which
 replace them by proper ones in such a way that the languages
  accepted do not change.

\subsection*{Initial and final inclusions}

Let $X$ be an ipc-set.
A ipc-subset $Y\subseteq X$ is \emph{initial}
if it is down-closed with respect to the reachability preorder $\preceq$ in $X$.
Equivalently, $Y$ is initial in $X$
if $\delta^1_B(x)\in Y$ implies $x\in Y$ for all $x\in X[U]$ and
$B\subseteq U\setminus T_U$.
(Since $Y$ is an ipc-set, the implication $x\in Y\implies \delta^0_A(x)\in Y$ follows.)
By reversal, $Y$ is \emph{final} if it
is up-closed with respect to $\preceq$ or, equivalently,
$\delta^0_A(x)\in Y$ implies $x\in Y$.
An \emph{initial} (\emph{final}) \emph{inclusion} is an injective
ipc-map whose image is an initial (final) ipc-subset.

\begin{lem}
  \label{l:ClPreservesInitialFinal}
  Let $f:Y\to X$ be an initial (final) inclusion of ipc-sets.  Then
  its closure $\Cl(f):\Cl(Y)\to\Cl(X)$ is an initial (final) inclusion
  of pc-sets.
\end{lem}
\begin{proof}
  Suppose $f$ is an initial inclusion, $x\in X[U]$, and $\delta^1_D([x;A,B])\in \im(\Cl(f))$.  Then
  $\delta^1_{D\setminus T_U}(x)\in\im(f)$ because
 \[
 	\delta^1_D([x;A,B])= [\delta^1_{D\setminus T_U}(x);A,B\cup (D\cap
  T_U)] \in \im(\Cl(f)).
  \]
  Thus $x\in \im(f)$ because $f$ is initial, and
  $[x;A,B]\in\im(\Cl(f))$ follows.
  The proof for final inclusions is similar.
\end{proof}

\subsection*{Proper iHDAs}

The \emph{start} and \emph{accept maps} of an iHDA $X$ are the ipc-maps
\begin{equation*}
  \iota_\bot^X= \bigsqcup_{x\in X_\bot}\!\! \iota_x:
  \bigsqcup_{x\in X_\bot}\! \isq^{\Iev(x)}\to X
  \qquad\text{ and }\qquad
  \iota^\top_X=\bigsqcup_{x\in X^\top}\!\! \iota_x:
  \bigsqcup_{x\in X^\top}\! \isq^{\Iev(x)}\to X.
\end{equation*}

An iHDA is \emph{start proper} if its start map is an initial
inclusion and \emph{accept proper} if its accept map is a final
inclusion. An iHDA is \emph{proper} if it is start proper, accept
proper and the images of the start map and the accept map are
disjoint.

The following lemma and example explain the structure of start and
accept proper iHDAs.

\begin{lem}
  \label{l:InitialProperCells}
  All start cells of start proper iHDAs are $\preceq$-min\-i\-mal. All
  accept cells of accept proper iHDAs are $\preceq$-maximal.
\end{lem}
\begin{proof}
  Let $x_\bot\in X_\bot$ and $U=\Iev(x_\bot)$. Then obviously
  $S_U=U$.  The top cell $c=[\emptyset|U|\emptyset]$ of $\isq^{U}$ is
  thus $\preceq$-minimal, in particular when regarded as a cell in
  $\bigsqcup_{x\in X_\bot}\isq^{\Iev(x)}$.  But $x_\bot=i_\bot^X(c)$
  and initial inclusions preserve $\preceq$-minimal elements.  So
  $x_\bot$ is $\preceq$-minimal.  The claim for accept cells follows
  by reversal.
\end{proof}

\begin{exa}
The condition of Lemma~\ref{l:InitialProperCells} is not sufficient
for properness.  The following diagrams show examples of iHDAs that
are not start proper:
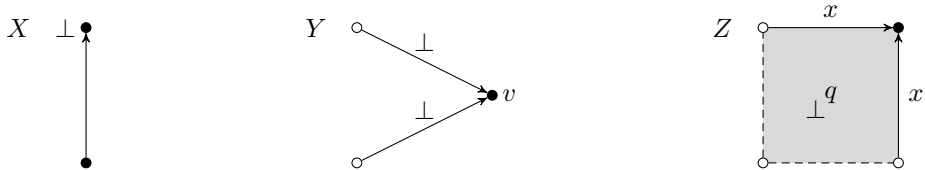
\begin{equation*}
  \hfill
  \begin{tikzpicture}[x=1.8cm, y=1.8cm]
    \begin{scope}
      \node[state] (a1) at (0,1) {};
      \node[state] (a0) at (0,0) {};
      \node[left] at (a1) {$\bot$};
      \path (a0) edge (a1);	
      \node at (-0.5,1) {$X$};
    \end{scope}
    \begin{scope}[shift={(2,0)}]
      \node[state, fill=white] (b1) at (0,1) {};
      \node[state, fill=white] (b0) at (0,0) {};
      \node[state] (b2) at (1,0.5) {};
      \path (b0) edge node[above] {$\bot$} (b2);		
      \path (b1) edge node[above] {$\bot$}(b2);		
      \node at (-0.3,1) {$Y$};
      \node[right] at (b2) {$v$};
    \end{scope}
    \begin{scope}[shift={(5,0)}]
      \path[fill=black!15] (0,0) to (1,0) to (1,1) to (0,1);
      \node[state, fill=white] (00) at (0,0) {};
      \node[state, fill=white] (10) at (1,0) {};
      \node[state, fill=white] (01) at (0,1) {};
      \node[state] (11) at (1,1) {};
      \path (00) edge[-, densely dashed] (10);
      \path (01) edge node[above] {$x$} (11);
      \path (00) edge[-, densely dashed]  (01);
      \path (10) edge node[right] {$x$} (11);
      \node at (-0.3,1) {$Z$};
      \node (q) at (0.5,0.5) {$q$};
      \node[below left=-1mm] at (q) {$\bot$};
    \end{scope}
  \end{tikzpicture}
  \hfill\mbox{}
\end{equation*}
Edges marked with $x$ have been identified.  In the first diagram, the
start map $\iota^X_\bot$ is an inclusion, but not initial.  In the
second and third diagram, neither $\iota^Y_\bot$ nor $\iota^Z_\bot$ is
an inclusion: $\iota^Y_\bot$ maps two different vertices to $v$;
$\iota^Z_\bot$ maps two different edges of $\isq^{\Iev(q)}$ to $x$.
\end{exa}

\subsection*{Lifting properties}

An ipc-map $f:Y\to X$ has the \emph{future lifting property}
(\emph{FLP}) if for every up-step $\alpha=(\delta^0_A(x)\arrO{A}x)$ in
$X$ and every $y\in Y$ such that $f(y)=\delta^0_A(x)$ there is an
up-step $\beta=(y\arrO{A}z)$ in $Y$ such that $f(\beta)=\alpha$. The
\emph{past lifting property} (\emph{PLP}) is defined by reversal.

FLP and PLP are equivalent to the lifting properties in the following
diagrams:
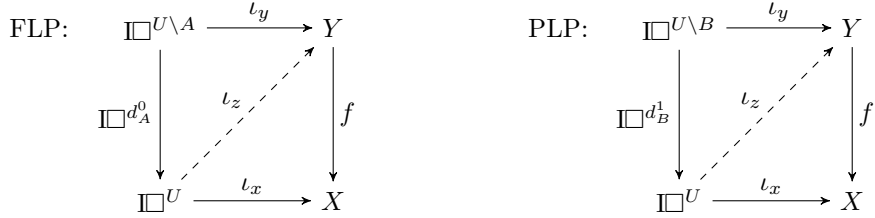
\begin{equation*}
  \hfill
\begin{tikzpicture}[x=2.3cm, y=2.3cm]
\begin{scope}[shift={(0,0)}]
\node at (-0.7,1) {FLP:};
	\node(sqUA) at (0,1) {$\isq^{U\setminus A}$};
	\node(Y) at (1,1) {$Y$};
	\node(sqU) at (0,0) {$\isq^U$};
	\node(X) at (1,0) {$X$};
	\path (sqUA) edge node[left=-0.5mm] {$\isq^{d^0_A}$} (sqU);
	\path (sqUA) edge node[above=-0.5mm] {$\iota_y$} (Y);
	\path (sqU) edge node[above=-0.5mm] {$\iota_x$} (X);
	\path (Y) edge node[right=-0.5mm] {$f$}  (X);
	\path (sqU) edge[dashed] node[above left=-1mm] {$\iota_z$} (Y);
\end{scope}
\begin{scope}[shift={(3,0)}]
\node at (-0.7,1) {PLP:};
	\node(sqUA) at (0,1) {$\isq^{U\setminus B}$};
	\node(Y) at (1,1) {$Y$};
	\node(sqU) at (0,0) {$\isq^U$};
	\node(X) at (1,0) {$X$};
	\path (sqUA) edge node[left=-0.5mm] {$\isq^{d^1_B}$} (sqU);
	\path (sqUA) edge node[above=-0.5mm] {$\iota_y$} (Y);
	\path (sqU) edge node[above=-0.5mm] {$\iota_x$} (X);
	\path (Y) edge node[right=-0.5mm] {$f$}  (X);
	\path (sqU) edge[dashed] node[above left=-1mm] {$\iota_z$} (Y);
\end{scope}
\end{tikzpicture}
\hfill\mbox{}
\end{equation*}

The next lemma states that path lifting properties allow to lift paths along $f$,
given that their source or target cells can be lifted.
It is immediate from the definitions.

\begin{lem}\label{l:LP}
  An ipc-map $f:Y\to X$ has the FLP if and only if for every $\alpha\in\Qath{X}$
  and $y\in f^{-1}(\src(\alpha))$ there exists a path
  $\beta\in \Qath{Y}$ such that $\src(\beta)=y$ and $f(\beta)=\alpha$.
  An analogous property holds for PLP. \qed
\end{lem}

Let $f:Y\to X$ be an ipc-map and $S,T\subseteq X$, that is, these are
subsets, but not necessarily ipc-subsets.  Then $f$ has the
\emph{total lifting property} (\emph{TLP}) with respect to $S$ and $T$
if for every path $\alpha\in\Qath{X}$ with $\src(\alpha)\in S$ and
$\tgt(\alpha)\in T$ and every $y\in f^{-1}(\src(\alpha))$ and
$z\in f^{-1}(\tgt(\alpha))$, there exists a path
$\beta\in\Qathft{Y}{y}{z}$ such that $f(\beta)=\alpha$.

\begin{prop}
  \label{p:LiftingProperties}
  Let $f:Y\to X$ be an iHDA map such that the functions
  $Y_\bot\to X_\bot$ and $Y^\top\to X^\top$ induced by $f$ are
  surjective.  Suppose at least one of the following holds:
	\begin{enumerate}
	\item $f$ has the future lifting property and $Y^\top=f^{-1}(X^\top)$,
	\item $f$ has the past lifting property and $Y_\bot=f^{-1}(X_\bot)$,
	\item $f$ has the total lifting property with respect to
          $X_\bot$ and $X^\top$.
	\end{enumerate}
	Then $f$ is a weak equivalence.
\end{prop}

\begin{proof}
  For (1), there exists an $y\in Y_\bot$ such that $f(y)=\src(\alpha)$
  by assumption, and $\beta\in \Qath{Y}$ such that $\src(\beta)=y$
  and $f(\beta)=\alpha$ by Lemma~\ref{l:LP}.  Moreover,
	\[
		\tgt(\beta)\in f^{-1}(\tgt(\alpha))\subseteq f^{-1}(X^\top)=Y^\top.
	\]
	\noindent
Item (2)
	follows from (1) by reversal.
	
	\noindent
	For (3), there are $y\in Y_\bot$, $z\in Y^\top$ such that
        $f(y)=\src(\alpha)$, $f(z)=\tgt(\alpha)$ by assumption.  Since
        $f$ has the TLP, there exists a $\beta\in \Qathft Y y z$ such
        that $f(\beta)=\alpha$.
\end{proof}

\subsection*{Cylinders}

Let $X, Y, Z\in \isq\Set$ and $f:Y\to X$, $g:Z\to X$ be
ipc-maps. Assume further that $f$ and $g$ have disjoint images.  This
is not directly used in the construction, but crucial in proofs.

The \emph{cylinder} $\Cyl(f,g)$ is the ipc-set such that
$\Cyl(f,g)[U]$ is the set of $(x,K,L,\phi,\psi)$, where
\begin{itemize}
\item
	$x\in X[U]$;
\item
	$K$ is an initial ipc-subset of $\isq^U$;
\item
	$L$ is a final ipc-subset of $\isq^U$;
\item
	$\phi:K\to Y$ is an ipc-map such that $f\circ \phi=\iota_x|_{K}$;
\item
	$\psi:L\to Z$ is an ipc-map such that $g\circ\psi=\iota_x|_{L}$.
\end{itemize}
For
$d_{A,B}\in\isq(V,U)$ and $(x,K,L,\phi,\psi)\in \Cyl(f,g)[U]$, we put
\begin{equation*}
\delta_{A,B}(x,K,L,\phi,\psi) = ( \delta_{A,B}(x),
(\isq^{d_{A,B}})^{-1}(K), (\isq^{d_{A,B}})^{-1}(L), \phi\circ
\isq^{d_{A,B}}, \psi\circ \isq^{d_{A,B}}).
\end{equation*}

Equivalently, $\Cyl(f,g)[U]$ is the set of commutative diagrams of
solid arrows in Figure \ref{fi:cylinder} and the face map
$\delta_{A,B}$ composes the diagram with the dashed arrows.
The following is then clear (recall that $f(Y)\cap g(Z)=\emptyset$).

\begin{lem}\label{p:CellsOfCylinderEasy}
  Let $(x,K,L,\phi,\psi)\in\Cyl(f,g)$. Then
  $K\subseteq (\iota_x)^{-1}(f(Y))$ and
  $L\subseteq (\iota_x)^{-1}(g(Z))$.  Thus $K\cap L=\emptyset$,
  $x\in f(Y)$ implies $L=\emptyset$, and $x\in g(Z)$ implies
  $K=\emptyset$.
\end{lem}

\begin{figure}
\begin{tikzpicture}[x=.8cm, y=.9cm]
\begin{scope}
	\draw[-,fill=lightgray!50!white] (0,0)--(0,1.2)--(1.2,1.7).. controls (1.4,1.3) and (2.2,1.0) .. (2.4,1.6)--(4.0,1.0)--(4.0,0.0);
	\draw[-, fill=lightgray] (0,0) .. controls (0,0.4)  and (4.0,0.4) .. (4,0)
	.. controls (4.0,-0.4)  and (0.0,-0.4) .. (0,0);
	\draw[-, fill=lightgray] (0,-1.4) .. controls (0,-1.0)  and (4.0,-1.0) .. (4,-1.4)
	.. controls (4.0,-1.8)  and (0.0,-1.8) .. (0,-1.4);
	\draw[-, fill=lightgray] (0,1.2) .. controls (0,1.7)  and (1.1,2.1) .. (1.2,1.7)
	 .. controls (1.2,1.0)  and (0.1,0.9) .. (0,1.2);
	\draw[-, fill=lightgray] (-2,1.2) .. controls (-2,1.7)  and (-0.9,2.1) .. (-0.8,1.7)
	 .. controls (-0.8,1.0)  and (-1.9,0.9) .. (-2,1.2);
	 \node at (0.6, 1.4) {$\tilde{f}(Y)$};
	 \node at (-1.4, 1.4) {$Y$};
	 \draw[-, fill=lightgray] (2.4,1.6) .. controls (2.5,2.0)  and (4.0,1.9) .. (4.0,1.0)
	  .. controls (4.0,0.6)  and (2.2,0.9) .. (2.4,1.6);
	 \draw[-, fill=lightgray] (4.8,1.6) .. controls (4.9,2.0)  and (6.4,1.9) .. (6.4,1.0)
	  .. controls (6.4,0.6)  and (4.6,0.9) .. (4.8,1.6);
	 \path (-0.7,1.4) edge[right hook->] node[above] {$\tilde{f}$} (-0.1,1.4);
	 \path (4.6,1.4) edge[left hook->] node[above] {$\tilde{g}$} (4.0,1.4);
	 \path (2.2,-1.0) edge[left hook->] node[right] {$j$} (2.2,-0.4);
	 \path (1.8,-0.4) edge node[left] {$p$} (1.8,-1.0);
	 \path (-1.1,0.9) edge node[left] {$f$} (0.0,-1.1);
	 \path (5.1,0.9) edge node[right] {$g$} (4.0,-1.1);
	 \node at (3.2, 1.3) {$\tilde{g}(Z)$};
	 \node at (5.6, 1.3) {$Z$};
 	 \node at (2, 0) {$j(X)$};
 	 \node at (2, -1.4) {$X$};
 	 \node at (1.8, 0.8) {$\Cyl(f,g)$};
\end{scope}
\begin{scope}[shift={(8.9,-2)}, scale=2.5]
 	 \node(Kv) at (0,1.6) {$(\isq^{d_{A,B}})^{-1}(K)$};
	\node(sqv) at (1,1.6) {$\isq^V$};
	\node(Lv) at (2,1.6) {$(\isq^{d_{A,B}})^{-1}(L)$};
	\node(K) at (0,0.8) {$K$};
	\node(sq) at (1,0.8) {$\isq^U$};
	\node(L) at (2,0.8) {$L$};
	\node(Y) at (0,0) {$Y$};
	\node(X) at (1,0) {$X$};
	\node(Z) at (2,0) {$Z$};
	\path (Kv) edge[dashed] node[left] {$\isq^{d_{A,B}}$} (K);
	\path (sqv) edge[dashed] node[right] {$\isq^{d_{A,B}}$} (sq);
	\path (Lv) edge[dashed] node[right] {$\isq^{d_{A,B}}$} (L);
	\path (K) edge node[left] {$\varphi$} (Y);
	\path (sq) edge node[right] {$\iota_x$} (X);
	\path (L) edge node[right] {$\psi$} (Z);
	\path (K) edge [right hook->]  (sq);
	\path (L) edge [left hook->]  (sq);
	\path (Kv) edge [right hook->,dashed]  (sqv);
	\path (Lv) edge [left hook->,dashed]  (sqv);
	\path (Y) edge node[above] {$f$} (X);
	\path (Z) edge node[above] {$g$} (X);
\end{scope}
\end{tikzpicture}
 \caption{The cylinder $\Cyl(f,g)$ and a diagram defining its cell.}
 \label{fi:cylinder}
\end{figure}

$\Cyl(f,g)$ is equipped with the ipc-maps shown in Figure \ref{fi:cylinder}.
They are defined by 
\[j(x)=(x,\emptyset,\emptyset, \emptyset, \emptyset),
\qquad
p(x,K,L,\phi,\psi)=x,
\]
\[
\tilde{f}(y)=(f(y),\isq^{\Iev(y)},\emptyset,\iota_y,\emptyset),
\qquad
\tilde{g}(z)=(g(z),\emptyset,\isq^{\Iev(z)},\emptyset,\iota_z).
\]

Intuitively, $\Cyl(f,g)$ may be regarded as a result of the following procedure:
in the disjoint union of $Y\sqcup X\sqcup Z$,
add $\varepsilon$-transitions $y\to f(y)$ for all cells $y\in Y$
and $g(z)\to z$ for all $z\in Z$.
Next, resolve all $\varepsilon$-transitions.
(Alas, we know no satisfactory definition of $\varepsilon$-transitions for ipc-sets).
See Figure \ref{fi:cylexample} for an example.

Next we collect some of properties of cylinders.
\begin{lem}
  \label{l:DoubleCylinderIsProper}
  {\ }
\begin{enumerate}
\item
	$p\circ \tilde{f}=f$, $p\circ \tilde{g}=g$, and $p\circ j=\id_X$.
\item
	$\tilde{f}$ is an initial inclusion and
	$	
		\tilde{f}(Y)=\{(x,K,L,\varphi,\psi)\in \Cyl(f,g)\mid K=\isq^{\Iev(x)}, L=\emptyset \}.
	$
\item
	$\tilde{g}$ is a final inclusion and 
	$
		\tilde{g}(Z)=\{(x,K,L,\varphi,\psi)\in \Cyl(f,g)\mid K=\emptyset, L=\isq^{\Iev(x)}\}.
	$
\item
	$j$ is an inclusion and $j(X)=\{(x,\emptyset,\emptyset,\emptyset,\emptyset)\in\Cyl(f,g)\}$.
\item
	$\tilde{f}(Y)$, $\tilde{g}(Z)$ and $j(X)$ are pairwise disjoint.
\end{enumerate}
\end{lem}

\begin{proof}
	Item (1) is straightforward from the definition.

\noindent
For (2), let $Y'\subseteq \Cyl(f,g)$ be the right-hand side of the
equation.  We show that
	\[
		h:Y'\ni (x,\isq^{\Iev(x)},\emptyset,\varphi,\emptyset)\mapsto \varphi([\emptyset|\emptyset])\in Y
	\]
	is the inverse of $\tilde{f}:Y\to Y'$.
	Indeed, 
	\[
		h(\tilde{f}(y))
		=h(f(y),\isq^{\Iev(y)},\emptyset,\iota_y,\emptyset)
		=\iota_y([\emptyset|\emptyset])
		=y
	\]
        and
	\begin{align*}
		\tilde{f}(h(x,\isq^{\Iev(x)},\emptyset,\varphi,\emptyset))
		=
		\tilde{f}(\varphi([\emptyset|\emptyset]))
		&=
		(f(\varphi([\emptyset|\emptyset]), \isq^{\Iev(x)},\emptyset,\iota_{\varphi([\emptyset|
		\emptyset])},\emptyset)
		\\
		&=
                (\iota_x([\emptyset|\emptyset]), \isq^{\Iev(x)},\emptyset,\varphi,\emptyset)\\
		&=
		(x, \isq^{\Iev(x)},\emptyset,\varphi,\emptyset).
	\end{align*}
	In the above, $f\circ \varphi=\iota_x$ holds by definition,
        and $\varphi=\iota_{\varphi([\emptyset|\emptyset])}$ since
        ipc-maps from $\isq^{\Iev(x)}$ are determined by their values
        on $[\emptyset|\emptyset]$.  Thus, $\tilde{f}$ defines an
        isomorphism between $Y$ and its image $Y'$.  It remains to
        show that $Y'\subseteq \Cyl(f,g)$ is an initial subset.
        Suppose $\delta^1_B(x,K,L,\varphi,\psi)\in Y'$.  Then
        $(\isq^{d^1_B})^{-1}(K)=\isq^{U\setminus B}$ and
        $(\isq^{d^1_B})^{-1}(L)=\emptyset$.  Hence, $K$ is an initial
        subset of $\isq^U$ containing an upper face $[\emptyset|B]$, so
        also $[\emptyset|\emptyset]\in K$ and then $K=\isq^{\Iev(x)}$.
        Therefore, $L=\emptyset$ and
        $(x,K,L,\varphi,\psi)\in Y'$.
	
Finally, (3) 
	follows from (2) by reversal, (4) 
	is obvious from the definition and (5) 
	follows from (2)--(4) since $\isq^{\Iev(x)}\neq\emptyset$ for all $x\in X$.
      \end{proof}

\begin{figure}
\begin{tikzpicture}
\def\colA{magenta!80!black}
\def\colB{cyan!80!black}
\begin{scope}[shift={(1,0)}]
	\node at (-0.7,4) {$Y$};
	\node at (-0.7,1.6) {$X$};
      \node[state,color=\colA] (a0) at (-0.5,3) {};
      \node[state,color=\colA] (a1) at (1,4) {};
      \path (a0) edge[color=\colA] node[above left,color=black] {$d$} (a1);
      \node[left] at (a0) {$p$};
      \node[right] at (a1) {$q$};
      \node[state,color=\colB] (b0) at (0,1) {};
      \node[state,color=\colB] (b1) at (2,1) {};
      \path (b0) edge[bend left=30,color=\colB] node[above, color=black] {$b$} (b1);
      \path (b1) edge[bend left=30,color=\colB] node[below, color=black] {$c$} (b0);
      \path (b0) edge [out=270,in=170,looseness=40,color=\colB] node[color=black] {$a$} (b0);
      \node[above right] at (b0) {$r$};
      \node[right] at (b1) {$s$};
\end{scope}
\begin{scope}[shift={(7,0)}, xscale=3]
      \node[state,color=\colA] (a0) at (-0.5,3) {};
      \node[state,color=\colA] (a1) at (1,4) {};
      \path (a0) edge[color=\colA] node[color=black,above] {$(a,\sq^{e},d)$} (a1);
      \node[left] at (a0) {$(r,\sq^\emptyset,p)$};
      \node[right] at (a1) {$(r,\sq^\emptyset,q)$};
      \node[state,color=\colB] (b0) at (0,1) {};
      \node[state,color=\colB] (b1) at (2,1) {};
      \path (b0) edge[bend left=20,color=\colB] node[below, color=black] {$(b,\emptyset,\emptyset)$} (b1);
      \path (b1) edge[bend left=80,color=\colB] node[below, color=black] {$(c,\emptyset,\emptyset)$} (b0);
      \path (b0) edge [out=270,in=170,looseness=40,color=\colB] node[color=black] {$(a,\emptyset,\emptyset)$} (b0);
      \node[below,yshift=-10pt] at (b0) {$(r,\emptyset,\emptyset)$};
      \path (a0) edge node[left] {$(a,0,p)$} (b0);
      \path (a0) edge node[above right, yshift=-5pt] {$(b,0,p)$} (b1);
      \path (a1) edge node[above left,pos=0.3, yshift=-8pt] {$(a,0,q)$} (b0);
      \path (a1) edge node[above right, yshift=-3pt] {$(b,0,q)$}  (b1);
      \node[right] at (b1) {$(s,\emptyset,\emptyset)$};
\end{scope}
\end{tikzpicture}
\caption
	{
	The cylinder $\Cyl(f,\emptyset)$ for a map $f:Y\to X$ determined by $f(d)=a$.
	All edges are labelled by $e$. Cells are marked by sequences $(x,K,\varphi(y))$,
	where $y$ is the top cell of $K$.
	$K=0$ stands for $\{[e|\emptyset]\}\subseteq\sq^e$.}
\label{fi:cylexample}
\end{figure}

\subsection*{Lifting properties of cylinders}

The following proposition allows us to use cylinders for replacing
iHDAs with proper iHDAs that accept the same languages.

\begin{prop}
  \label{p:DoubleCyllinderLifting}
  The projection $p:\Cyl(f,g)\to X$ has the future and past lifting
  property, as well as the total lifting property with respect to
  $f(Y)$ and $g(Z)$.
\end{prop}

\begin{proof}
  Suppose $x\in X[U]$, $y\in \Cyl(f,g)[U\setminus A]$ and
  $p(y)=\delta_A^0(x)$.  Then $y=(\delta^0_A(x),K,L,\phi,\psi)$ for
  some $K$, $L$, $\phi$, $\psi$.  For
  $z=(x,\isq^{d^0_{A}}(K),\isq^{d^0_{A}}(L), \phi \circ
  (\isq^{d^0_A})^{-1}, \psi \circ (\isq^{d^0_A})^{-1})$ we have
  $p(z)=x$ and $\delta^0_A(z)=y$. Thus $p$ has the FLP, and the PLP
  follows by reversal.

  For the TLP, let $\alpha=(x_0,\omega_1,\dotsc,x_n)\in\Qath{X}$,
  $y,z\in \Cyl(f,g)$, and suppose $p(y)=x_0$, $p(z)=x_n$.  Then
  $y=(x_0,K_0,\emptyset,\phi_0,\emptyset)$ and
  $z=(x_n,\emptyset,L_n,\emptyset,\psi_n)$ for some $K_0$, $\phi_0$,
  $L_n$, $\psi_n$, and the remaining items are empty by Lemma
  \ref{p:CellsOfCylinderEasy}.  We abbreviate $U_k=\Iev(x_k)$.

  Define $K_k\subseteq \isq^{U_k}$ and $\phi_k:K_k\to Y$ inductively:
  \begin{itemize}
  \item if $\omega_k=d^1_B$, $B\subseteq U_{k-1}\setminus T_{U_{k-1}}$, then
    $K_k=(\isq^{d^1_B})^{-1}(K_{k-1})$,
    $\phi_k=\phi_{k-1}\circ \isq^{d^1_B}$;
  \item if $\omega_k=d^0_A$, $A\subseteq U_k\setminus S_{U_k}$, then
    $K_{k}=\isq^{d^0_A}(K_{k-1})$,
    $\phi_k=\phi_{k-1}\circ (\isq^{d^0_A})^{-1}$.
  \end{itemize}
  We further define $L_k$ and $\psi_k$ inductively in backwards
  fashion.  Now
  \[
  	\beta=((x_0,K_0,L_0,\phi_0,\psi_0),\omega_1,\dotsc,(x_n,K_n,L_n,\phi_n,\psi_n))
 \]
 is a path in $\Cyl(f,g)$ such that $p(\beta)=\alpha$.  Moreover,
 $L_0=\emptyset=K_n$ by Lemma~\ref{p:CellsOfCylinderEasy}, and
 therefore $\src(\beta)=y$ and $\tgt(\beta)=z$.  The TLP with respect
 to $f(Y)$ and $g(Z)$ is thus established.
\end{proof} 

Using cylinders we can now show that it suffices to focus on start or
accept proper iHDAs as recognisers of regular languages.  Recall that
by Lemma~\ref{l:RegIsUnionOfSimple}, every regular language is
a finite union of simple regular languages.  This prepares us for the gluing
compositions of iHDAs in the following sections.

\begin{prop}
  \label{p:MakingSimpleSourceProper}
  Every simple regular language is recognised by a start simple and
  start proper iHDA, and by an accept simple and accept proper iHDA.
\end{prop}

\begin{proof}
  We prove only the first claim; the second then follows by reversal.
  Suppose $L$ is recognised by the simple iHDA $X$ with start cell
  $x_\bot\in X[U]$.  Let $Y$ be the iHDA with underlying precubical
  set $\Cyl(\iota^X_\bot,\emptyset)$, where $\emptyset:\emptyset\to X$
  is the empty map.  Further, let
  $y_\bot=(x_\bot,\isq^U,\emptyset,\id_{\isq^U},\emptyset)$ be the
  only start cell of $Y$ and $Y^\top=p^{-1}(X^\top)$. Since
  $\iota_\bot^Y=\widetilde{\iota_\bot^X}$, $Y$ is start proper by
  Lemma\ \ref{l:DoubleCylinderIsProper}(b).  Moreover, the projection
  $p:Y\to X$ has the FLP by
  Proposition~\ref{p:DoubleCyllinderLifting}. Thus
  $\Lang(Y)=\Lang(X)=L$ by Proposition~\ref{p:LiftingProperties}(a).
\end{proof}

\begin{prop}
  \label{p:IdSubtract}
  If the language $L$ is regular, then so is $L\setminus \Id$.
\end{prop}
\begin{proof}
  Suppose first that $L$ is simple.  Let $X$ be a start simple and
  start proper iHDA recognising $L$, owing to
  Proposition~\ref{p:MakingSimpleSourceProper}, and let $Y$ be the
  iHDA with the same underlying ipc-set and start cells as $X$, and
  with accept cells $Y^{\top}=X^\top\setminus X_\bot$.  By Lemma
  \ref{l:InitialProperCells}, an accepting path $\alpha\in\Qath{X}$ is
  accepting in $Y$ if and only if it has positive length
  ($\ev(\alpha)$ is not an identity). Thus
  $\Lang(Y)=\Lang(X)\setminus\Id=L\setminus\Id$ is regular.  If $L$ is
  not simple, then let $L=\bigcup_{i} L_i$ be a finite sum of simple
  languages.  Then
  $L\setminus \Id=(\bigcup_i L_i)\setminus \Id=\bigcup_i (L_i\setminus
  \Id)$ is regular by the first case and Proposition~\ref{p:RegUnion}.
\end{proof}

\section{Gluing toolbox}
\label{s:Toolbox}

We now present several constructions of composite HDAs from simpler
pieces and study properties of their paths.

\subsection*{Sequential gluing}

We fix precubical sets $X_1,\dotsc,X_n$ and $Y_1,\dotsc,Y_{n-1}$,
initial inclusions $f_k:Y_k\to X_{k+1}$ and final inclusions
$g_k:Y_k\to X_k$ ($1\leq k<n$).  We also assume that the
$f_{k-1}(Y_{k-1}), g_{k}(Y_{k})\subseteq X_k$ are disjoint subsets and
that the $Y_k$ are acyclic, which means by definition that they are
partially ordered by the reachability preorder $\preceq$ (see
Section~\ref{s:languages}).

The \emph{sequential gluing} given by this data forms the precubical set
\[
	Z=\ZZ_n((X_k)_{k=1}^n, (Y_k)_{k=1}^{n-1}, (f_k)_{k=1}^{n-1},(g_k)_{k=1}^{n-1}),
\]
which is the colimit of the diagram
\begin{equation}\label{e:ZigZag}
  \vcenter{\hbox{
\begin{tikzpicture}[x=1.5cm, y=1.5cm]
	\node (x1) at (0,1) {$X_1$};
	\node (y1) at (1,0) {$Y_1$};
	\node (x2) at (2,1) {$X_2$};
	\node (y2) at (3,0) {$Y_2$};
	\node (yn) at (5,0) {$Y_{n-1}$};
	\node (xn) at (6,1) {$X_{n}$};
	\node (d) at (4,1) {$\dots$};
	\path (y1) edge node[below left=-1mm] {$g_1$} (x1);
	\path (y1) edge node[above left=-1mm] {$f_1$} (x2);
	\path (y2) edge node[below left=-1mm] {$g_2$} (x2);
	\path (y2) edge node[above left=-1mm] {$f_2$} (d);
	\path (yn) edge node[below left=-1mm] {$g_{n-1}$} (d);
	\path (yn) edge node[above left=-1mm] {$f_{n-1}$} (xn);
      \end{tikzpicture}
    }}%
  \hspace{-.2cm}.
\end{equation}

Colimits of presheaves can be calculated pointwise.  So $Z$ is
obtained from $\bigsqcup_{k=1}^n X_n$ by identifying cells
$g_k(y)\in X_{k}$ and $f_{k}(y)\in X_{k+1}$ for all $1\le k\le n-1$
and $y\in Y_k$.  Let $j_k:X_k\to Z$ and $i_k:Y_k\to Z$ denote the
structural maps.

\begin{lem}
  \label{l:snake1}
  The maps $i_k$ and $j_k$ are injective for all $1\leq
  k<n$. Moreover, 
  $i_k(Y_k)=j_{k}(X_k)\cap j_{k+1}(X_{k+1})$.
\end{lem}

\begin{proof}
  As $f_{k-1}(Y_{k-1})\cap g_{k}(Y_k)=\emptyset$, each cell
  $x_k\in X_k$ can be identified with at most one other cell in either
  $X_{k-1}$ or $X_{k+1}$.  Consequently, different cells of $X_k$ are
  never identified.  Since $j_{k+1}\circ f_k=i_k=j_k\circ g_k$, we
  have $i_k(Y_k)\subseteq j_k(Y_k)\cap j_{k+1}(Y_{k+1})$.  Further, if
  $j_k(x_k)=j_{k+1}(x_{k+1})$ then $x_k$ and $x_{k+1}$ represent the
  same element in $Z$.  Equivalence classes have at most two elements,
  so $x_k=g_k(y)$ and $x_{k+1}=f_k(y)$ for some $y\in Y_k$.
\end{proof}

It follows that $X_k$ is isomorphic to its image $j_k(X_k)$ in $Z$ and
$Y_k$ is isomorphic to $i_k(Y_k)$.  For simplicity, we regard $X_k$
and $Y_k$ as precubical subsets of $Z$ such that
$Y_k=X_k\cap X_{k+1}$ is a final subset of $X_k$ and an initial subset
of $X_{k+1}$.  We can then turn $Z$ into an HDA with $Z_\bot=(X_1)_\bot$,
$Z^\top=(X_n)^\top$ for any HDAs $X_1$, $X_n$.

A \emph{checkpoint sequence} for
an accepting path $\alpha=(z_1,\varphi_1,\dotsc,z_m)\in\Qath{Z}$
is a sequence $(r_0,r_1,\dotsc,r_n)$ such that
$r_0=z_1$, $r_n=z_m$
and, for every $0<k<n$, $r_k\in Y_k$ and $r_k=z_{l}$ for some $l=l(k)$.
See Figure \ref{fi:Checkpoint} for an example.

\begin{figure}
  \centering
	\begin{tikzpicture}[x=1.4cm, y=1.4cm]
		\fill[fill=lightgray!50!white,semitransparent] (0,0) ellipse [x radius=1.5, y radius=1];
		\fill[fill=lightgray!50!white,semitransparent] (1.6,-0.7) ellipse [x radius=1.2, y radius=0.8, rotate=-20];
		\fill[fill=lightgray!50!white,semitransparent] (3.4,-0.2) ellipse [x radius=1.8, y radius=1.2, rotate=30];
		\fill[fill=lightgray!50!white,semitransparent] (5.5,0) ellipse [x radius=1.6, y radius=1.2];
		\draw (0,0) ellipse [x radius=1.5, y radius=1];
		\draw (1.6,-0.7) ellipse [x radius=1.2, y radius=0.8, rotate=-20];
		\draw (3.4,-0.2) ellipse [x radius=1.8, y radius=1.2, rotate=30];
		\draw (5.5,0) ellipse [x radius=1.6, y radius=1.2];
		\node at (-1.2,1) {$X_1$};
		\node at (0.9,-1.6) {$X_2$};
		\node at (2.5,1.0) {$X_3$};
		\node at (6.8,-1.1) {$X_4$};
    	\node[circle,draw=black,fill=black,inner sep=0pt,minimum size=3pt]
	    (a) at (-1,0) {};	
	    \node[below] at (a) {$z_1$};
	    \node[above] at (a) {$r_0$};
	    \node[below left] at (a) {$\bot$};
    	\node[circle,draw=black,fill=black,inner sep=0pt,minimum size=3pt]
	    (b) at (0,0.2) {};
	    \draw (a) -- (b);
	    \node[above] at (b) {$z_2$};
    	\node[circle,draw=black,fill=black,inner sep=0pt,minimum size=3pt]
	    (c) at (0.8,-0.4) {};		
   	    \draw (b) -- (c);	
	    \node[below] at (c) {$z_3$};
	    \node[above] at (c) {$r_1$};
    	\node[circle,draw=black,fill=black,inner sep=0pt,minimum size=3pt]
	    (d) at (2.3,-0.6) {};		
   	    \draw (c) -- (d);	
	    \node[below] at (d) {$z_4$};
	    \node[above] at (d) {$r_2$};
    	\node[circle,draw=black,fill=black,inner sep=0pt,minimum size=3pt]
	    (e) at (3.2,-0.7) {};		
   	    \draw (d) -- (e);	
	    \node[below right] at (e) {$z_5=z_8$};
    	\node[circle,draw=black,fill=black,inner sep=0pt,minimum size=3pt]
	    (f) at (2.8,0.3) {};		
   	    \draw (e) -- (f);	
	    \node[above] at (f) {$z_6$};
    	\node[circle,draw=black,fill=black,inner sep=0pt,minimum size=3pt]
	    (g) at (3.4,0.5) {};		
   	    \draw (f) -- (g);
   	    \draw (g) -- (e);
	    \node[above] at (g) {$z_7$};
    	\node[circle,draw=black,fill=black,inner sep=0pt,minimum size=3pt]
	    (h) at (4.2,0.0) {};		
   	    \draw (e) -- (h);	
	    \node[below] at (h) {$z_9$};
    	\node[circle,draw=black,fill=black,inner sep=0pt,minimum size=3pt]
	    (i) at (4.7,0.0) {};		
   	    \draw (h) -- (i);	
	    \node[below] at (i) {$z_{10}$};
	    \node[above] at (i) {$r_3$};
    	\node[circle,draw=black,fill=black,inner sep=0pt,minimum size=3pt]
	    (j) at (5.7,-0.3) {};		
   	    \draw (i) -- (j);	
	    \node[below] at (j) {$z_{11}$};
    	\node[circle,draw=black,fill=black,inner sep=0pt,minimum size=3pt]
	    (k) at (6.5,0.1) {};		
   	    \draw (j) -- (k);	
	    \node[below] at (k) {$z_{12}$};
	    \node[above] at (k) {$r_4$};
	    \node[above right] at (k) {$\top$};
	\end{tikzpicture}
	\caption{Example of checkpoint sequence.}
	\label{fi:Checkpoint}
\end{figure}

\begin{lem}
  \label{l:CheckpointSequence}
	Every accepting path $\alpha\in\Qath{Z}$ admits a checkpoint sequence.
	If $(r_k)_{k=0}^{n}$ is a checkpoint sequence for $\alpha$, then
        \begin{enumerate}
	\item
		the indices $l(k)$ are uniquely determined by $(r_k)_{k=0}^n$ and increasing,
	\item
		$z_{l(k-1)},z_{l(k-1)+1},\dotsc,z_{l(k)}\in X_k$,
	\item
		$\alpha=\beta_1*\dotsc*\beta_n$,
	where 
	$
		\beta_k=(z_{l(k-1)},\varphi_{l(k-1)},\dotsc,z_{l(k)})\in \Qathft{X_k}{r_{k-1}}{r_k}.
	$
	\end{enumerate}
\end{lem}
\begin{proof}
	Define the function $h:Z\to \mathbb{N}$ by
	\[
		h(z)=
		\begin{cases}
			2k & \text{if $z\in Y_k$,}\\
			2k-1 & \text{if $z\in X_k\setminus (Y_{k-1}\cup Y_k)$.}
		\end{cases}
	\]
	If $z\in Y_k$, then $\delta^0_A(z),\delta^1_B(z)\in Y_k$.  If
        $z\in X_k\setminus(Y_{k-1}\cup Y_k)$, then
        $\delta^0_A(z)\in X_k\setminus Y_{k}$, since
        $Y_k\subseteq X_k$ is a final subset, and
        $\delta^1_B(z)\in X_k\setminus Y_{k-1}$, since
        $Y_{k-1}\subseteq X_k$ is an initial subset.  Therefore, if
        $(z,\varphi,z')$ is a step in $Z$, then either $h(z)=h(z')$ or
        $h(z)=h(z')-1$.
	
	Let $\alpha=(z_1,\varphi_1,\dotsc,z_m)$.  The sequence
        $h(z_1),\dotsc,h(z_m)$ is then increasing by steps of size 0 and 1, and
        $h(z_1)\in\{1,2\}$, $h(z_m)\in\{2n-2,2n-1\}$.  Thus, there is
        a sequence $l(1)<\dotsm<l(n-1)$ of indices such that
        $h(z_{l(k)})=2k$, that is, $r_k=z_{l(k)}\in Y_k$ and $(r_k)$ is
        a checkpoint sequence for $\alpha$.
	
	\noindent For (1), fix a checkpoint sequence $(r_k)$ for
        $\alpha$.  Suppose $r_k=z_l=z_{l'}\in Y_k$ for $l\leq l'$.
        Then $h(z_j)=2k$ for all $l\leq j\leq l'$ and
        $(z_{l},\varphi_l,\dotsc,z_{l'})$ is a cycle in $Y_k$.  By
        assumption, $Y_k$ is acyclic, and therefore $l=l'=l(k)$.  The
        sequence $l(k)$ is increasing because the sequence $h(z_i)$
        is.
	
	\noindent For (2), note that
        $2k-2=h(z_{l(k-1)})\leq h(z_j)\leq h(z_{l(k)})=2k$ for every
        $l(k-1)\leq j\leq l(k)$. Thus, $z_j\in X_k$.
	
	\noindent Finally, (3) follows from (2) and injectivity of the
        maps $j_k$.
\end{proof}

\subsection*{Vertical decomposition}

We reuse the notation from the previous subsection. We assume that
$Y_k=\bigsqcup_{q\in C_k} Y_{k,q}$ is written as a disjoint union
of components.  For any sequence $\mathbf{q}=(q_k)_{k=1}^{n-1}$,
$q_k\in C_k$ we write
\[
	Z(\mathbf{q})=
	\ZZ_n((X_k)_{k=1}^n, (Y_{k,q_k})_{k=1}^{n-1}, (f_k|_{Y_{k,q_k}})_{k=1}^{n-1},(g_k|_{Y_{k,q_k}})_{k=1}^{n-1}).
\]
This is the sequential gluing of the $X_k$'s in which we do not glue
along the whole pc-sets $Y_k$, but only along their chosen components.
By taking the union indexed by all possible choices we obtain an
HDA that is weakly equivalent to the original sequential gluing:

\begin{prop}
\label{p:ToolVerticalUnfolding}
The map
$\bigsqcup_{\mathbf{q}\in C_1\times\dotsm\times C_{n-1}}
Z(\mathbf{q})\to Z$ induced by the identities on $X_k$ and the
inclusions $Y_{k,q_k}\subseteq Y_k$ is a weak equivalence.
\end{prop}
\begin{proof}
  Suppose $\alpha\in \Qath{Z}$ is an accepting path and $(r_k)$ a
  checkpoint sequence for $\alpha$.  Let $q_k$ be the index of the
  component $Y_{k,q_k}$ containing $r_k$.  The
  representation $\alpha=\beta_1*\dotsm*\beta_n$
  associated to $(r_k)$ also defines a path $\alpha'$ in
  $Z(\mathbf{q})$, which obviously maps to $\alpha$ under the
  canonical map.
\end{proof}

\subsection*{Self-gluing}

Let $X$ be an HDA, $Y$ a precubical set, $f:Y\to X$ an initial
inclusion and $g:Y\to X$ a final inclusion.  Suppose $Y$ is acyclic
and the sets $f(Y)$, $g(Y)$, $X_\bot$ and $X^\top$ are pairwise
disjoint.
Below we identify $f(Y)$, the ``initial'' copy of $Y$ in $X$, with the ``final'' copy $g(Y)$.
Then we show that the resulting automaton is weakly equivalent to the union of the sequence of the finite gluing compositions.

Define an HDA
\begin{equation}\label{e:Coeq}
  V=\VV(X,Y,f,g)=\colim\left(
    Y\overset{f}{\underset{g}{\longrightrightarrows}} X
  \right).
\end{equation}
Let $j:X\to V$, $i:Y\to V$ be the structural maps, let
$V_\bot=j(X_\bot)$ and $V^\top=j(X^\top)$.  Hence $V$ is obtained from
$X$ by identifying cells $f(y)$ and $g(y)$ for $y\in Y$.  Every cell
$z\in V$ is either represented by a single cell $x\in X$ if $z\not\in i(Y)$
or by a pair of cells $(f(y), g(y))$, $y\in Y$.

Unlike for sequential gluings, we cannot assume that $X$ is proper.
Hence $j$ is not necessarily injective and $X$ need not be a
precubical subset of $V$.

For $n\geq 1$ let
\[
	\ZZ_n(X,Y,f,g)
	=
	\ZZ_n((X_k)_{k=1}^n, (Y_k)_{k=1}^{n-1}, (f_k)_{k=1}^{n-1},(g_k)_{k=1}^{n-1})
\]
where we take $X_k=X$, $Y_k=Y$,
$f_k=f$, $g_k=g$ for all $k$.  As before, we let
\[
	\ZZ_n(X,Y,f,g)_\bot=(X_1)_\bot,\qquad \ZZ_n(X,Y,f,g)^\top=(X_n)^\top.
\]
The transformation
from diagram (\ref{e:ZigZag}) to (\ref{e:Coeq}) that maps 
all $X_k$'s into $X$, $Y_k$'s into $Y$, $f_k$'s into $f$ and $g_k$'s
into $g$ induces a map $\pi_n:\ZZ_n(X,Y,f,g)\to \VV(X,Y,f,g)$.

\begin{prop}
\label{p:ToolHorizontalUnfolding}
	The map 
	\[
		\pi=\bigsqcup\pi_n:\ZZ_n(X,Y,f,g)\to\VV(X,Y,f,g)
	\]
	is a weak equivalence.
\end{prop}

\begin{proof}
  Suppose $\alpha=(z_1,\varphi_1,\dotsc,z_m)\in\Qath{V}$ is an
  accepting path. We can choose a sequence $1<l(1)< \dotsc<l(n-1)<m$
  of indices such that $z_{l(k)}\in i(Y)$. Then there exists $b(k)$
  with $l(k-1)<b(k)<l(k)$ such that $z_{b(k)}\not\in i(Y)$ ($1<k<n-1$)
  and $k$ is maximal.
  Further, let $b(0)=1$, $b(n)=m$.  Equivalently,
  we can choose one representative $l(k)$ from every sequence of
  consecutive cells of $\alpha$ that belong to $i(Y)$, and we can
  choose representatives $b(k)$ from all sequences of cells that not
  belong to $i(Y)$.  For $1<k<n-1$ we choose cells
  $x^k_s\in j^{-1}(z_s)$ for $l(k-1)\leq s\leq l(k)$.  If
  $z_s\not\in i(Y)$, then $x^k_s$ is the only element of
  $j^{-1}(z_s)$.  Otherwise, $j^{-1}(z_s)=\{f(y_s),g(y_s)\}$ for some
  $y_s\in Y$ and we put $x^k_s=f(y_s)$ for $s<b(k)$ and $x^k_s=g(y_s)$
  for $s>b(k)$.
	
  We show that
  $\beta_k=(x^k_{l(k-1)},\varphi_{l(k-1)},\dotsc,x^k_{l(k)})$ is a
  path in $X$.  Pick an arbitrary index $s\in\{l(k-1),\dotsc,l(k)-1\}$
  and assume that $\varphi_s=d^1_B$.  We must check that
  $\delta^1_B(x^k_s)=x^k_{s+1}$.  It is clear that both
  $\delta^1_B(x^k_s)$ and $x^k_{s+1}$ belong to $j^{-1}(z_{s+1})$
  because $z_{s+1}=\delta^1_B(z_s)$.  If $z_{s+1}\not\in i(Y)$, then
  $j^{-1}(z_{s+1})$ has one element and there is nothing to prove.
  Otherwise, $j^{-1}(z_{s+1})=\{f(y),g(y)\}$ for some $y\in Y$.
\begin{itemize}
\item
	If $z_s\not\in i(Y)$, then $s+1>b(k)$,
	and thus, by definition, $x^k_{s+1}=g(y)$.
	Further, $\delta^1_B(x^k_s)\not\in f(Y)$
	(since $x^k_s\not\in f(Y)$ and $f(Y)$ is an initial subset of $X$).
	Hence $\delta^1_B(x^k_s)=g(y)=x^k_{s+1}$.
      \item If $z_s\in i(Y)$, then either $s>b(k)$ and
        $x^k_s,x^k_{s+1}\in g(Y)$, or $s+1<b(k)$ and
        $x^k_s,x^k_{s+1}\in f(Y)$.  In both cases
        $\delta^1_B(x^k_s)=x^k_{s+1}$.
\end{itemize}
The case $\varphi_k=d^0_A$ follows by reversal.
Finally, $\beta_k\in\Qath{X}$.

Let $\alpha_k=(x^k_{l(k-1)},\varphi_{l(k-1)},\dotsc,x^k_{l(k)})\in\Qath{X}$. Then
clearly $j(\beta_k)=\alpha_k$ and there are cells $y_k\in Y$ such
that $\tgt(\beta_k)=g(y_k)$ and $\src(\beta_{k+1})=f(y_k)$.  Then
$ \gamma=j_1(\beta_1)*\dotsm*j_n(\beta_n), $ where
$j_k:X\cong X_k\to \ZZ_n(X,Y,f,g)$ is the structural map, is a well-defined
concatenation.  This yields
\[
	\pi_n(\gamma)
	=\pi_n(j_1(\beta_1)*\dotsm*j_n(\beta_n))
	= j(\beta_1)*\dotsm*j(\beta_n)
	=\alpha
\]
and the claim holds.
\end{proof}

\section{Sequential composition of simple iHDAs}
\label{s:SeqComp}

Let $X_1,\dotsc,X_n\in\IHDA$.  Suppose that $X_1$ is accept simple and
accept proper, that $X_2,\dotsc,X_{n-1}$ are simple and proper, and
that $X_n$ is start simple and start proper.  Let $x_\bot^k$ ($k>1$)
and $x^\top_k$, for $k<n$, be the only start and accept cells of
$X_k$, and assume that $\ev(x^\top_k)=\ev(x_\bot^{k+1})=U_k$.  The
domains of both maps $\Cl(\iota_{\smash[b]{x^\top_k}})$ and $\Cl(\iota_{\smash[b]{x_\bot^{k+1}}})$ are thus
equal to $\sq^{U_k}$ by Lemma~\ref{l:ClosureOfCube}.  We do not require
that $\Iev(x^\top_k)=\Iev(x_\bot^{k+1})$.

The \emph{sequential composition} of $X_1,\dotsc,X_n$ is the HDA
\begin{equation}
\label{e:ZSeqComp}
	X_1*\dotsm* X_n
	=
	\ZZ_n(\Cl(X_k)_{k=1}^n,(\sq^{U_k})_{k=1}^{n-1},\Cl(\iota_{x_\bot^k})_{k=1}^{n-1},\Cl(\iota_{x^\top_k})_{k=1}^{n-1})
\end{equation}
with $(X_1*\dotsm*X_n)_\bot = \Cl(X_1)_\bot$,
$(X_1*\dotsm*X_n)^\top = \Cl(X_n)^\top$.  The assumptions required by
\eqref{e:ZigZag} are thus satisfied: the cubes $\sq^{U_k}$ are acyclic
and the images of $\Cl(\iota_{x_\bot^k})$ and $\Cl(\iota_{x^\top_k})$
are disjoint by properness of the $X_k$.  Note that the sequential
composition considered is an $n$-ary operation: it maps a sequence of
iHDAs to an HDA.  For short, we henceforth write $Z$ for the HDA in
(\ref{e:ZSeqComp}).

We have $\Iev(x^\top_k)=\ilo {S_k} {(U_k)} {U_k}$ and
$\Iev(x_\bot^{k+1})=\ilo {U_k} {(U_k)} {T_k}$ for some
$S_k, T_k\subseteq U_k$, which are not necessarily disjoint.  The
$i_k:\sq^{U_k}\to Z$, $j_k:\Cl(X_k)\to Z$ are structural maps, which
we sometimes omit, regarding $\Cl(X_k)$ and $\sq^{U_k}$ as precubical
subsets of $Z$.  Let
$u_k=i_k([\emptyset|\emptyset])=j_k([x_k^\top;\emptyset,\emptyset])=j_{k+1}([x^{k+1}_\bot;\emptyset,\emptyset])$.

\begin{prop}\label{p:LangIncl}
	$\Lang(X_1)*\dotsm*\Lang(X_n) \subseteq \Lang(X_1*\dotsm*X_n)$.
\end{prop}
\begin{proof}
  If $P\in \Lang(X_1)*\dotsm*\Lang(X_n)$, then
  $P\subsu Q_1*\dotsm* Q_n$ for some $Q_k\in\Lang(X_k)$.  Choose
  accepting paths $\beta_k\in\Qath{\Cl(X_k)}$ such that
  $\ev(\beta_k)=Q_k$, according to
  Proposition~\ref{l:ClPreservesLanguage}.  Since
  $\tgt(\beta_k)=[x_k^\top;\emptyset,\emptyset]$ and
  $\src(\beta_{k+1})=[x^{k+1}_\bot;\emptyset,\emptyset]$ represent the
  same cell $u_k$ in $Z$, the $\beta_k$ can be composed. Then
  \begin{equation*}
    P\subsu Q_1*\dotsm* Q_n = \ev(\beta_1)*\dotsm*\ev(\beta_n)
    =
    \ev(\beta_1*\dotsc*\beta_n) \in \Lang(X_1*\dotsm* X_n). \qedhere
  \end{equation*}
\end{proof}

The \emph{defect} of an accepting path $\alpha\in\Qath{Z}$ is
\[
	\min\left\lbrace \sum_{k=1}^{n-1} (|U_k|-\dim(r_k))
	\mid
	\text{$(r_k)_{k=0}^{n}$ is a checkpoint sequence for $\alpha$}
	\right\rbrace.
\]
The defect of $\alpha$ measures how far $\alpha$ passes from the cells $u_k$.
It is non-negative since
$\dim(\sq^{U_k})=|U_k|$ and equal to $0$ if and only if $r_k=u_k$
for all $1\leq k<n$
since $u_k$ are the only cells in $\sq^U$ of dimension $|U_k|$.
A checkpoint sequence for which the minimum is obtained is called
\emph{maximal}, because it  consists of cells with the maximal dimensions
possible.  

\begin{lem}
  \label{l:DefZero}
  Let $\alpha\in\Qath{Z}$ be a path that is not subsumed by another
  path with a smaller defect. Then $(u_k)_{k=0}^n$ is a checkpoint sequence
  for $\alpha$ (we assume $u_0=\src(\alpha)$, $u_k=\tgt(\alpha)$).
\end{lem}

\begin{proof} 
  Let $(r_k)$ be a maximal checkpoint sequence for $\alpha$.  By
  assumption, $r_k=\delta_{E_k,F_k}(u_k)$ and $E_k,F_k$ are uniquely
  determined by injectivity of $i_k$.  We wish to show that
  $E_k=F_k=\emptyset$.
	
  Let $r_0=\src(\alpha)$, $r_n=\tgt(\alpha)$.  Let
  $\alpha=\beta_1*\dotsm*\beta_n$,
  $\beta_k\in \Qathft{X_k}{r_{k-1}}{r_k}$, be the representation
  determined by $(r_k)_{k=0}^n$ according to
  Lemma~\ref{l:CheckpointSequence}.(c).  By Lemma~\ref{l:Red}, every
  path $\beta_k$ is subsumed by a path $\gamma_k$ of the form
  \[
    (r_{k-1}=[y;C_k\cup A_k,D_k] \arrO{A_k} [y;C_k,D_k]) *\beta'_k* ([z;C_k,D_k] \arrI{B_k}
    [z;C_k,D_k\cup B_k]=r_k).
  \]
  We show that $B_k=\emptyset$.  For $k=n$,
  $[z;C_k,D_k\cup B_k]=[x_n^\top;\emptyset,\emptyset]\in
  \Cl(X_n)^\top$ and thus $C_n=D_n\cup B_n=\emptyset$.  For $k<n$, the
  cell $r_k$ is a face of~$u_k$.  Then $[z;C_k,D_k\cup B_k]$ is a face
  of~$[x^\top_k;\emptyset,\emptyset]$ by injectivity of $j_k$, $z$ is
  a face of~$x^\top_k$ by Lemma~\ref{l:FacesInClosures} and
  $[z;C_k,D_k]$ is a face of~$[x^\top_k;\emptyset,\emptyset]$ by the
  same lemma.  Thus $[z;C_k,D_k]$ is a face of~$u_k$, or equivalently,
  $[z;C_k,D_k]\in \sq^{U_k}$.  If $B\neq \emptyset$, then
  $[z;C_k,D_k]$ is of higher dimension than $r_k=[z;C_k,D_k\cup B_k]$.
  Then,
  \[
  	(r_1,\dotsc,r_{k-1},[z;C_k,D_k],r_{k+1},\dotsc,r_{n-1})
  \]
  is a checkpoint sequence that contradicts the maximality of $(r_k)$
  and shows that $B_k=\emptyset$.  Similarly, $A_k=\emptyset$ follows
  by reversal.  This shows that every $\beta_k$ has the form
  $([z^k_j;C_k,D_k],\varphi^k_j)_{j=1}^{m_k}$, that is, the second and
  third coordinates in $\beta_k$ remain constant.

  Next, we relate the sets $C_k$, $D_k$, $E_k$ and $F_k$.  For every
  $k=1,\dotsc,n-1$,
\begin{equation*}
	r_k
	=
	\delta_{E_k,F_k}([x^\top_k;\emptyset,\emptyset])
	=
	[\delta^0_{E_k\setminus S_k}(x^\top_k);E_k\cap S_k,F_k],
\end{equation*}
and hence $C_k=E_k\cap S_k$, $D_k=F_k$.  Moreover,
\begin{equation*}
	r_k
	=
	\delta_{E_k,F_k}([x_\bot^{k+1};\emptyset,\emptyset])
	=
	[\delta^0_{F_k\setminus T_k}(x_\bot^{k+1});E_k, F_k\cap T_k],
\end{equation*}
and thus $C_{k+1}=E_k$, $D_{k+1}=F_k\cap T_k$.  Therefore,
$C_k\subseteq C_{k+1}$ and $D_k\supseteq D_{k+1}$.  Furthermore,
$C_1=D_1=\emptyset$ since $\src(\beta_1)\in \Cl(X_1)_\bot$ and
$C_n=D_n=\emptyset$ since $\tgt(\beta_n)\in \Cl(X_n)^\top$.  The sets
$C_k$, $D_k$, $E_k$ and $F_k$ are therefore empty,
$r_k=\delta_{E_k,F_k}(u_k)=u_k$ and $\alpha$ has defect $0$.
\end{proof}

\begin{lem}
  \label{l:MaxCheckpointSequence}
  Every accepting path $\gamma\in\Qath{Z}$ is subsumed by a path
  $\alpha$ with checkpoint sequence $(\src(\alpha),u_1,\dots,u_{n-1},\tgt(\alpha))$.  Thus
  $\alpha$ traverses the interiors of the gluing cubes $\sq^{U_k}$.
\end{lem}

\begin{proof}
  Let $\alpha\in\Qath{Z}$ be a path with minimal defect among all
  paths subsuming $\gamma$. The conclusion then follows by
  Proposition\ \ref{l:DefZero}.
\end{proof}

\begin{prop}\label{p:IHDAGluing}
  $\Lang(X_1*\dotsm*X_n)=\Lang(X_1)*\dots*\Lang(X_n)$.
\end{prop}

\begin{proof}
  Let $P\in\Lang(Z)$ and $\gamma\in\Qath{Z}$ be an accepting path
  such that $\ev(\gamma)=P$.  Then $\gamma$ is subsumed by an
  $\alpha\in\Qath{Z}$ with checkpoint sequence $(u_k)_{k=0}^n$ by Lemma\
  \ref{l:MaxCheckpointSequence} and the $\beta_k\in\Qath{\Cl(X_k)}$
  determined by $(u_k)_{k=0}^n$, according to
  Lemma~\ref{l:CheckpointSequence}, are accepting.  Therefore
  \begin{align*}
    P = \ev(\gamma)
    &\subsu \ev(\alpha)\\
    &=
    \ev(j_1(\beta_1)*\dotsm*j_n(\beta_n))\\    
    &=
   	\ev(\beta_1)*\dotsm*\ev(\beta_n)
    \in
    \Lang(\Cl(X_1))*\dots*\Lang(\Cl(X_n))\\
    &=
    \Lang(X_1)*\dots*\Lang(X_n)
\end{align*}
and thus $\Lang(Z)\subseteq\Lang(X_1)*\dots*\Lang(X_n)$.
The converse inclusion follows from Proposition\ \ref{p:LangIncl}.
\end{proof}

After all these preparations we can finally show that gluing
compositions of regular languages are regular.

\begin{proof}[Proof of Proposition {\ref{p:RegConcat}}]
  Let $L$ and $M$ be regular languages.  If they are simple, then
  there exist simple iHDAs $X$ and $Y$ that recognise $L$ and $M$.
  Proposition\ \ref{p:MakingSimpleSourceProper} allows us to assume
  that $X$ is accept simple and accept proper and that $Y$ is start
  simple and start proper.  Proposition\ \ref{p:IHDAGluing} then
  implies that
  \[
    \Lang(X*Y)=\Lang(X)*\Lang(Y)=L*M
  \]
  is regular.  Otherwise, if $L$ and $M$ are not simple, then
  $L=\bigcup_i L_i$ and $M=\bigcup_j M_j$ for simple regular languages
  $L_i$ and $M_j$ by Lemma~\ref{l:RegIsUnionOfSimple}.  In this case, 
\[  
  L*M = \big(\bigcup_i L_i\big)*\big(\bigcup_j M_j\big)
  =\bigcup_{i}\bigcup_j L_i*M_j
\]
	 is regular by Proposition \ref{p:RegUnion}.
\end{proof}

\section{Kleene plus}
\label{s:Spider}

Suppose $X$ is an iHDA such that $\Lang(X)$ is separated.  We
construct an HDA $X^+$ such that $\Lang(X^+)=\Lang(X)^+$.  We wish to
identify accept cells with start cells that have the same events. But
identifying all of them would produce too many accepting paths.  For
every accept cell $y$ we thus add a copy for every start cell
compatible with $y$, and an extra one to replace the original one --
and likewise for start cells. We must then ensure that the iHDA
constructed is proper.  The following fact ensures that we can
construct the cylinder below.

\begin{lem}
  \label{l:SeparatediHDA}
  Let $X\in \IHDA$ and $\Lang(X)$ be separated. Then
  $\im(\iota^X_\bot)\cap \im(\iota_X^\top)=\emptyset$.
\end{lem}

\begin{proof}
  Suppose $y\in\im(\iota^X_\bot)\cap \im(\iota_X^\top)$ for some
  $y\in X[V]$. Then $\delta^1_B(x)=y=\delta^0_A(z)$ for some
  $U,W\in\isq$, $x\in X_\bot[U]$, $z\in X^\top[W]$, $B\subseteq U$,
  $A\subseteq W$ and $U\setminus B=V=W\setminus A$ because
  start/accept cells have only upper/lower faces.  Then every event of
  the ipomset
	\[
		\ev\left(x\arrI{B}y\arrO{A}z\right)
		=
		\ilo U U V*\ilo V W W
	\]
	in $\Lang(X)$ lies in one of the interfaces $U$, $W$ and
        $\Lang(X)$ is not separated.
\end{proof}

Let $G=\{(y,x)\mid y\in X^\top,\; x\in X_\bot,\; \ev(x)=\ev(y)\}$, let
$G_\bot=\{(\bot,x)\mid x\in X_\bot\}$ and
$G^\top=\{(y,\top)\mid y\in X^\top\}$.  Define
\[
	J_\bot = \bigsqcup_{(y,x)\in G\cup G_\bot}
	\isq^{\Iev(x)}
	\qquad\text{ and }\qquad
	J^\top = \bigsqcup_{(y,x)\in G\cup G^\top}
	\isq^{\Iev(y)}.
\]
Further, let $w_\bot=\bigsqcup \iota_x :J_\bot\to X$,
$w^\top=\bigsqcup \iota_y:J^\top\to X$.  The maps $w_\bot$ and
$w^\top$ are similar to the start and accept maps, but more than one
cube can be mapped into a start or accept cell.

We define the \emph{spider} of $X$ as the cylinder
$\Spider(X)=\Cyl(w_\bot,w^\top)$.  It is well-defined because
$\im(w_\bot)\cap \im(w^\top)=\im(\iota^X_\bot)\cap
\im(\iota_X^\top)=\emptyset$, since $\Lang(X)$ is separated and by Lemma~\ref{l:SeparatediHDA}.  The spider is equipped with an initial
inclusion $\widetilde{w}_\bot:J_\bot\to\Spider(X)$ and a final
inclusion $\widetilde{w}^\top:J^\top\to\Spider(X)$ (Lemma\
\ref{l:DoubleCylinderIsProper}).  We write
\begin{align*}
\spO y x&=\widetilde{w}_\bot((y,x),[\emptyset|\emptyset])\in \Spider(X)[\Iev(x)],
\\
\spI y x&=\widetilde{w}^\top((y,x),[\emptyset|\emptyset])\in \Spider(X)[\Iev(y)]
\end{align*}
for $(y,x)\in G\cup G_\bot$ in the upper line and $(y,x)\in G\cup
G^\top$ in the lower line.

Each cell $\spO y x$ is a ``copy'' of $x$ that is to be connected to a
copy $\spI y x$ of $y$.  Such copies serve as start and accept cells
of $\Spider(X)$: we put
$\Spider(X)_\bot=\{\spO \bot x\mid x\in X_\bot\}$ and
$\Spider(X)^\top=\{\spI y \top \mid y\in X^\top\}$.  See Figure
\ref{fi:spider} for an example.

We also use $\Spider(X)$ in combination with other sets of start and
accept cells: for each subset $S,T\subseteq\Spider(X)$,
$\Spider(X)_S^T$ denotes the iHDA with underlying ipc-set
$\Spider(X)$, start cells $S$ and accept cells $T$.  We use a similar
convention for $X$.

\begin{figure}
  \centering
	\begin{tikzpicture}[x=1.4cm, y=1.4cm]
		\filldraw[-,fill=lightgray]
		(-1.5,0) .. controls (-1.3,0.2) ..
		(-1.5,0.4) .. controls (-1.2,0.9) ..
		(-2.0,0.8) .. controls (-1.8,1.0) ..
		(-2.0,1.2) .. controls (-1.7, 1.5) ..
		(-2.0,1.8) .. controls (-1.8,1.9) ..
		(-2.0,2.2) .. controls (0.0,1.8) ..
		(2.0,2.2) .. controls (1.8,1.9) ..
		(2.0,1.8) .. controls (1.7, 1.5) ..
		(2.0,1.2) .. controls (1.8,1.0) ..
		(2.0,0.8) .. controls (1.2,0.9) ..
		(1.5,0.4) .. controls (1.3,0.2) ..
		(1.5,0) .. controls (0,0.3) ..
		(-1.5,0);
		\fill (-1.5,0) circle (2pt); \node[left] at (-1.5,0) {$\spO \bot x$};
		\fill (-1.5,0.4) circle (2pt); \node[left] at (-1.5,0.4) {$\spO \bot w$};
		\fill (1.5,0) circle (2pt); \node[right] at (1.5,0) {$\spI y \top$};
		\fill (1.5,0.4) circle (2pt); \node[right] at (1.5,0.4) {$\spI z \top$};
		\fill (-2.0,0.8) circle (2pt); \node[left] at (-2.0,0.8) {$\spO y x$};
		\fill (-2.0,1.2) circle (2pt); \node[left] at (-2.0,1.2) {$\spO z x$};
		\fill (2.0,0.8) circle (2pt); \node[right] at (2.0,0.8) {$\spI y w$};
		\fill (2.0,1.2) circle (2pt); \node[right] at (2.0,1.2) {$\spI y x$};
		\fill (-2.0,1.8) circle (2pt); \node[left] at (-2.0,1.8) {$\spO y w$};
		\fill (-2.0,2.2) circle (2pt); \node[left] at (-2.0,2.2) {$\spO z w$};
		\fill (2.0,1.8) circle (2pt); \node[right] at (2.0,1.8) {$\spI z w$};
		\fill (2.0,2.2) circle (2pt); \node[right] at (2.0,2.2) {$\spI z x$};
	\end{tikzpicture}
	\caption{A specimen of spider}
        \label{fi:spider}
\end{figure}

\begin{lem}
  \label{l:SpiderLifting}
  For $(y',x)\in G \cup G_\bot$,
  $(y,x')\in G \cup G^\top$,
  the projection map
  \begin{equation*}
    p:\Spider(X)_{\spO {y'} x}^{\spI y {x'}}\to X_x^y
  \end{equation*}
  is a weak equivalence.
\end{lem}

\begin{proof}
  We have $p(\spO {y'}x)=x\in \im(w_\bot)$ and
  $p(\spI y {x'})=y\in \im(w^\top)$.  Since $p$ has the TLP with
  respect to $x$, $y$ by Proposition~\ref{p:DoubleCyllinderLifting},
  it is a weak equivalence by Proposition\
  \ref{p:LiftingProperties}(c).
\end{proof}

Every accepting path $\alpha\in\Qathft X x y$ can therefore be lifted
to a path from any copy of $x$ to any copy of $y$.

Next, we identify ``start copies'' with ``accept copies''.  To achieve
this we need to pass to HDAs.  We abbreviate
$\CSpider(X)=\Cl(\Spider(X))$. We regard cells $\spO y x$ and
$\spI y x$ as cells of $\CSpider(X)$ -- formally, we should write
$[\spO y x;\emptyset,\emptyset]$ and $[\spI y x;\emptyset,\emptyset]$
instead.

Let $\Weld_\bot(X)\subseteq J_\bot$, $\Weld^\top(X)\subseteq J^\top$
be the unions of cubes indexed by $G$ (we omit $G_\bot$ and $G^\top$, respectively).
The closures of  $\Weld_\bot(X)$ and $\Weld^\top(X)$ are naturally isomorphic to
\begin{equation*}
	\Weld(X)=\bigsqcup_{(x,y)\in G} \sq^{\ev(x)}=\bigsqcup_{(x,y)\in G} \sq^{\ev(y)}
\end{equation*}
by Lemma~\ref{l:ClosureOfCube}. Let $f,g:\Weld(X)\to \CSpider(X)$ be
the closures of the restrictions
$f=\Cl(\widetilde{w}_\bot|_{\Weld_\bot(X)})$ and
$g=\Cl(\widetilde{w}^\top|_{\Weld^\top(X)})$.  Since
$\Weld_\bot(X)\subseteq J_\bot$ and $\Weld^\top(X)\subseteq J^\top$
are initial and final subsets, $f$ is an initial inclusion and $g$ a
final one by Lemma~\ref{l:ClPreservesInitialFinal}.  Now we are ready
to define the HDA
\[
	X^+=\VV(\CSpider(X),\Weld(X),f,g).
\]
It remains to show that $\Lang(X^+)=\Lang(X)^+$. The map
\begin{equation}
  \label{e:Weld}
		\bigsqcup_{n\geq 1}
		\ZZ_n(\CSpider(X), \Weld(X),f,g)
		\to
		\VV(\CSpider(X), \Weld(X),f,g)	
\end{equation}
is a weak equivalence by Proposition~\ref{p:ToolHorizontalUnfolding}.
For any sequence $\Gamma=(x_k,y_k)_{k=1}^{n-1}\in G^{n-1}$ and
$U_k=\ev(x_k)=\ev(y_k)$ we define
\begin{align*}
		\Snake_n(X;\Gamma)
		&=
		\ZZ_n(\CSpider(X)_{k=1}^n,
		(\sq^{U_k})_{k=1}^{n-1},
		\Cl(\iota_{\spO {y_k}{x_k}})_{k=1}^{n-1},
		\Cl(\iota_{\spI {y_k}{x_k}})_{k=1}^{n-1})\\
		&=
		\Spider(X)_{[\bot\backslash X_\bot]}^{[y_1/ x_1]}
	*
	\Spider(X)_{[y_1\backslash x_1]}^{[y_2/x_2]}
	*
	\dotsm
	*
	\Spider(X)_{[y_{n-1}\backslash x_{n-1}]}^{[X^\top/\top]},
\end{align*}
where 
$[\bot\backslash X_\bot]=\{[\bot\backslash x]\mid x\in X_\bot\}$,
$[X^\top/\top]=\{[y/\top]\mid y\in X^\top\}$.

Proposition~\ref{p:ToolVerticalUnfolding},
applied to  the decomposition (\ref{e:Weld}),
shows that 
\[
		\bigsqcup_{\Gamma\in G^{n-1}}
		\Snake_n(X;\Gamma)
		\to
		\ZZ_n(\CSpider(X), \Weld(X),f,g)	
\]
is a weak equivalence. As weakly equivalent HDAs have equal languages
(Lemma~\ref{l:EqHDA}), we obtain the following fact.

\begin{lem}
  \label{l:SpLang}
  $\Lang(X^+)=\bigcup_{n\geq 1} \bigcup_{\Gamma\in
    G^{n-1}}\Lang(\Snake_n(X;\Gamma))$.
\end{lem}

Proposition~\ref{p:IHDAGluing} and Lemma~\ref{l:SpiderLifting} then imply
that
\begin{align}
  \label{e:LanguageOfSnakes}
  \Lang(\Snake_n(X;\Gamma)) \hspace*{-5em} \notag \\
  &=
  \Lang\left(\Spider(X)_{\spO \bot {X_\bot}}^{[y_1/ x_1]}\right) *
  \Lang\left(\Spider(X)_{[y_1\backslash x_1]}^{[y_2/x_2]}\right) * \dotsm *
  \Lang\left(\Spider(X)_{[y_{n-1}\backslash
    x_{n-1}]}^{[{X^\top}/\top]}\right) \notag
  \\
  &=
  \Lang(X_{X_\bot}^{y_1})*\Lang(X_{x_1}^{y_2})*\dotsm*\Lang(X_{x_{n-1}}^{X^\top}).
\end{align}

\begin{lem}
\label{l:SwarmLang}
	$\bigcup_{\Gamma\in G^{n-1}}\Lang(\Snake_n(X;\Gamma))=\Lang(X)^{n}$.
\end{lem}

\begin{proof}
	We have
	\begin{align*}
		\Lang(X)^{n} 
		&=\Lang(X)*\dotsm*\Lang(X)\\
		&=
		\Big({\bigcup_{y_1\in X^\top}} \Lang(X_{X_\bot}^{y_1})\Big)
		*
		\Big(\underset{y_2\in X^\top}{\bigcup_{x_1\in X_\bot}} \Lang(X_{x_1}^{y_2})\Big)
		*\dotsm*
		\Big({\bigcup_{x_{n-1}\in X_\bot}} \Lang(X_{x_{n-1}}^{X^\top})\Big)\\
		&=
		\underset{y_1,\dotsc,y_{n-1}\in X^\top}{\bigcup_{x_1,\dotsc,x_{n-1}\in X_\bot}}
		\Lang(X_{X_\bot}^{y_1})*\Lang(X_{x_1}^{y_2})*\dotsm*\Lang(X_{x_{n-1}}^{X^\top})
		\\
		&\overset{(\dagger)}=
		\bigcup_{\Gamma=(x_k,y_k)\in G^{n-1}}
		\Lang(X_{X_\bot}^{y_1})*\Lang(X_{x_1}^{y_2})*\dotsm*\Lang(X_{x_{n-1}}^{X^\top})
		\\
		&\overset{\eqref{e:LanguageOfSnakes}}=
		\bigcup_{\Gamma\in G^{n-1}}
		\Lang(\Snake_n(X;\Gamma)).
	\end{align*}
	In $(\dagger)$ we use the fact that
        $\Lang(X_{x_{k-1}}^{y_k})*\Lang(X_{x_{k}}^{y_{k+1}})=\emptyset$
        whenever $\ev(y_k)\neq \ev(x_k)$.
\end{proof}

\begin{prop}
  \label{p:RegSepPlus}
  The Kleene plus of a separated regular language is regular.
\end{prop}
\begin{proof}
  From Lemmas \ref{l:SpLang} and \ref{l:SwarmLang} it follows that
  \[
  	\Lang(X^+)=\bigcup_{n\geq 1}\Lang(X)^n=\Lang(X)^+.\qedhere
  \]
\end{proof}

Finally we can prove that the Kleene plus of any regular
language is regular.

\begin{proof}[Proof of Proposition \ref{p:RegPlus}]
  Suppose $L$ is regular.  If $L\cap \Id=\emptyset$, then $L^{n}$ is
  separated for sufficiently large $n$ by Lemma\
  \ref{l:SeparatingPower}.  In this case,
  \[
  	L^+ = \bigcup_{i=1}^n L^i \cup \Big( \bigcup_{i=1}^n L^i \Big) *  (L^n)^+
  \]
   is regular by Propositions \ref{p:RegUnion},
  \ref{p:RegConcat} and \ref{p:RegSepPlus}. 
  If $L\cap \Id\ne \emptyset$, then
  \[
  	L^+ =((L\cap \Id) \cup (L\setminus\Id))^+ =(L\cap \Id) \cup
  (L\setminus \Id)^+
  \]
   is regular by
  Propositions~\ref{p:RegUnion} and \ref{p:IdSubtract}.
\end{proof}

As outlined in Section~\ref{s:KleeneTheorem}, the proof of the Kleene
theorem for higher-dimensional automata is now complete.

\section{Conclusion}

Automata accept languages, but higher-dimensional automata have for a
long time been an exception to this rule.  Here, we have proved a
Kleene theorem for HDAs, connecting models to behaviours through an
equivalence between regular and rational languages.

Showing that regular languages are rational was quite direct, while the
converse direction required some effort.  One reason is that HDAs may
be glued not only at states, but also at higher-dimensional cells.
This led us to consider languages of ipomsets and to equip HDAs with
interfaces, yielding iHDAs.  After showing that HDAs and iHDAs
recognise the same languages, we used constructions inspired by
topology to glue (i)HDAs and show that rational operations on
languages can be reflected by operations on them.

Kleene theorems build bridges between machines and languages, and
there is a vast literature on this subject.  In non-interleaving
concurrency, one school considers Mazurkiewicz trace
languages. Zielonka introduces asynchronous automata and shows that
languages are regular if and only if they are recognisable
\cite{DBLP:journals/ita/Zielonka87}.  Droste's automata with
concurrency relations have similar properties
\cite{DBLP:conf/icalp/Droste94}.  Yet not all rational trace languages
that are generated from singletons using union, concatenation and
Kleene star are recognisable \cite{DBLP:journals/mst/ChoffrutG95}.
Trace languages use a binary notion of independence and already
2-dimensional HDAs may exhibit behaviour that cannot be captured by
trace languages \cite{Goubault02-cmcim}.

Another school studies Kleene theorems for series-parallel pomset
languages and automata models for these, such as branching and pomset
automata \cite{DBLP:journals/tcs/LodayaW00,
  DBLP:journals/jlp/KappeBLSZ19}, and Petri automata
\cite{DBLP:journals/lmcs/BrunetP17}. Series-parallel pomsets are
incomparable to the interval orders accepted by Petri nets or
HDAs~\cite{DBLP:books/sp/Vogler92,
  DBLP:journals/iandc/FahrenbergJSZ22}.

HDAs have been developed first of all with a view on operational,
topological and geometric aspects of concurrency, see
\cite{DBLP:books/sp/FajstrupGHMR16} and the extensive bibliography in
\cite{DBLP:journals/tcs/Glabbeek06}.  But languages for HDAs have only
been introduced recently \cite{Hdalang}. Topological intuitions have
also guided our present work, for example in the cylinder
construction.

Our formalisation of (i)HDAs as presheaves on a category of labelled
posets opens up connections to presheaf automata
\cite{DBLP:journals/jcss/Sobocinski15}, coalgebra, and open maps
\cite{DBLP:journals/iandc/JoyalNW96}, which we intend to explore.
Finally, our introduction of iHDA morphisms akin to cofibrations and
trivial fibrations hints at factorisation systems.  Weak factorisation
systems and model categories have been considered in a bisimulation
context, for example in \cite{KurzR05-weak}, and we wonder about the
connection.

\subsection*{Acknowledgments}

We are very grateful to the anonymous
reviewers who helped us to improve the presentation of our results
significantly. 

\bibliographystyle{alphaurl}
\bibliography{mybib}

\appendix

\section{Definitions of HDAs}
\label{ap:hda}

Precubical sets and HDAs appear in different incarnations in the
literature, all of them more or less equivalent.  We discuss some of
these in this Appendix to relate them with our own approach, but make
no claim as to completeness.

\subsection*{Precubical sets}

Precubical sets à la Grandis \cite{GrandisM03-Site,
  book/Grandis09} are presheaves on a small category $\sqgra$, defined
by the following data:
\begin{itemize}
\item objects are $\{0, 1\}^n$ for $n\ge 0$;
\item elementary coface maps $d_i^\nu: \{0, 1\}^n\to \{0, 1\}^{n+1}$,
  for $i=1,\dotsc, n+1$ and $\nu=0, 1$, are given by
  $d_i^\nu(t_1,\dotsc, t_n)=(t_1,\dotsc, t_{i-1}, \nu, t_i,\dotsc,
  t_n)$.
\end{itemize}
Elementary coface maps compose to coface maps
$\{0, 1\}^m\to \{0, 1\}^n$ for $n\ge m$ in the standard way.

$\sqgra$-sets, that is, elements $X\in \Set^{\sqgra^\op}$, are then
graded sets $X=\{X_n\}_{n\ge 0}$, where $X_n=X[\{0, 1\}^n]$, together
with face maps $X_n\to X_m$ for $n\ge m$.  The elementary face maps
are denoted $\delta_i^\nu=X[d_i^\nu]$. They must satisfy the
\emph{precubical identity}, for any $\nu,\mu\in\{0,1\}$ and $i<j$:
\begin{equation}
  \label{e:precub}
  \delta_i^\nu \delta_j^\mu = \delta_{j-1}^\mu \delta_i^\nu.
 \end{equation}

 This description of $\sqgra$-sets may be taken as a definition
 without using presheaves.  For example, van Glabbeek
 \cite{DBLP:journals/tcs/Glabbeek06} defines a precubical set
 $Q=(Q,s,t)$ as a family of sets $(Q_n)_{n\geq 0}$ and maps
 $s_i:Q_n\to Q_{n-1}$, $1\leq i\leq n$, such that
 $\alpha_i\circ \beta_j =\beta_{j-1}\circ\alpha_i$ for all
 $1\leq i<j\leq n$ and $\alpha,\beta\in\{s,t\}$. This is equivalent to
 the above.

 We have previously introduced another base category, $\fullsqz$,
 defined as follows \cite{Hdalang}:
\begin{itemize}
\item objects are totally ordered sets $(S, {\intord_S})$;
\item morphisms $S\to T$ are pairs $(f, \epsilon)$, where
  $f: S\hookrightarrow T$ is an order preserving injection and
  $\epsilon: T\to \{0, \exec, 1\}$ satisfies
  $f(S)=\epsilon^{-1}(\exec)$.
\end{itemize}
The element $\exec$ stands for ``active'', a notation previously used
by van Glabbeek. Writing $A=\epsilon^{-1}(0)$ and $B=\epsilon^{-1}(1)$
makes the above notion of morphism equivalent to the triples
$(f, A, B)$ consisting of $f:S\hookrightarrow T$ (order preserving and
injective) and $A, B\subseteq T$ such that $T=A\sqcup f(S)\sqcup B$
(disjoint union).  Except for the labels, this is our definition of
$\fullsq$ in Section \ref{s:hdas}.

Using this definition, it can be shown that the full subcategory of
$\fullsqz$ spanned by the objects $\emptyset$ and $\{1,\dotsc, n\}$
for $n\ge 1$ is skeletal and equivalent to $\fullsqz$ \cite{Hdalang}.
Moreover, this subcategory, $\sqz$, is isomorphic to $\sqgra$, and the
presheaf categories on $\fullsqz$ and on $\sqz$ (and thus also on
$\sqgra$) are uniquely naturally isomorphic.  It is clear that $\sqz$
is a representative of the quotient of $\fullsqz$ with respect to
isomorphisms, so, except for the labelling, this is again our category
$\sq$ from Section \ref{s:hdas}.

The advantage of $\fullsqz$ and $\fullsq$ over the skeletal versions
is that the precubical identity \eqref{e:precub} is automatic and that
there is a built-in notion of events and actions, that is, in a
$\fullsqz$-set $X$, each cell $x\in X[U]$ has events $U$.

\subsection*{HDAs}

Higher-dimensional automata are $\Sigma$-labelled precubical sets with
specified start and accept cells.  The labelling may be obtained using
the labelling object $\bang \Sigma$ \cite{Goubault02-cmcim}.  This is
the precubical set with $\bang \Sigma_n=\Sigma^n$ and
$\delta_i^\nu((a_1,\dotsc, a_n))=(a_1,\dotsc, a_{i-1}, a_{i+1},\dotsc,
a_n)$. A labelled precubical set is then a precubical map
$X\to \bang \Sigma$, that is, an object of the slice category of
precubical sets over $\bang \Sigma$.

Each labelling function $\lambda: X\to \bang \Sigma$ induces a function
$\lambda_1: X_1\to \Sigma$ satisfying
$\lambda_1(\delta_1^0(x))=\lambda_1(\delta_1^1(x))$ and
$\lambda_1(\delta_2^0(x))=\lambda_1(\delta_2^1(x))$ for all
$x\in X_2$.  Conversely, each such function extends uniquely to a
precubical map $X\to \bang \Sigma$ \cite[Lemma 14]{Hdalang}, so that
$\lambda_1$ may be taken as the primary definition instead.  This is
the approach in \cite{DBLP:journals/tcs/Glabbeek06}, where HDAs are
defined as precubical sets $Q$ equipped with functions
$\lambda_1\to \Sigma$ such that $\lambda_1(s_i(q))=\lambda_1(t_i(q))$
for all $q\in Q_2$ and $i=1,2$, and subsets of start and accept states
$I,F\subseteq Q_0$.

Regarded as a presheaf, $\bang \Sigma( S)=\Set( S, \Sigma)$. Hence
$\bang \Sigma$ is representable in $\Set$ via the forgetful functor
$\fullsqz\to \Set$ \cite{Hdalang}.  Labels can thus be integrated into
the base category, which turns $\fullsqz$ into our category $\fullsq$,
with labelled totally ordered sets as objects.  Using $\fullsq$
instead of $\fullsqz$ allows working in a labelled setting ab
initio instead of taking a slice category.

To summarise, starting from an HDA $X$ as defined in this article, an
HDA $(Q,s,t,\lambda_1,I,F)$ à la van Glabbeek
\cite{DBLP:journals/tcs/Glabbeek06} can be obtained as follows:
\begin{itemize}
\item $Q_n=\bigsqcup_{U\in\sq,\;|U|=n} X[U]$.
\item If $x\in X[U]$, then $s_i(x)=\delta^0_u(x)$ and $t_i(x)=\delta^1_u(x)$,
  where $u\in U$ is the $i$-th smallest element of $U$ in the order
  $\intord_U$.
\item If $x\in X[U]\subseteq Q_1$ with
  $U=(\{e\},\emptyset,\lambda(e)=a)$, then $\lambda_1(x)=a$.
\item $I=X_\bot$, $F=X^\top$.
\end{itemize}

Conversely, let $(Q,s,t,\lambda_1,I,F)$ be an HDA à la van Glabbeek.
Then there are unique labelling functions $\lambda_n:Q_n\to \Sigma^n$
that satisfy $\lambda_{n-1}(\alpha_i(q))=\delta_i(\lambda_{n}(q))$
\cite[Lemma\ 14]{Hdalang}, where $\alpha\in\{s,t\}$ and $\delta_i$
discards the $i$-th element of a sequence.  We can then construct an HDA $X$ in
the sense of this article as follows:
\begin{itemize}
\item $X[U]=\{q\in Q_n\mid \lambda_n(q)=U\}$ for $U\in \fullsq$ and
  $|U|=n$.
\item $\delta^0_a(q)=s_i(q)$ and $\delta^1_a(q)=t_i(q)$ for
  $q\in X[U]$ and $a\in U$ the $i$-th smallest element of $U$ in the
  order $\intord_U$.  The remaining face maps are composites of these.
\item $X_\bot=I$ and $X^\top=F$.
\end{itemize}

\end{document}